%% file: main.tex
\documentclass[11pt,preprint]{elsarticle}
\usepackage[utf8]{inputenc}

\usepackage{microtype}
\usepackage{xspace}
\usepackage{url}
\biboptions{sort&compress}
\biboptions{numbers,sort&compress}
\usepackage[margin=1in]{geometry}
\usepackage{amsmath,amsthm,amssymb}
\usepackage{bm}
\usepackage{tcolorbox}
\usepackage{enumitem}
\usepackage{subcaption}
\usepackage{float}
\usepackage{tabularx}
\usepackage{booktabs}
\usepackage{graphicx}
\usepackage{mdframed}
\usepackage{adjustbox}
\usepackage{hyperref}\hypersetup{pdfborder={0 0 0}}
\usepackage{miller}
\usepackage[capitalise]{cleveref}
\usepackage{multicol}
\usepackage{placeins}

\definecolor{C0}{HTML}{1F77B4}
\definecolor{C1}{HTML}{FF7F0E}
\definecolor{C2}{HTML}{2ca02c}
\definecolor{C3}{HTML}{d62728}
\definecolor{C4}{HTML}{9467bd}
\definecolor{C5}{HTML}{8c564b}

\usepackage[commentmarkup=todo,authormarkup=none,final]{changes}
\definechangesauthor[color=C0]{brandon}
\definechangesauthor[color=C1]{dan}
\definechangesauthor[color=C2]{ilia}
\definechangesauthor[color=C3]{brendon}
\definechangesauthor[color=C4]{ellad}
\definechangesauthor[color=C5]{ryan}

\definechangesauthor[color=C0]{R2}
\makeatletter
\@namedef{Changes@AuthorColor}{C2}
\colorlet{Changes@Color}{C2}
\renewcommand{\Changes@Markup@comment}[3]{%
  \IfStrEq{\Changes@optioncommentmarkup}{todo}%
		{\colorlet{Changes@todocolor}{authorcolor}\todo[color=Changes@todocolor!10, bordercolor=Changes@todocolor, linecolor=Changes@todocolor!70, nolist]{\textbf #1}}{}}
\makeatother
\renewcommand\_{\textunderscore\allowbreak}
\graphicspath{{./figures/}{./results/}{./resultsold/}}

\def\etal{{et al.}\xspace}                      
\def\abinitio{{ab initio}\xspace}

\def\IP{IP\xspace}                              
\def\IPs{IPs\xspace}
\def\kimmodel{Model\xspace}
\def\kimmodels{Models\xspace}
\def\kimmodeldriver{Model Driver\xspace}

\def\kimtest{Test\xspace}
\def\kimtests{Tests\xspace}
\def\kimtestdriver{Test Driver\xspace}
\def\kimtestdrivers{Test Drivers\xspace}
\def\kimtestdriverscap{Test Drivers\xspace}    

\makeatletter
\def\ps@pprintTitle{%
  \let\@oddhead\@empty
  \let\@evenhead\@empty
  \def\@oddfoot{\reset@font\hfil\thepage\hfil}
  \let\@evenfoot\@oddfoot
}
\makeatother

\begin{document}
\title{Automated determination of grain boundary energy and potential-dependence using the OpenKIM framework}
\author[umn]{Brendon~Waters}
\author[umn]{Daniel~S.~Karls}
\author[umn]{Ilia~Nikiforov}
\author[umn]{Ryan~S.~Elliott}
\author[umn]{Ellad~B.~Tadmor}
\author[uccs]{Brandon~Runnels\corref{cor1}}
\cortext[cor1]{Corresponding author}
\address[umn]{Department of Aerospace Engineering and Mechanics, University of Minnesota, Minneapolis, MN, USA}
\address[uccs]{Department of Mechanical and Aerospace Engineering, University of Colorado, Colorado Springs, CO, USA}

\begin{abstract}
We present a systematic methodology, built within the Open Knowledgebase of Interatomic Models (OpenKIM) framework (\url{https://openkim.org}), for quantifying properties of grain boundaries (GBs) for arbitrary interatomic potentials (\IPs), GB character, and lattice structure and species.
The framework currently generates results for symmetric tilt GBs in cubic materials, but can be readily extended to other types of boundaries.
In this paper, GB energy data are presented that were generated automatically for 
Al, Ni, Cu, Fe, and Mo with 225 \IPs; the system is installed on openkim.org and will continue to generate results for all new \IPs uploaded to OpenKIM.
The results from the atomistic calculations are compared to the lattice matching model, which is a semi-analytic geometric model for approximating GB energy.
It is determined that the energy predicted by all \IPs (that are stable for the given boundary type) correlate closely with the energy from the model, up to a multiplicative factor.
It thus is concluded that the qualitative form of the GB energy versus tilt angle is dominated more by geometry than the choice of \IP, but that the \IP can strongly affect the energy level.
The spread in GB energy predictions across the ensemble of \IPs in OpenKIM provides a measure of uncertainty for GB energy predictions by classical \IPs.
\end{abstract}


\ifdefined\usetodonotes
\begin{tcolorbox}[title={\bf Todonote convention}]
  Please use the following convention when making notes:
  \begin{center}\verb-\yourname{Addresseename, we need to XYZ}-\end{center}
  In other words use your name macro to make any comments, then address specific people in the text.
  For instance, if Brandon wants Bob to run more results, he should express this in the following way.
  \begin{center}
    \verb-\brandon{Bob, we need to run more results}-
  \end{center}
\end{tcolorbox}
\listoftodos
\setcounter{page}{0}
\fi

\maketitle

\section{Introduction}

As exploration of the behavior of materials progresses towards ever smaller scales, the influence of grain boundaries (GBs) on their mechanical properties has become increasingly relevant.
In particular, the phenomenon of GB anisotropy, the dependence of GB excess energy~\cite{sutton1995interfaces} on the relative crystallographic orientation of the two adjacent grains, has been identified as a non-trivial effect in processes such as plasticity~\cite{van2013grain}, twinning and phase transformation~\cite{Beyerlein2013}, and solidification~\cite{Chen1998}.
However, while the orientation relationship between the grains (embodied in the coincident site lattice and $\Sigma$ value of the boundary) was originally thought to be predominant in determining the energy of a GB~\cite{brandon1964field}, it has become clear that it is strongly dependent upon the interface inclination, as well.
That is, it has been shown that the excess energy exhibits a marked dependence upon all five microscopic degrees of freedom that define a GB~\cite{smith1980can,sutton2015}, which we collectively refer to as its \emph{character}.

Because of the eminent role played by GBs in micromechanics, and because a knowledge of their stability will be critical in developing methods for GB engineering~\cite{watanabe1999control}, it is necessary to gather accurate data that relates the excess energy of a GB to its character.
Grain boundaries have been a subject of study for many decades.
The seminal analysis of GB energy as a function of misorientation was by Read~\cite{read1950dislocation}, resulting in the well-known Read--Shockley equation.
This was followed by insights from other investigators, such as the Frank--Bilby equation~\cite{frank1950resultant}
and the O-Lattice model~\cite{bollmann2012crystal}, all of which are restricted to GBs that are interpretable as arrays of geometrically necessary dislocations.
For the general case, many attempts have been made to capture GB behavior accurately at all points in GB character space
by means of analytical or semi-analytical models.
Some methods, such as that of Bulatov \etal~\cite{bulatov2014grain}, rely on an interpolation-based approach for constructing a model informed by GB data~\cite{olmsted2009survey} that reproduces the dependence of excess energy on orientation over a wide range of known GB configurations.
It has also been argued in \cite{bulatov2014grain} that a geometry or crystallography-driven method is sufficient for approximating GB energy up to a scaling factor.
Udler and Seidman suggested that GB energy scales with elastic moduli, implying a possible relationship; however, this has yet to be verified \cite{udler1996grain}. 
In a similar vein, some machine learning models (as well as potentials \cite{nishiyama2020application}) have proven useful in estimating GB energy \cite{ye2022universal,patala2019understanding}, although their ability to predict irregular GB energy features is debatable.

Advances from focused ion beam tomography (FIB)~\cite{inkson20013d} to nondestructive full 3D tomography~\cite{midgley20033d} provide a means for determining the effect of GB energy on microstructural evolution.
However, aside from a modest collection of observations of excess energy for specific subsets of GB character~\cite{gjostein1959absolute,miura1994temperature}, experimental data is scarce.
The lack of experimental GB energy data has been mitigated to some extent by advancements in computational hardware that have made it possible to perform realistic atomistic simulations of GBs with sufficient complexity to determine energetic trends.
Due to the computational cost, such simulations are typically performed using interatomic potentials (\IPs): approximate functional forms fitted to \abinitio and experimental data.
\IPs effect a significant reduction of order by solving only for atomic nuclei positions without resolving electron density.
The resulting simulations are more computationally efficient, but require that the functional form and parameters of the \IP be carefully selected.

Multiple wide-ranging atomistic studies of GB energy using \IPs have been conducted.
One of the first broad surveys was performed by Wolf~\cite{wolf1989structureI,wolf1989structureII,wolf1990structureIII,wolf1990structureIV,wolf1989correlationI,wolf1990correlationII}, who considered symmetric and asymmetric tilt and twist boundaries in face-centered cubic (fcc) and body-centered cubic (bcc) materials.
Still, exploration of GB energy for various materials, \IPs, and crystal structures is a topic of ongoing research~\cite{hahn2016symmetric}.
Of particular note are the evident similarities that have been identified between the relaxed GB energies of various materials with identical crystal lattice types, implying that their excess energy surfaces are the same up to a material-specific scaling constant~\cite{olmsted2009survey}.

Because the precise effect of \IP choice on GB energy is generally missing from current GB studies, the primary contribution of this work is to specifically analyze the correlation, if any, between \IPs and the relaxed GB energy they predict.
Doing so requires a systematic framework for cataloging the massive sets of data that result from this broad sweep of possible GB character and, to this end, we leverage the Open Knowledgebase of Interatomic Models (OpenKIM, KIM)~\cite{tadmor:elliott:2011,tadmor2013nsf}.
This system automates the process of computing the predictions of the many \IPs stored within it by allowing users to upload atomistic simulation codes that calculate material properties of interest.
We have developed this capability for computing the GB energy-versus-tilt angle relation for symmetric tilt GBs in cubic materials.
In this article, we report on the results for aluminum (Al), nickel (Ni), copper (Cu), iron (Fe), and molybdenum (Mo) simulated using 225 \IPs (as of December 2022).
Additional results are continuously generated as \IPs and material systems are added to OpenKIM, and current results are available online at \url{https://openkim.org}.

The remainder of this paper is structured as follows:
\Cref{sec:openkim} provides an overview of the OpenKIM framework as it relates to the problem of systematic GB energy calculations.
\Cref{sec:gbtestdriver} provides the details of the algorithm used to compute the GB energy and its implementation.
In \Cref{sec:results}, results are presented and discussed for the above mentioned systems.
The implications and limitations of the work are reviewed in \Cref{sec:conclusion}.

\section{The OpenKIM framework and grain boundary energy calculations}\label{sec:openkim}

OpenKIM~\cite{tadmor:elliott:2011,tadmor2013nsf} is a cyberinfrastructure hosted at \url{https://openkim.org} for archiving computer implementations of \IPs (referred to as \emph{\kimmodels} in KIM terminology) and testing their predictions for different material properties.
All KIM \kimmodels conform to an application programming interface (API)~\cite{kimapi} that allows them to be used seamlessly and without alteration with a number of major simulation packages that conform to the KIM API (for a current list, see \url{https://openkim.org/projects-using-kim}).
These \kimmodels are automatically coupled to a large number of ``\kimtests'' in the system that compute physical properties of interest.
Some examples for crystalline systems are cohesive energy, equilibrium lattice constant, phonon dispersion, stacking fault energy, surface energy, thermal expansion, and vacancy formation energy.
The \kimtest--\kimmodel matching and calculation process is handled by the ``KIM Processing Pipeline''~\cite{kim-pipeline:2020, kim-hpc-pipeline:2022}, which creates corresponding jobs and distributes them to high-performance computing clusters whenever a new \kimmodel or \kimtest is uploaded.
To reduce compute time and promote reusable code, \kimtests can access the results of other \kimtests in the system for properties that they depend on, such as the equilibrium lattice constant of the crystal or its cohesive energy.
These dependencies are handled automatically by the pipeline when determining the order in which sequences of \kimtest--\kimmodel pairs are executed.

The results of all compatible \kimtest--\kimmodel couplings run by the pipeline are stored online in the publicly accessible OpenKIM repository.
These results can be viewed through a user-extendable visualization system
integrated into the OpenKIM framework, as well as directly through standardized web queries~\cite{kimquery}.
This provides a wealth of information on the predictions of stored \kimmodels for a large number of material properties.
New \kimmodels and \kimtests are uploadable by the user community, making the OpenKIM system an evolving, adaptive system.

An individual \kimtest in the OpenKIM system is a fully specified calculation, e.g.\ the cohesive energy of Al in the fcc structure, or the ($111$) surface energy of Fe in the bcc structure.
In practice, it is more efficient to work with ``\kimtestdrivers''.
\kimtestdriverscap are analogous to pure abstract classes in object oriented programming; they are designed to be as general as possible to reduce code redundancy.
For example, a cohesive energy \kimtestdriver could take in as input the crystal structure (lattice vectors, basis atoms, and species) and compute the energy-per-unit cell for this system for any \kimmodel with which it is coupled.
Each set of inputs to the \kimtestdriver then constitutes a \kimtest.
A \kimtestdriver can be a stand-alone computer program written in any language supported by the KIM API (C, C++, Fortran 90 and greater) or can be a script that runs an external simulation code (called a ``simulator'' in KIM), such as the LAMMPS molecular dynamics (MD) package~\cite{thompson2022lammps}.
Consistent with this philosophy of modularity, KIM \kimmodels are handled similarly.
An individual KIM \kimmodel may either consist of a self-contained program or as a parameter file that is read by a \emph{\kimmodeldriver}.


To study GB energetics, we have developed a KIM \kimtestdriver that computes the GB energy versus angle for symmetric tilt GBs in cubic materials~\cite{TD410381120771003} using LAMMPS.
This \kimtestdriver, and several dozen \kimtests which use it, are installed within the OpenKIM system and, as mentioned above, will therefore automatically be run with any new \kimmodels uploaded to the system.
The specific algorithm used by the \kimtestdriver to compute the tilt GB energy is described in the next section.

\section{Symmetric tilt GB energy \kimtestdriver algorithm}\label{sec:gbtestdriver}

The symmetric tilt GB \kimtestdriver reads in the species, crystal structure, tilt axis and range of tilt angles, a spatial resolution for the grid search over translations along the GB in units of the lattice constant of the material, and the maximum allowable length of any dimension of the periodic cell.
Note that the interface inclination is not read in as a parameter because the current \kimtestdriver is intended to model only symmetric tilt GBs.
The equilibrium lattice constant and cohesive energy of the specified lattice are then automatically imported from the results of other \kimtests that were previously run against the \IP being tested, as well as interatomic energies across a range of non-equilibrium lattice spacings (i.e.\ the cohesive energy versus lattice spacing curve), which are used to determine the minimum atomic separation where the interatomic energy of the lattice exceeds the single-atom energy. Any pairs of atoms in the initial unrelaxed GB configuration that are within this distance from a neighbor have one member automatically deleted to avoid unphysical configurations.\footnote{For a typical \IP, the energy of a single isolated atom is zero, and hence the deletion criterion applies to pairs of atoms whose distance is less than the lattice constant at which the lattice energy becomes positive at short distances.}
The following section describes the algorithm used in the \kimtestdriver for constructing LAMMPS simulations.

\subsection{Definition of grain boundary excess energy}
Following established definitions, (cf.~\cite{gibbs1928collected,sutton1995interfaces,arya2003structure,jiang2005prediction,runnels2016analytical}) we use the following definition for grain boundary excess energy between two grains, denoted ``+'' and ``-''\deleted[id=R2,comment={2.1}]{}
\begin{align}
  \gamma = \lim_{L\to\infty}\frac{1}{L^2}\Big(\lim_{H\to\infty}\big(E_{\text{tot}}(\chi^+_{L,L,H},\chi^-_{L,L,H}) - (N^+_{L,L,H} + N^-_{L,L,H})E_{\text{coh}}\big)\Big),
\end{align}
where $\chi^+_{L,L,H},\chi^-_{L,L,H}$ are the relaxed atomic positions contained within the simulation domain, $[-\frac{L}{2},\frac{L}{2}]^2\times[-\frac{H}{2},\frac{H}{2}]$, and $N^+_{L,L,H}+N^-_{L,L,H}$ is the total number of atoms contained in the simulation domain.
$E_{\text{coh}}$ is the cohesive energy of the crystals in their ground state.
If the boundary is periodic within some finite cell, $[-\frac{L_1}{2},\frac{L_1}{2}]\times[-\frac{L_2}{2},\frac{L_2}{2}]$, the energy reduces to\deleted[id=R2,comment={2.1}]{}
\begin{align}
  \gamma = \frac{1}{L_1L_2}\Big(\lim_{H\to\infty}\big(E_{\text{tot}}(\chi^+_{L_1,L_2,H},\chi^-_{L_1,L_2,H}) - (N^+_{L_1,L_2,H} + N^-_{L_1,L_2,H})E_{\text{coh}}\big)\Big).
\end{align}
In the limit as $H$ becomes large, 
then the energy is sufficiently approximated by\deleted[id=R2,comment={2.1}]{}
\begin{align}\label{eq:excessenergy}
  \gamma \approx \frac{1}{L_1L_2}\Big(E_{\text{tot}}(\chi^+_{L_1,L_2,H},\chi^-_{L_1,L_2,H}) - (N^+_{L_1,L_2,H} + N^-_{L_1,L_2,H})E_{\text{coh}}\Big),
\end{align}
where $E_{\text{tot}}$ is computed via atomistic methods.

Nonnegativity of $\gamma$ requires that the first term in \cref{eq:excessenergy} be greater than the sum of the second two.
$E_{\text{coh}}$ should be the energy per atom of the crystal in its optimal lattice configuration, guaranteeing that the boundary configuration will incur an increase of energy.
However, if a cohesive energy for a higher-energy lattice configuration is chosen, e.g.\ a bcc cohesive energy used for a material with an fcc ground state, then the minimization may result in a phase transformation from the unstable lattice to the stable one.
Such a configuration would result in a meaningless, possibly negative value of $\gamma$, and would be said to be an unstable configuration.

In the present work, GB energy computations are considered comprehensively for all \IPs designed for a given material, regardless of actual ground state.
As a result, several unstable configurations are included that may be systematically dismissed.

\subsection{Domain generation}

\begin{figure}
  \begin{minipage}{0.35\linewidth}
    \centering
    \includegraphics{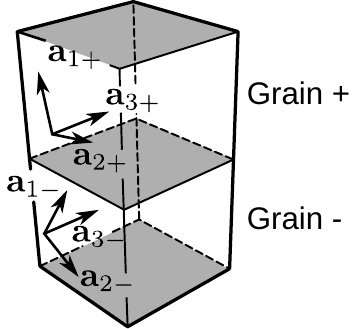}
    \caption{
      Computational simulation domain for a symmetric tilt GB.
      Periodicity requires the creation of two GBs, the first of which is shown in the center and the second of which is formed by effectively adjoining the top and bottom of the simulation domain.
    }
    \label{fig:periodiccella}
  \end{minipage}\hfill
  \begin{minipage}{0.6\linewidth}
    \centering
    \includegraphics{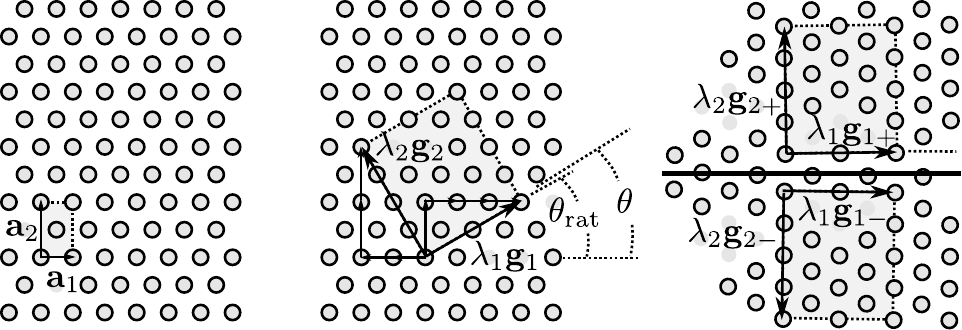}
    \caption{
      For the set of FCC [111] tilt boundaries, given a value for $\theta$, we seek the nearest angle $\theta_\mathrm{rat}$ so that a reasonably sized repeating cell can be constructed.
      The precise relative position of the two grains is determined by grid search, and overlapping atoms are deleted as described below.
    }
    \label{fig:periodiccellb}
  \end{minipage}
\end{figure}

When setting up an atomistic GB simulation, a periodic computational representative domain is generally used.
In order to respect this periodicity, a typical calculation of the GB energy necessitates the creation of two GBs (see \cref{fig:periodiccella}).
The algorithm implemented in the \kimtestdriver for generating the repeating cell for a prescribed tilt angle is based on the significant body of work already done on the subject (see, for example,~\cite{tschopp2007structure,tschopp2008atomistic,tschopp2007asymmetric}).
Let $\bm{g}_{1\pm},\bm{g}_{2\pm},\bm{g}_{3\pm}\in\mathbb{R}^3$ be the unit vectors defining the orientation of the periodic cell, and let $\bm{a}_1,\bm{a}_2,\bm{a}_3\in\mathbb{R}^3$ define the rectangular unit cell such that $\bm{a}_3$ is coincident with the axis of rotation.
(Note that this is not generally the same as the unit cell, nor is it unique for every choice of tilt axis.)
The vectors defining the periodic cell used to represent each grain, shown in \cref{fig:periodiccellb}, are defined as $\{\lambda_i \bm{g}_{i\pm}\}$,
which we express in terms of $\{\bm{a}_i\}$ as\begin{align}
  \lambda_1 \bm{g}_{1\pm} &= m\,\bm{a}_1\pm n\,\bm{a}_2, &
  \lambda_2 \bm{g}_{2\pm} &= \mp\,p\,\bm{a}_1 + q\,\bm{a}_2,
\end{align}
where $m,n,p,q\in\mathbb{Z}$ are integral coefficients, and we note that $\bm{a}_3=\bm{g}_3$ for tilt boundaries.
The task at hand is to compute the integral coefficients given a tilt angle $\theta$.
Generally, an arbitrary angle $\theta$ will be irrational (corresponding to periodically incommensurate boundaries), so it is necessary to locate the ``closest'' rational angle that defines an acceptably small periodic cell.
This is done using the following equations, derived by an elementary geometric calculation:
\begin{equation}
  (m,n) = \operatorname{rat}_d\Bigg[\frac{|\bm{a}_1|}{|\bm{a}_2|}\tan\theta\Bigg],
\qquad
  (p,q) = \operatorname{rat}_d\Bigg[\frac{|\bm{a}_2|}{|\bm{a}_1|}\tan\theta\Bigg],
\end{equation}
where $\operatorname{rat}_d:\mathbb{R} \to \mathbb{Z}\times\mathbb{Z}$ is a rationalization function written in terms of a ``maximum denominator'' $d$, defined as
\begin{equation}
  \operatorname{rat}_d(\xi) = \underset{a,b\in\mathbb{Z},|a|\le d}{\operatorname{arg\,min}}\,\Big|\frac{b}{a} - \xi\Big|.
  \label{eq:ratfcn}
\end{equation}
That is, $d$ is an integer that determines the maximum allowable size of the periodic cell in terms of integer multiples of the original cell width.
Increasing $d$ increases the density of data points in grain boundary space.
The rational angle yielded using the formulae above is given by
\begin{align}
  \theta_\mathrm{rat} = \tan^{-1}\Bigg[\frac{n|\bm{a}_2|}{m|\bm{a}_1|}\Bigg]
  \overset{!}{=} \tan^{-1}\Bigg[\frac{p|\bm{a}_1|}{q|\bm{a}_2|}\Bigg],
  \label{eq:enforcedequality}
\end{align}
where $\overset{!}{=}$ denotes enforced equality.
The enforced equality in \cref{eq:enforcedequality} is critical because, especially for non-cubic periodic cells, rationalizations are not necessarily equal for both axes.
The dimensions of the final unit cell used to construct the simulation domain are computed in terms of $m,n,p,q,$ and $\theta_\mathrm{rat}$ to be
\begin{align}
  \lambda_1 &= m|\bm{a}_1|\cos\theta + n|\bm{a}_2|\sin\theta, \\
  \lambda_2 &= p|\bm{a}_1|\sin\theta + q|\bm{a}_2|\cos\theta,
\end{align}
where $\lambda_3=1$ for tilt boundaries.
The $\Sigma$ value of the GB, a common classification used when describing tilt GBs, can be computed directly from these dimensions:
\begin{align}
  \Sigma &= \min_{a\in\mathbb{Z}}\frac{1}{2^a}\operatorname{int}\Bigg[\frac{\lambda_1\lambda_2\lambda_3}{|\bm{a}_1||\bm{a}_2||\bm{a}_3|}\Bigg] &
  &\text{ s.t. }&
                  \operatorname{int}\Bigg[\frac{\lambda_1\lambda_2\lambda_3}{|\bm{a}_1||\bm{a}_2||\bm{a}_3|}\Bigg] \operatorname{mod} 2^a &= 0.
\end{align}
This algorithm can readily be extended to other types of boundaries, but is used in its current form for the OpenKIM \kimtestdriver we have written for this work.
We emphasize that it allows the user to control the resolution of sampled boundaries in a tilt family by controlling only the maximum-denominator integer $d$.
Conversely, given a maximum allowable repeating cell size, it can be used to generate a boundary configuration for an arbitrary tilt angle.

\subsection{Estimation of ground state energy}

It has been well-documented that GBs may exist in a large number of metastable configurations at 0K \cite{han2016grain}. 
For a given \IP, the particular configuration obtained depends on the initial guess and the energy minimization (optimization) method used.
It is generally understood that these configurational parameters can substantially affect the result, making the ground state difficult to find \cite{han2016grain}.
Common practice \cite{homer2022examination,olmsted2009survey} is to perform a high resolution grid search over available parameters, taking the lowest value.
On the other hand, determination of the actual ground state demands computational resources well beyond those feasible for this high-throughput framework.
Therefore, this \kimtestdriver aims to strike an appropriate balance between management of computational resources and consistent, realistic test results across all platforms by performing a grid search with limited resolution.
\added[id=R2,comment={2.2a}]{
  Such a grid search is not sufficiently fine to converge to the actual ground state.
  Instead, we aim for a search that is likely to report the energy of a representative, low-energy state that is neither the ground state, nor an artificially high energy state.
}

\replaced[id=R2,comment={2.2a}]{In the design of a grid search algorithm}{In determining the ground state energy}, there are usually two primary considerations: the atom deletion criterion and grain offsets.
The first, atom deletion, is governed by a cutoff radius for eliminating atomic overlap.
When generating initial configurations of GBs, the algorithm can create unphysical arrangements with with one or more pairs of atoms located at identical or very close positions.
In some cases, a consistent cutoff radius (for example, half the lattice constant) is used to eliminate atoms that are initially too close to each other~\cite{hahn2016symmetric}.
However, this may result in undefined behavior for some \IPs.
As explained in \cref{sec:gbtestdriver}, the approach used here is to set the cutoff to the lattice spacing at which the lattice energy exceeds the single-atom energy for the \IP being used.
This critical radius is determined for each \IP based on values that are automatically referenced in the OpenKIM repository by the \kimtestdriver.

The second consideration is the set of grain offsets used as initial guesses for energy minimization.
The initial position of the top grain with respect to the bottom grain can substantially alter the energy, as reported in~\cite{tschopp2007asymmetric,tschopp2007structure,tschopp2007structures,tucker2010evolution} and others.
The standard approach is to perform a grid search over the space of initial translations: for each initial translation, allow the minimizer to find the local ground state, and then select the lowest of the returned results.
High-resolution grid searches are often used \added[id=R2,comment={2.2a}]{when finding the ground state energy}, but for \replaced[id=R2,comment={2.2a}]{efficiently estimating a representative metastable state}{computational tractability within this high-throughput system}, the grid search in the OpenKIM \kimtestdriver is limited to in-plane translations across the CSL\added[id=R2,comment={2.2c}]{, truncated according to symmetry,} in increments of 1/4th of the relevant lattice constant.\footnote{\added[id=R2]{In some previous works (\cite{olmsted2009survey,homer2022examination}) the lattice displacements were restricted to elements in the displacement-shift complete (DSC) lattice. We have opted instead to select offset vectors that are consistent between boundaries (i.e. irrespective of bicrystallography) in order to ensure consistent sampling of the ensemble of in-plane translations. }}\added[id=R2,comment={2.2c}]{}
\added[id=R2,comment={2.2b}]{Translations normal to the grain boundary are not explicitly considered.}
The number of translations considered, each of which requires a full energy minimization calculation, can range from 16 for a no-boundary case to thousands, depending on the misorientation.
For each shift vector on the grid, the atomic positions and the \added[id=R2]{dimensions of the} orthogonal simulation box are relaxed in all directions\added[id=R2,comment={2.2b}]{, including the direction normal to the grain boundary,} using the Polak--Ribi\`{e}re variant of the conjugate gradient method~\cite{polak2012optimization}.
\added[id=R2,comment={2.2b}]{This implicitly explores translations of the grains normal to the grain boundary.}

\section{Computational results for symmetric tilt grain boundaries}\label{sec:results}

\begin{figure}
  \centering
  \includegraphics[width=0.5\linewidth]{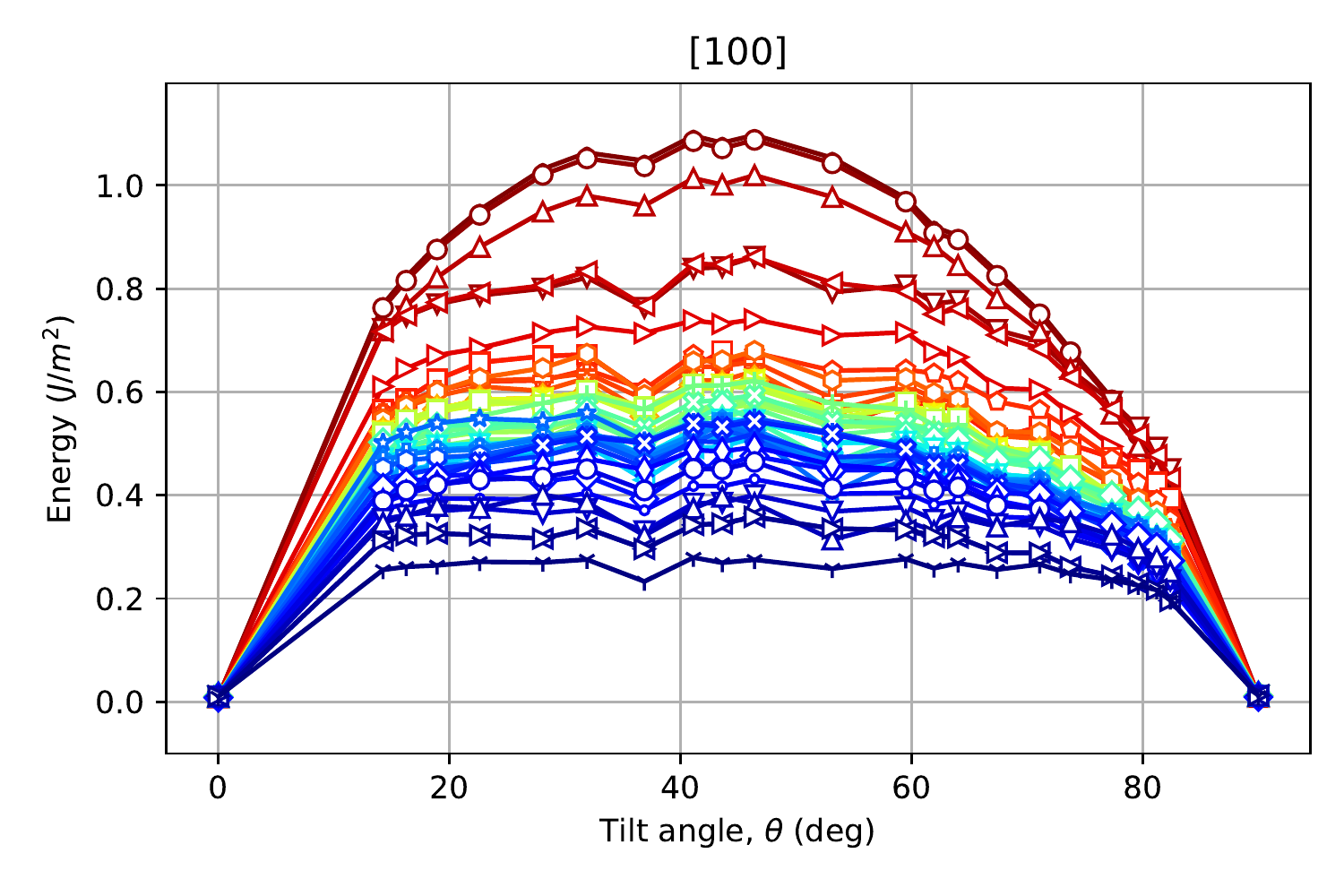}%
  \includegraphics[width=0.5\linewidth]{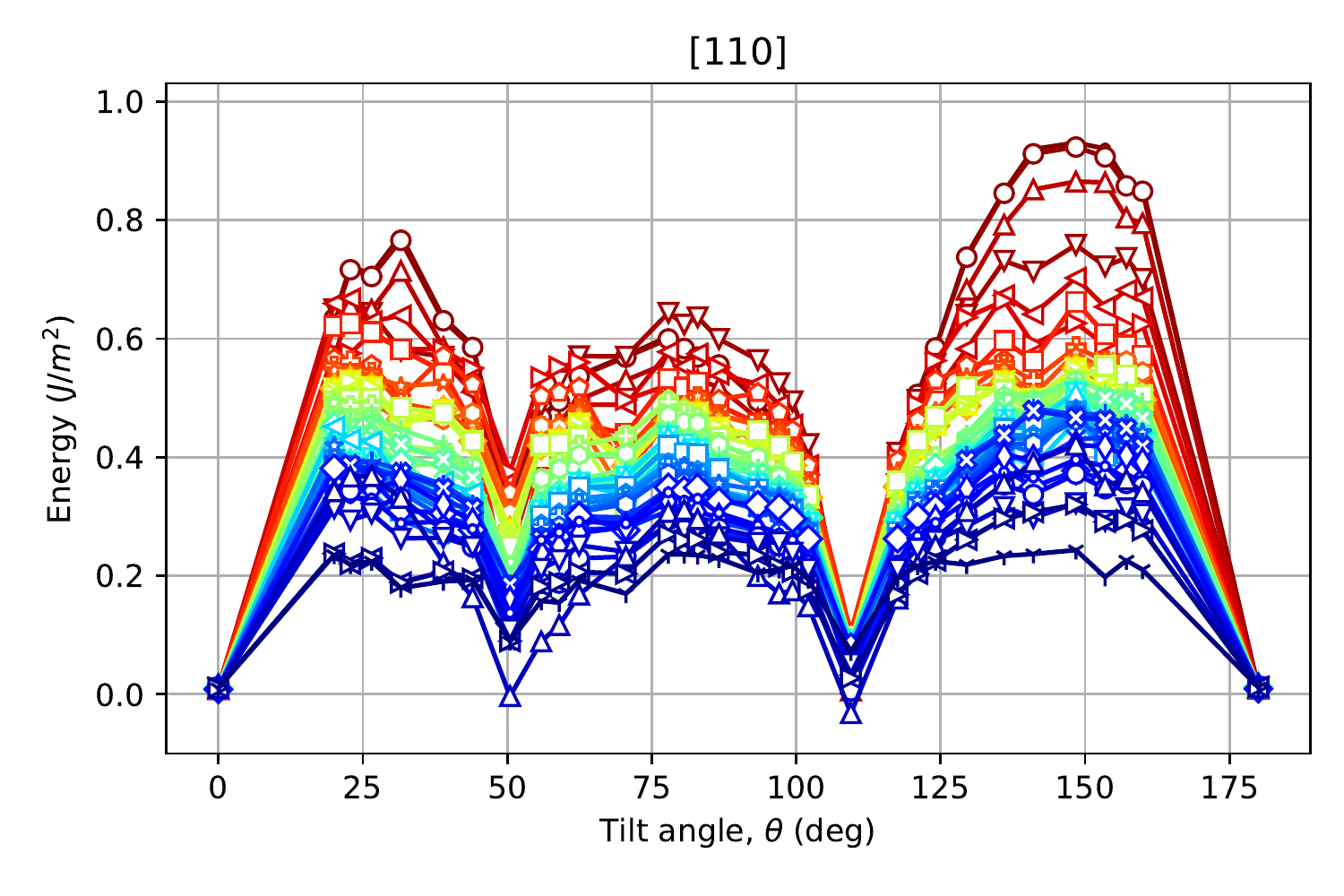}
  \includegraphics[width=0.5\linewidth]{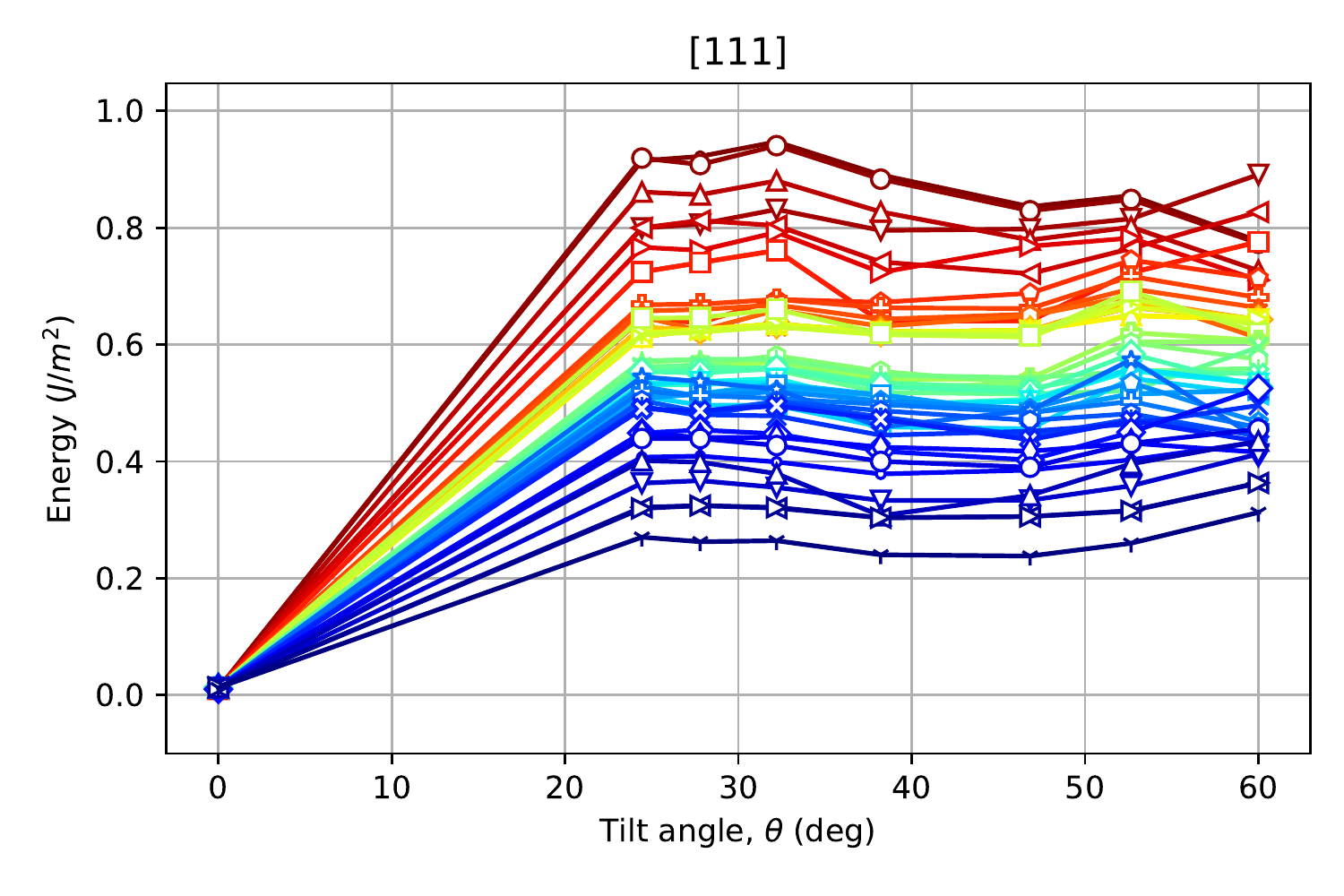}%
  \includegraphics[width=0.5\linewidth]{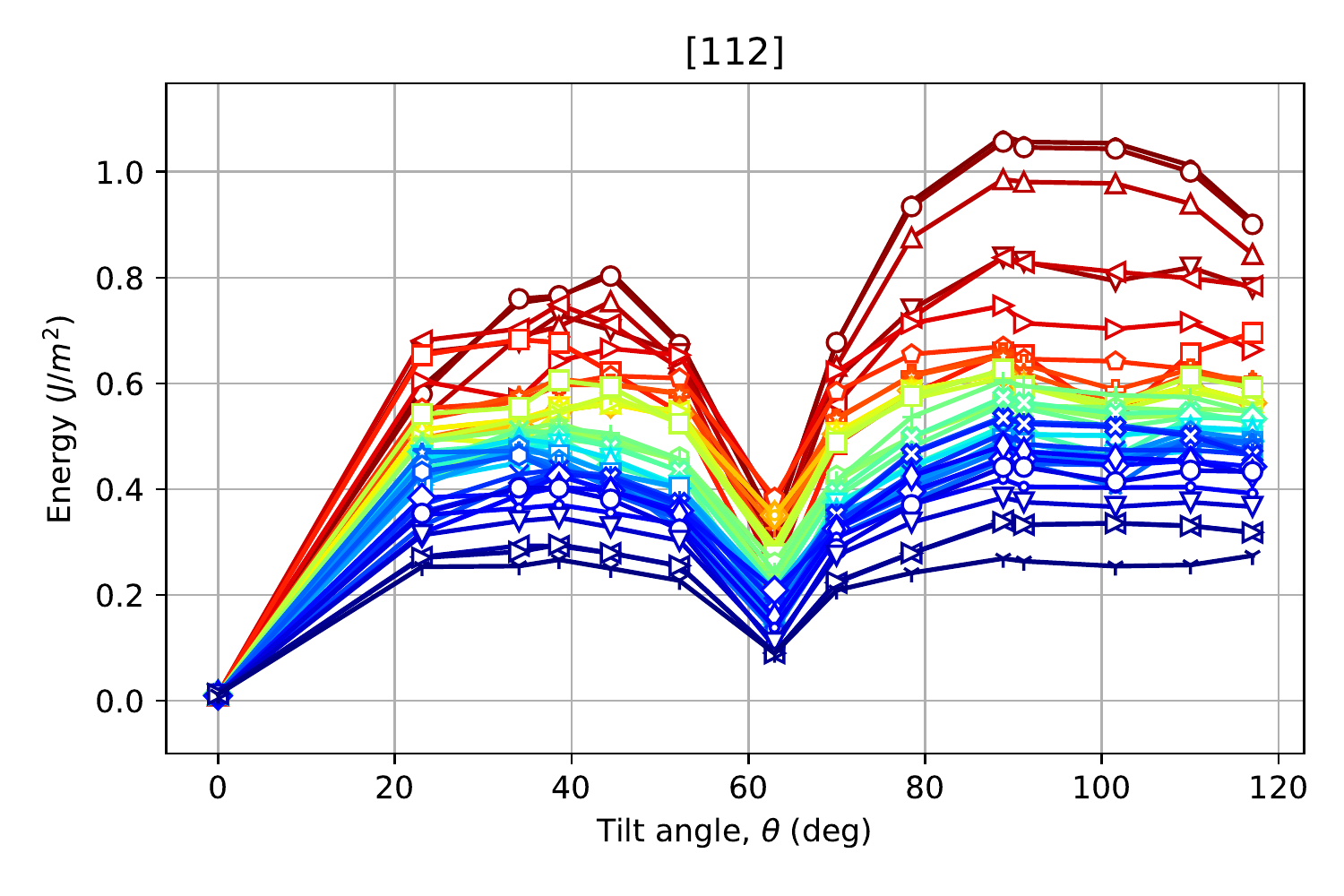}
  \centering\resizebox{\textwidth}{!}{\input{resultsv4/data/GrainBoundaryCubicCrystalSymmetricTiltRelaxedEnergyVsAngle_fcc100_Al__TE_918853243284_003/legend.pgf}}
  \caption{GB energies for fcc Al.}
  \label{fig:md_fcc_al}
\end{figure}
\begin{figure}
  \centering
  \includegraphics[width=0.5\linewidth]{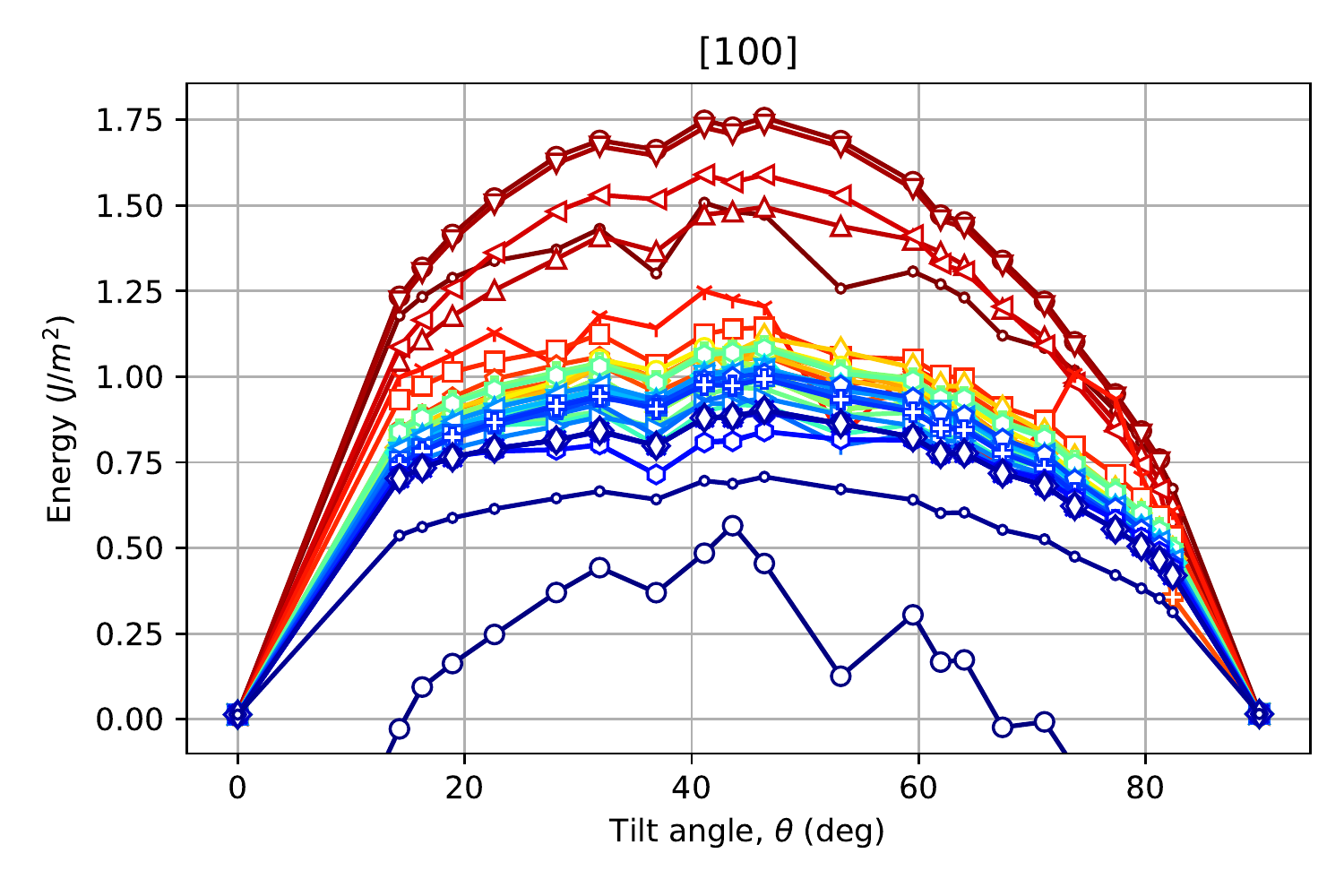}%
  \includegraphics[width=0.5\linewidth]{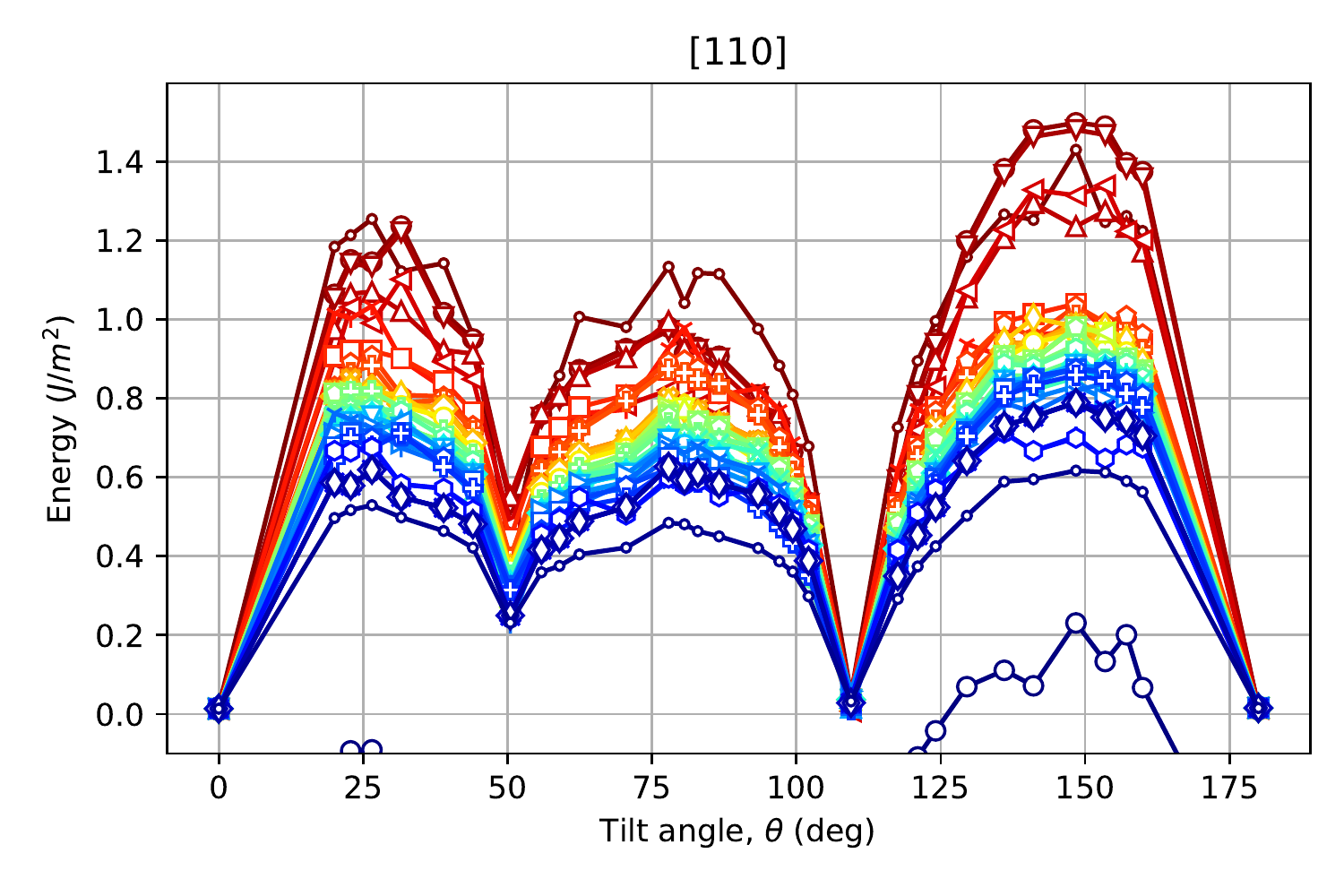}
  \includegraphics[width=0.5\linewidth]{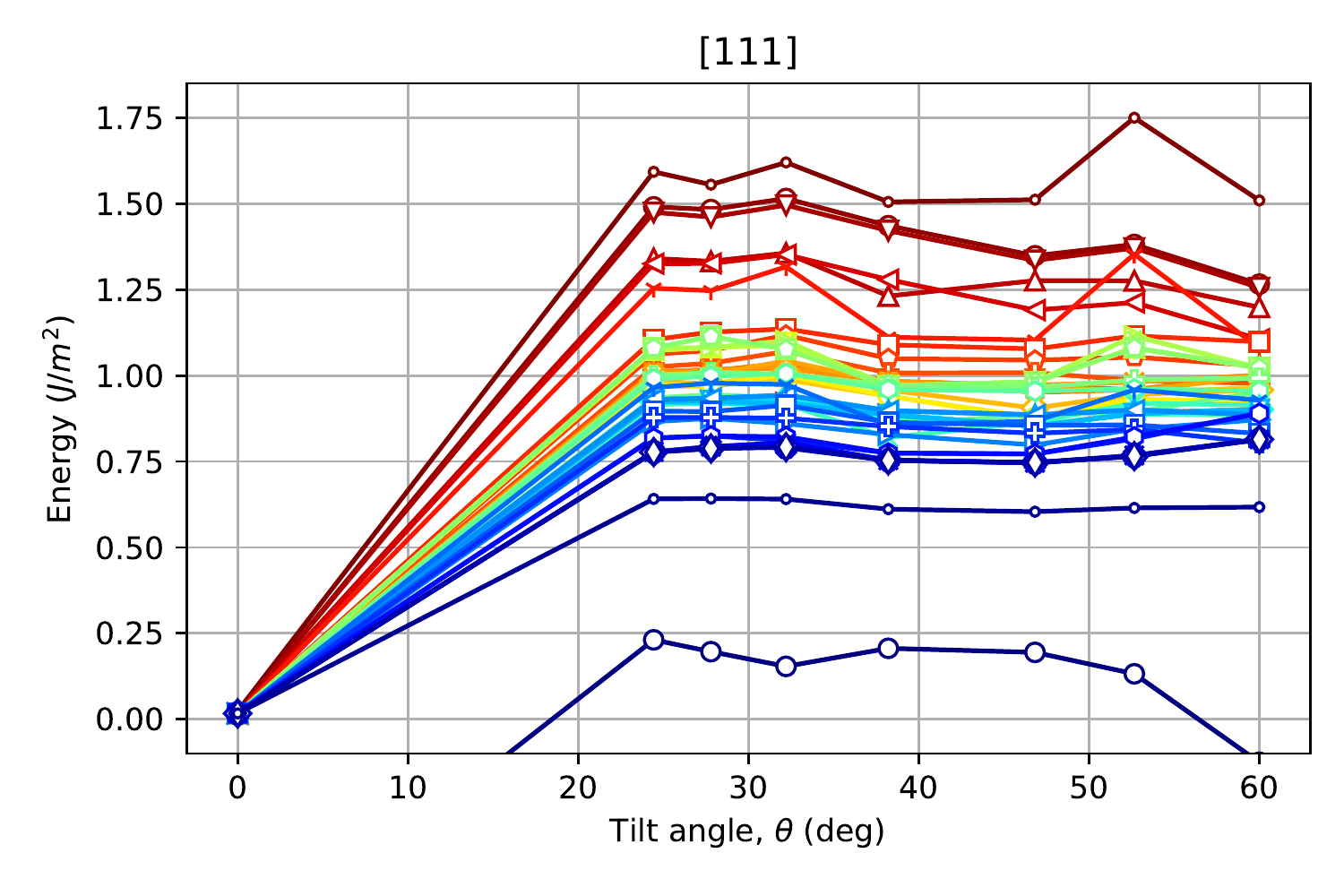}%
  \includegraphics[width=0.5\linewidth]{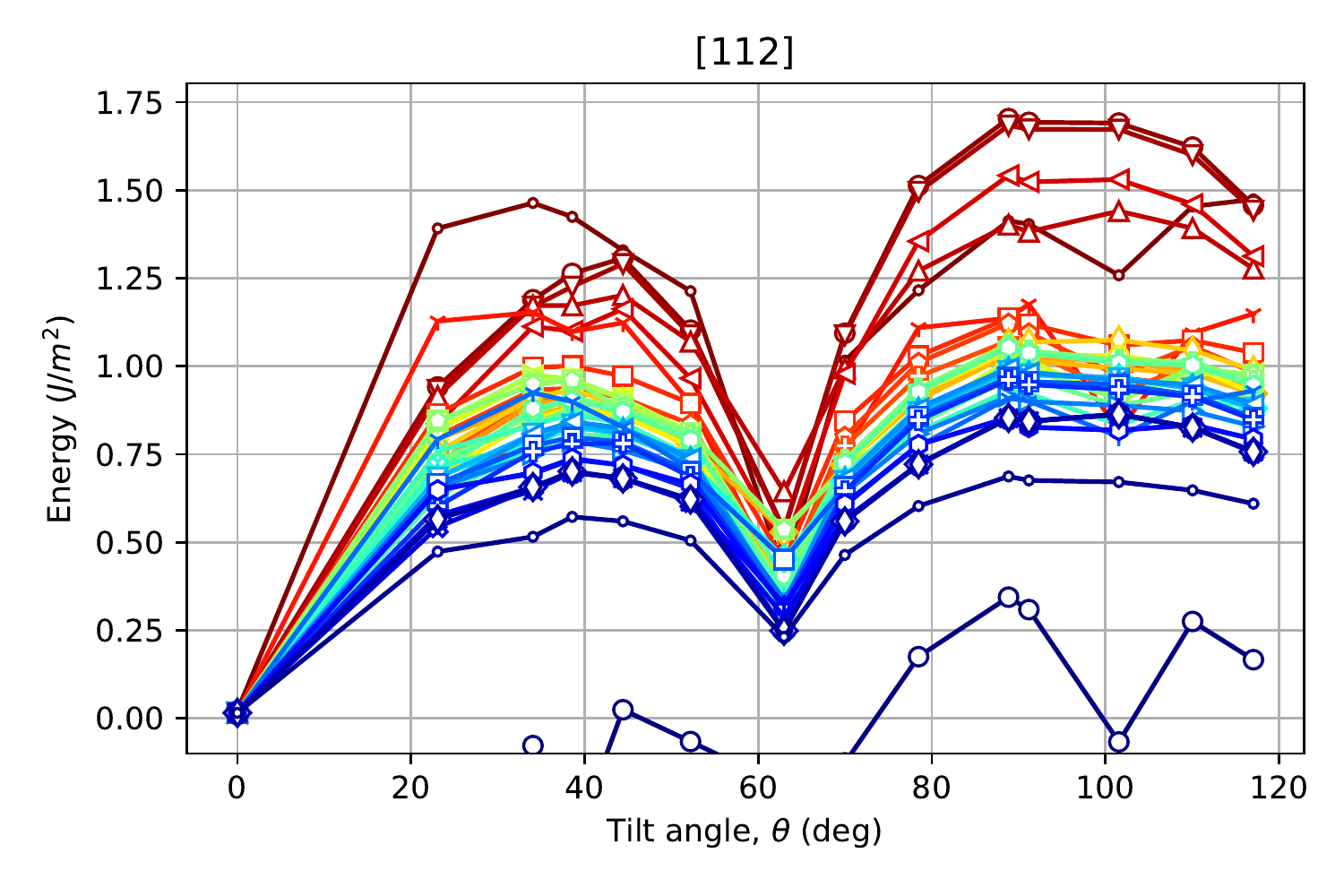}
  \centering\resizebox{\textwidth}{!}{\input{resultsv4/data/GrainBoundaryCubicCrystalSymmetricTiltRelaxedEnergyVsAngle_fcc100_Cu__TE_529988253259_001/legend.pgf}}
  \caption{GB energies for fcc Cu.}
  \label{fig:md_fcc_cu}
\end{figure}
\begin{figure}
  \centering
  \includegraphics[width=0.5\linewidth]{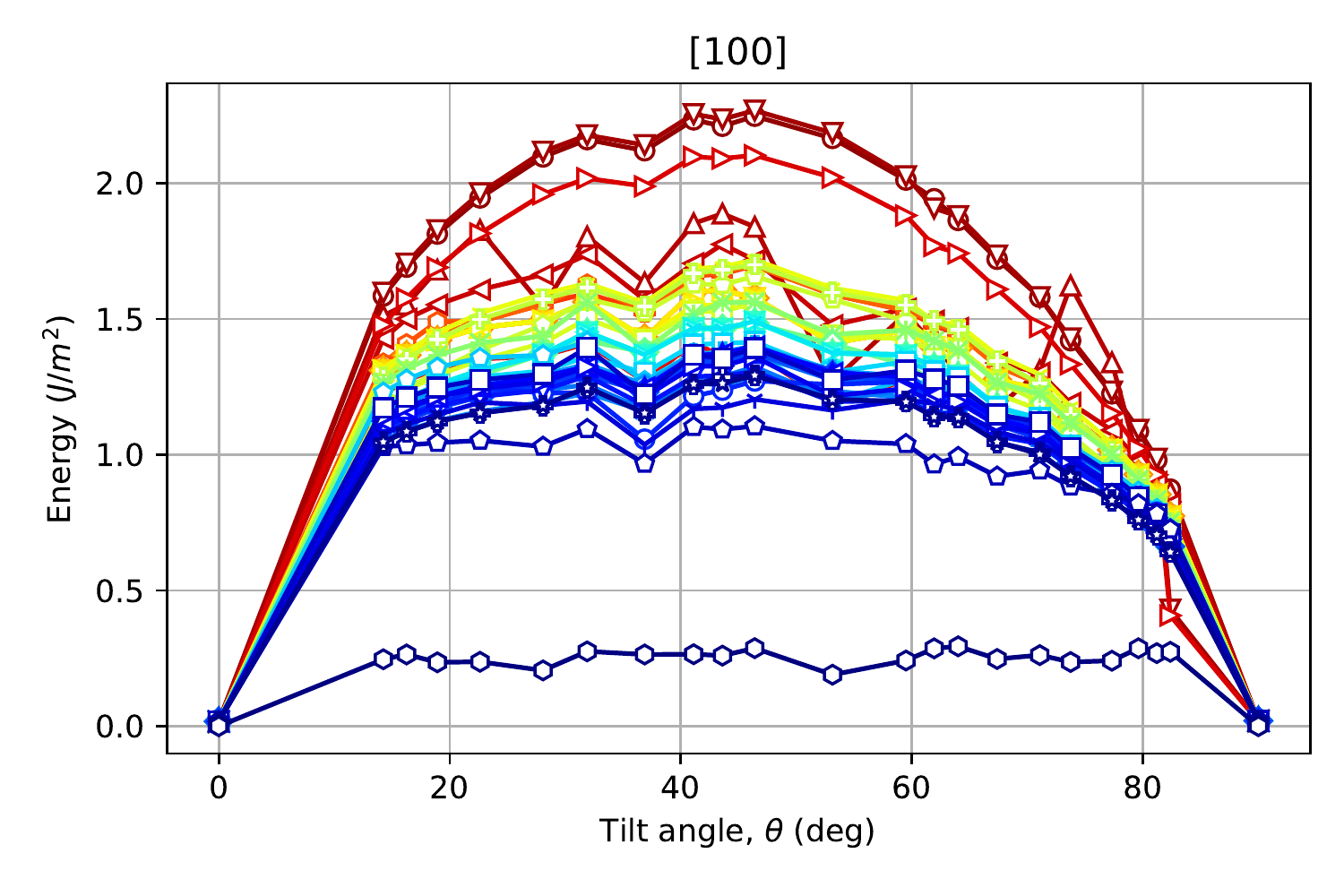}%
  \includegraphics[width=0.5\linewidth]{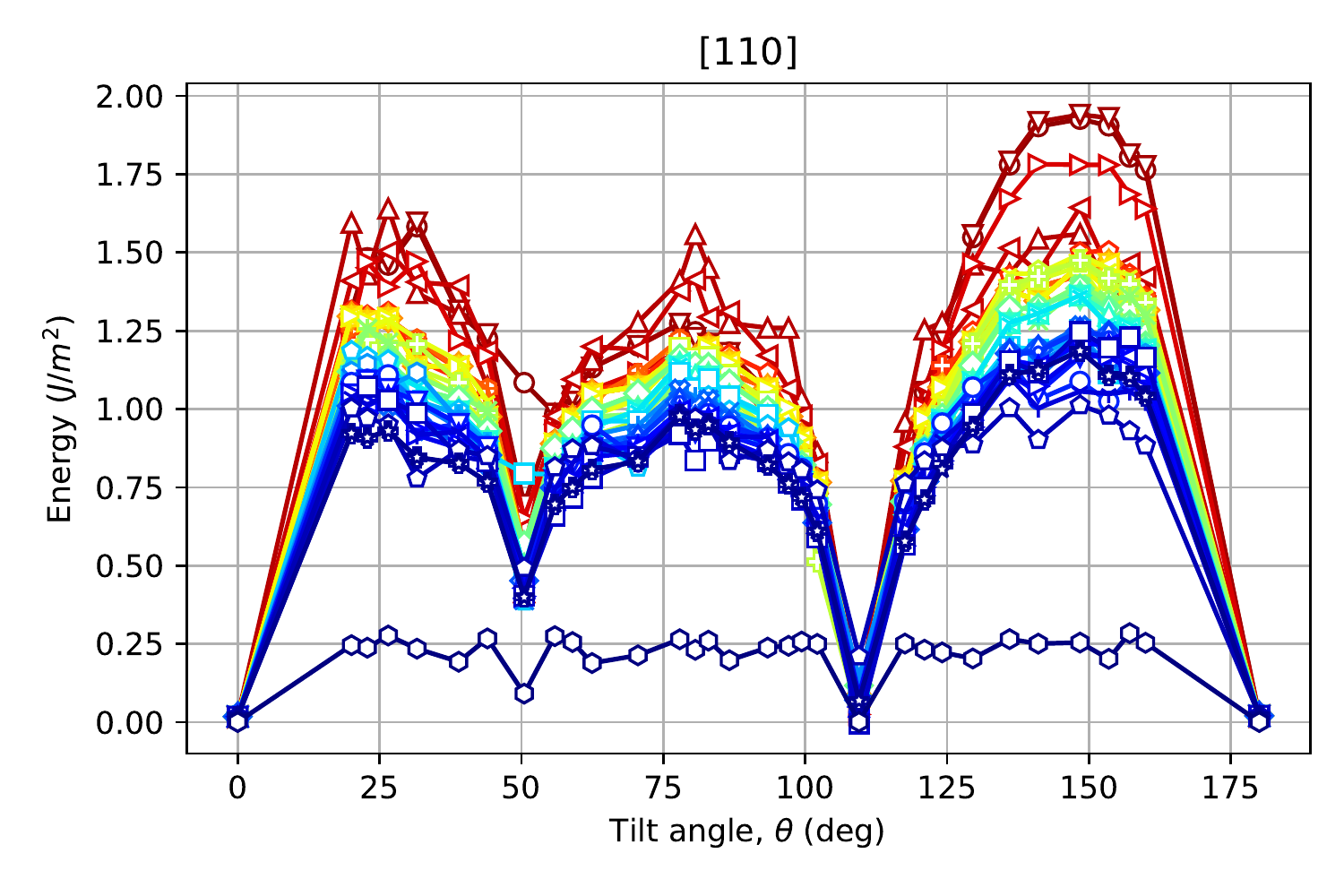}
  \includegraphics[width=0.5\linewidth]{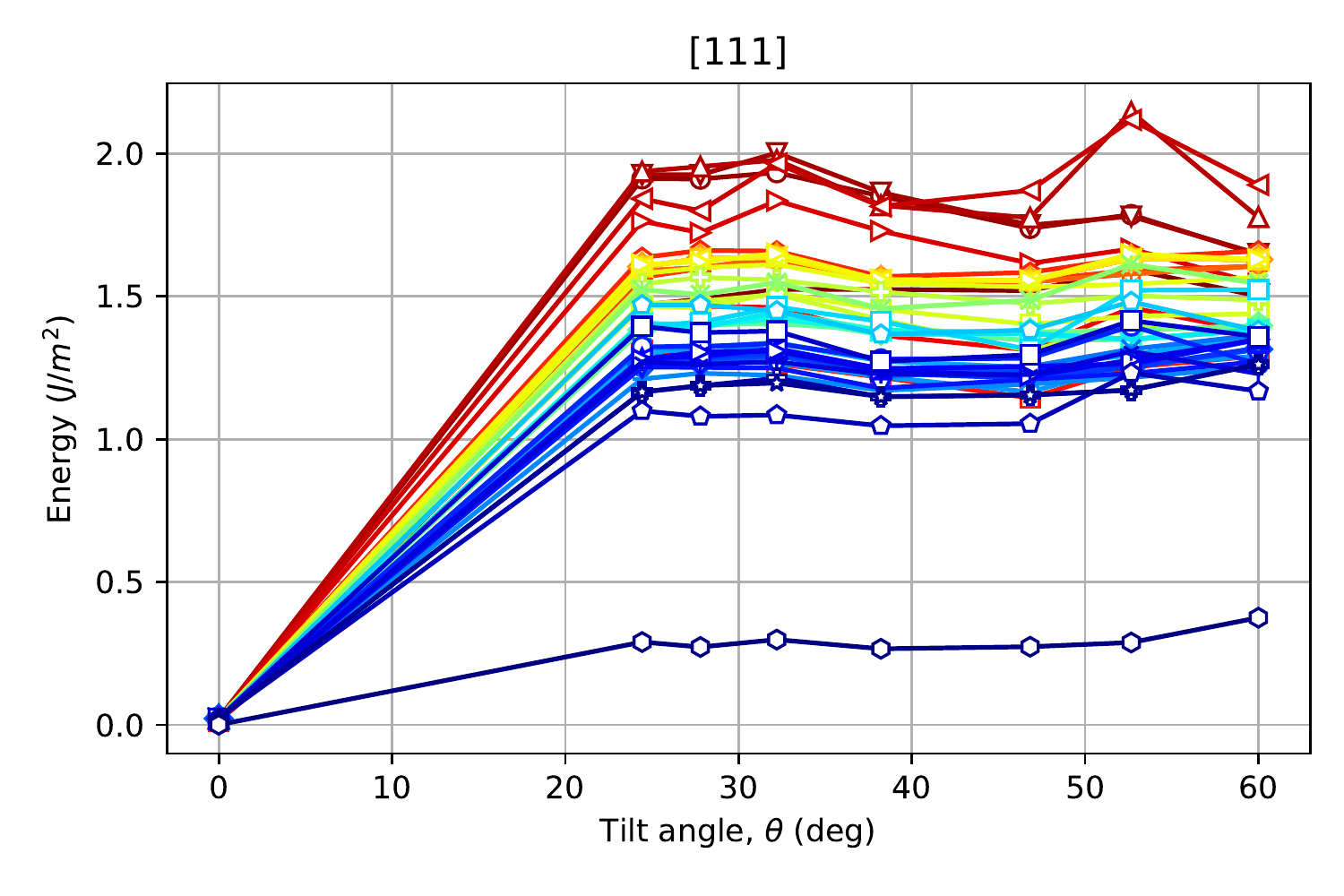}%
  \includegraphics[width=0.5\linewidth]{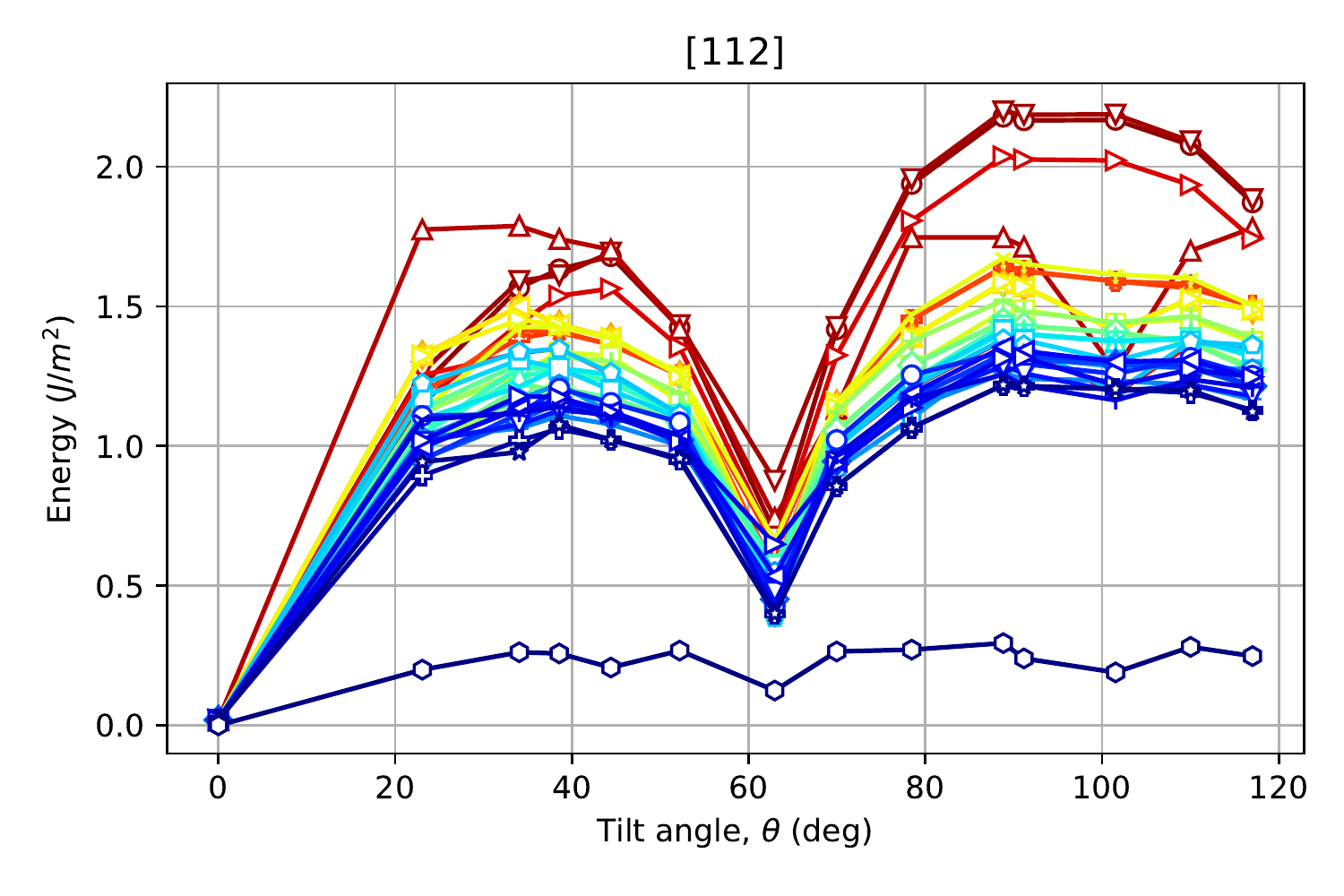}
  \centering\resizebox{\textwidth}{!}{\input{resultsv4/data/GrainBoundaryCubicCrystalSymmetricTiltRelaxedEnergyVsAngle_fcc100_Ni__TE_457754988992_001/legend.pgf}}
  \caption{GB energies for fcc Ni.}
  \label{fig:md_fcc_ni}
\end{figure}
\begin{figure}
  \centering
  \includegraphics[width=0.5\linewidth]{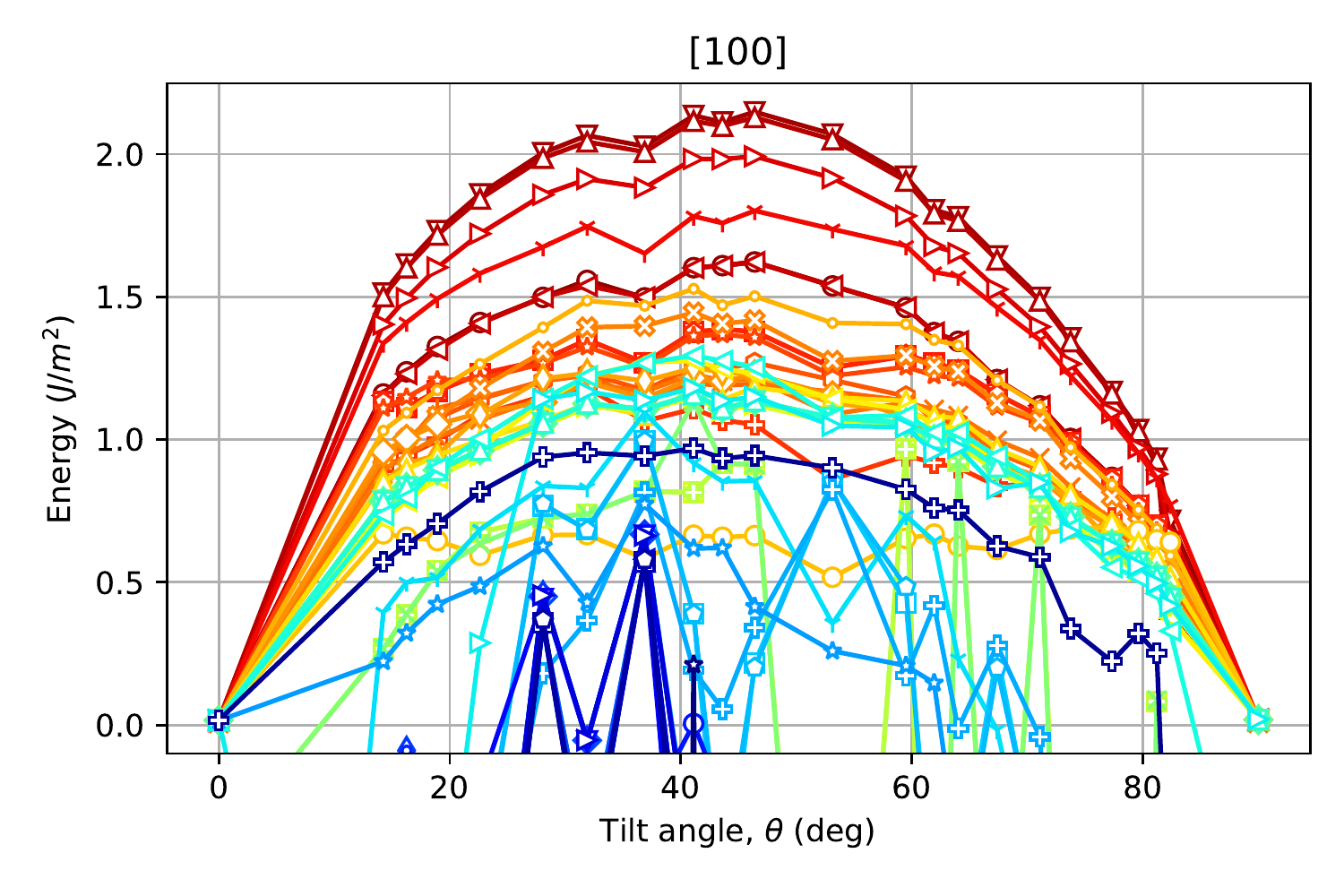}%
  \includegraphics[width=0.5\linewidth]{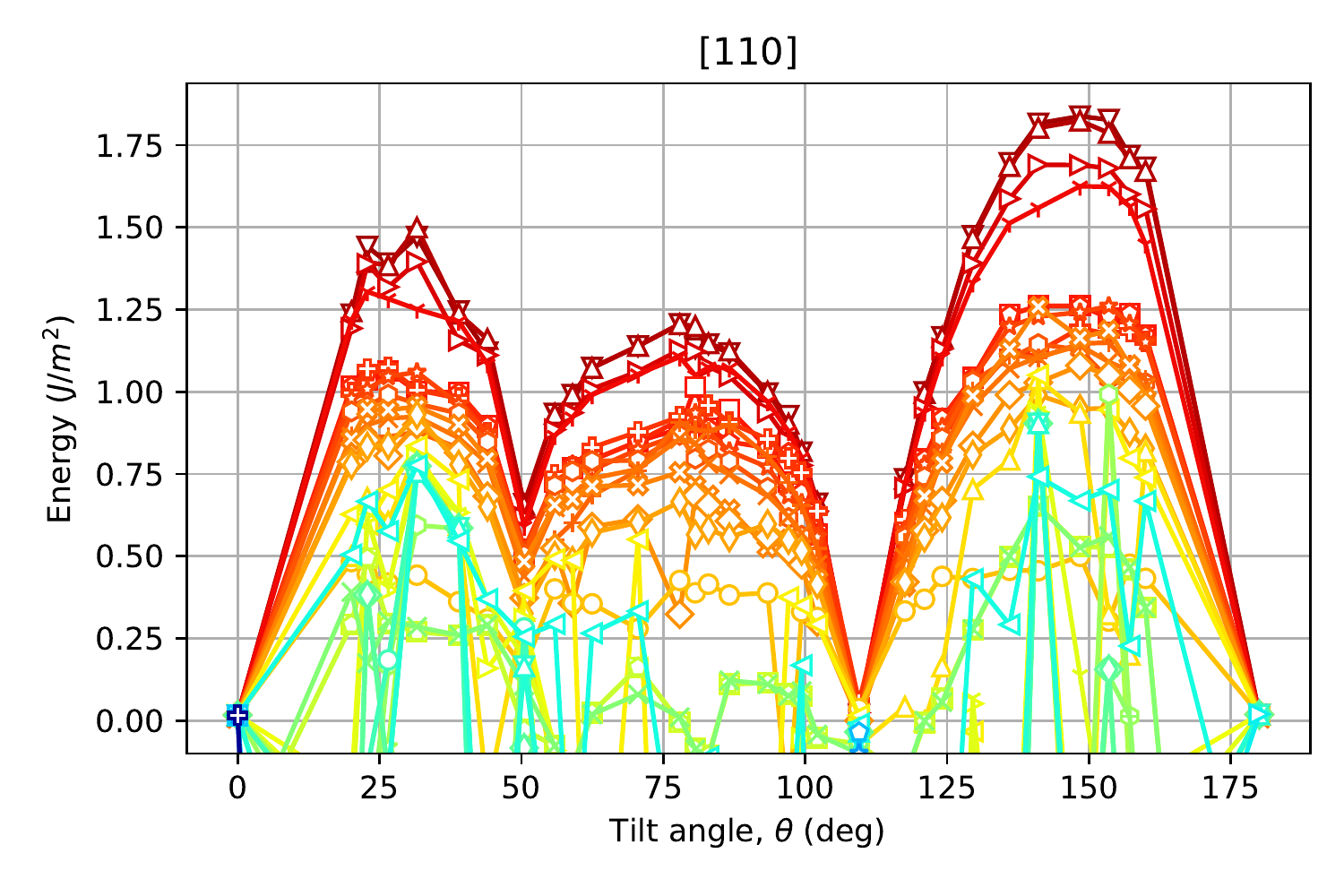}
  \includegraphics[width=0.5\linewidth]{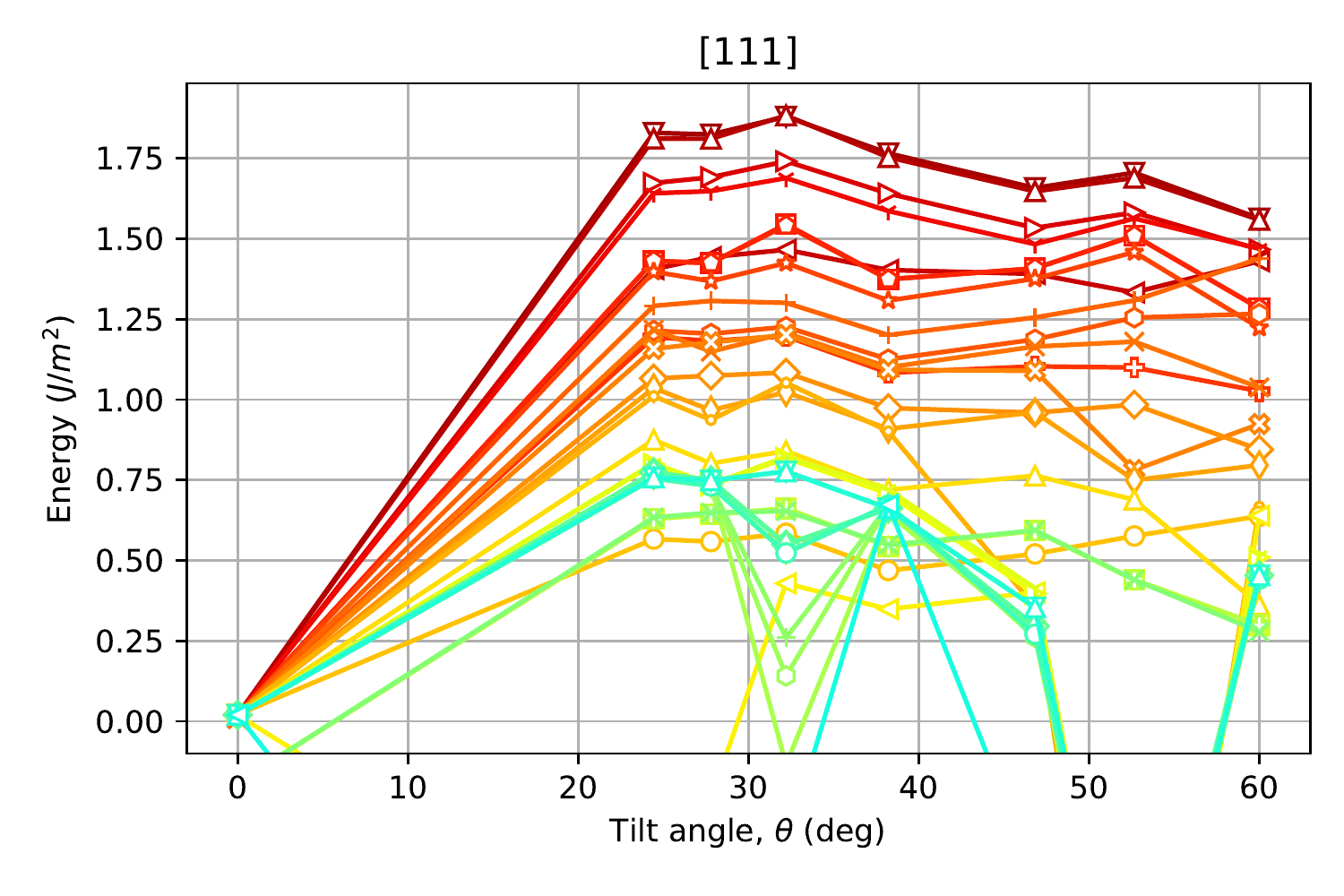}%
  \includegraphics[width=0.5\linewidth]{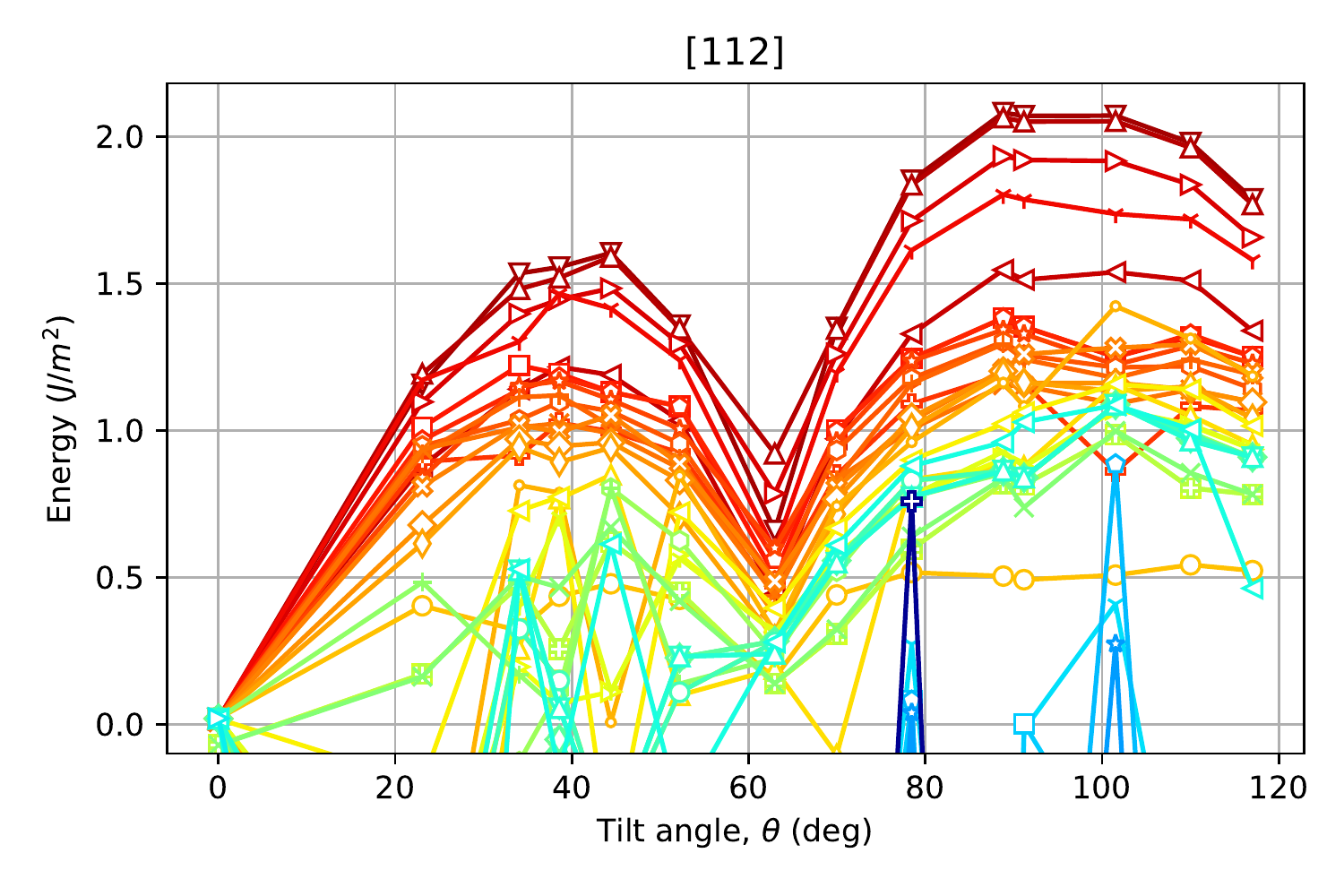}
  \centering\resizebox{\textwidth}{!}{\input{resultsv4/data/GrainBoundaryCubicCrystalSymmetricTiltRelaxedEnergyVsAngle_fcc100_Fe__TE_814353485766_001/legend.pgf}}
  \caption{GB energies for fcc Fe.}
  \label{fig:md_fcc_fe}
\end{figure}
\begin{figure}
  \centering
  \includegraphics[width=0.5\linewidth]{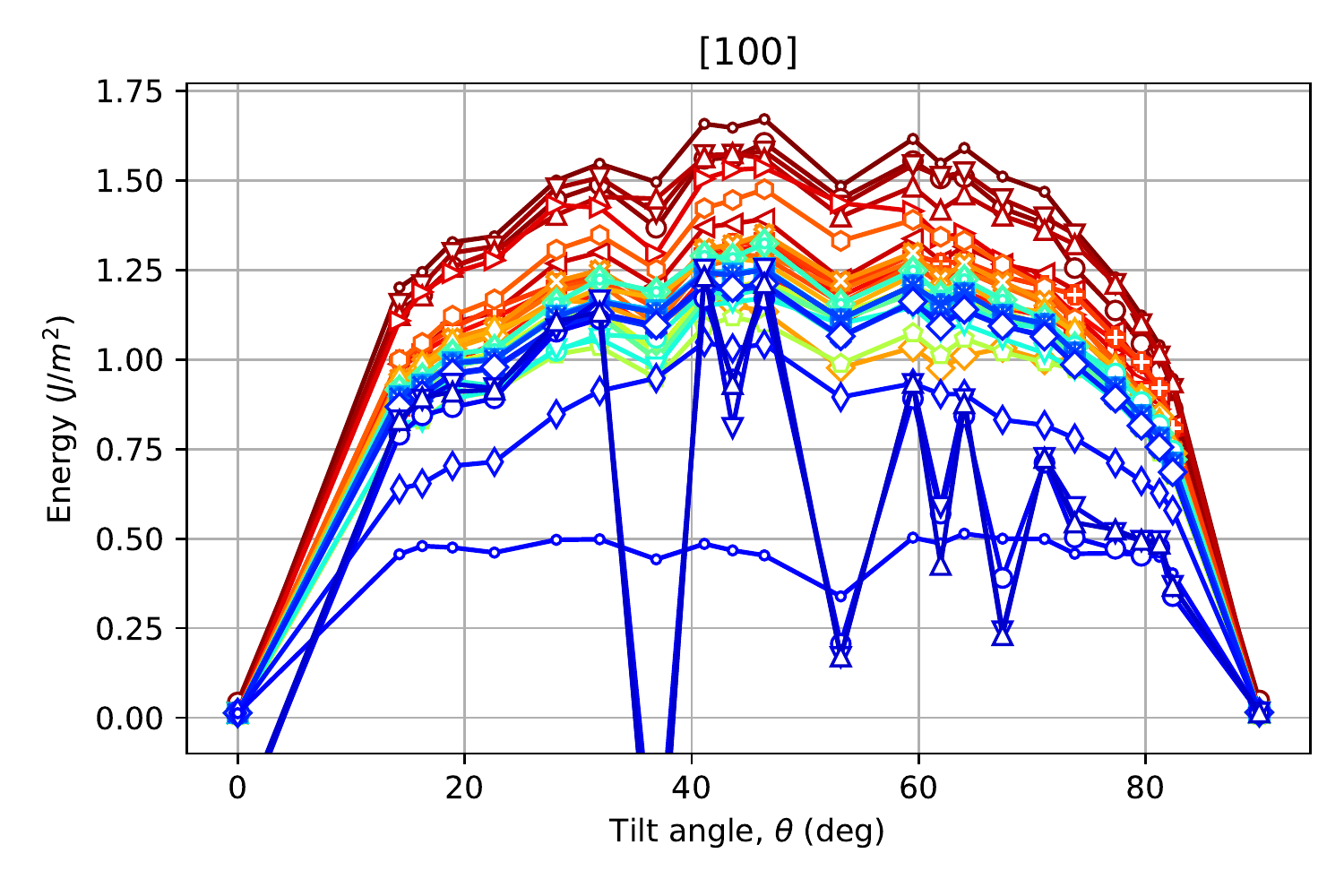}%
  \includegraphics[width=0.5\linewidth]{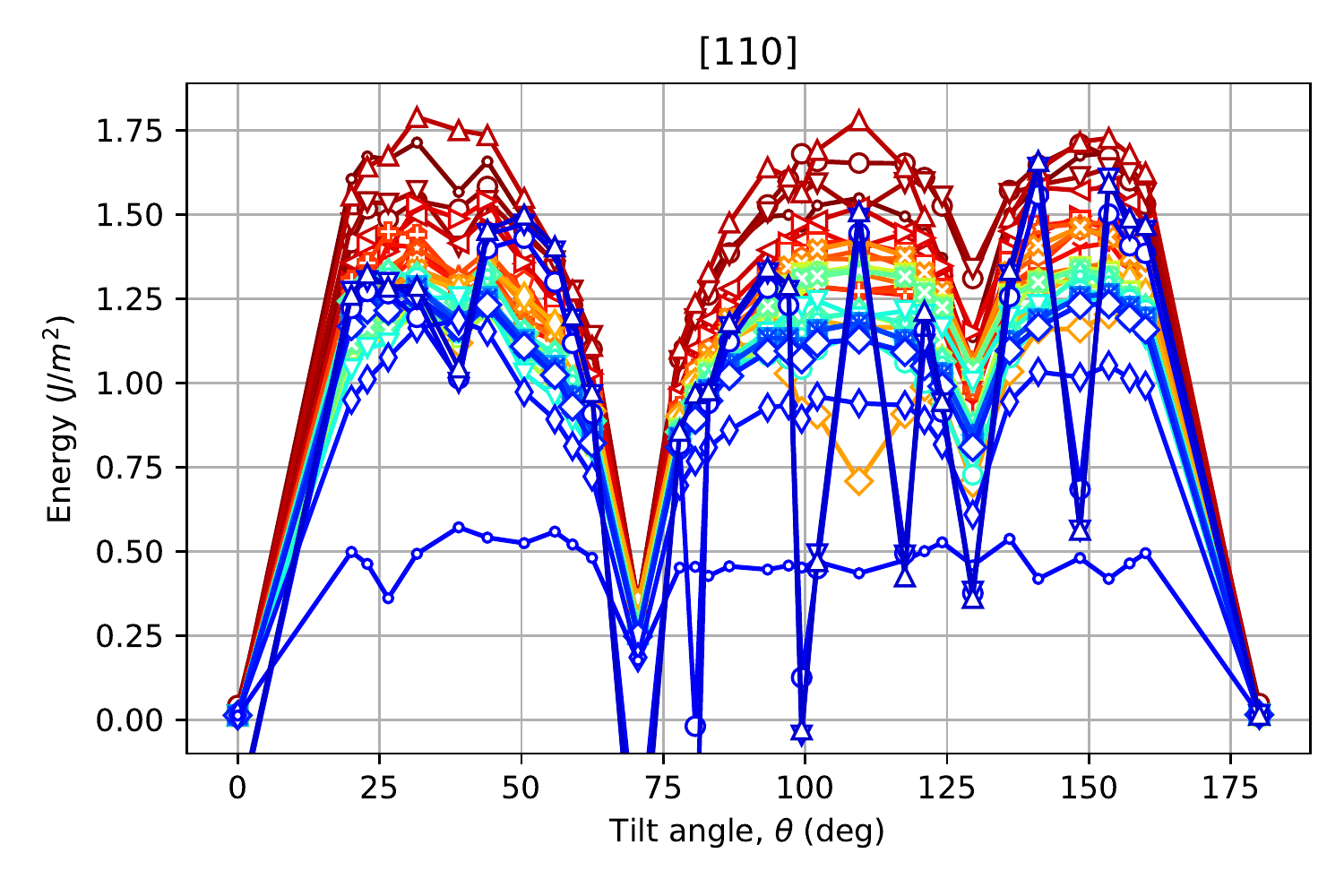}
  \includegraphics[width=0.5\linewidth]{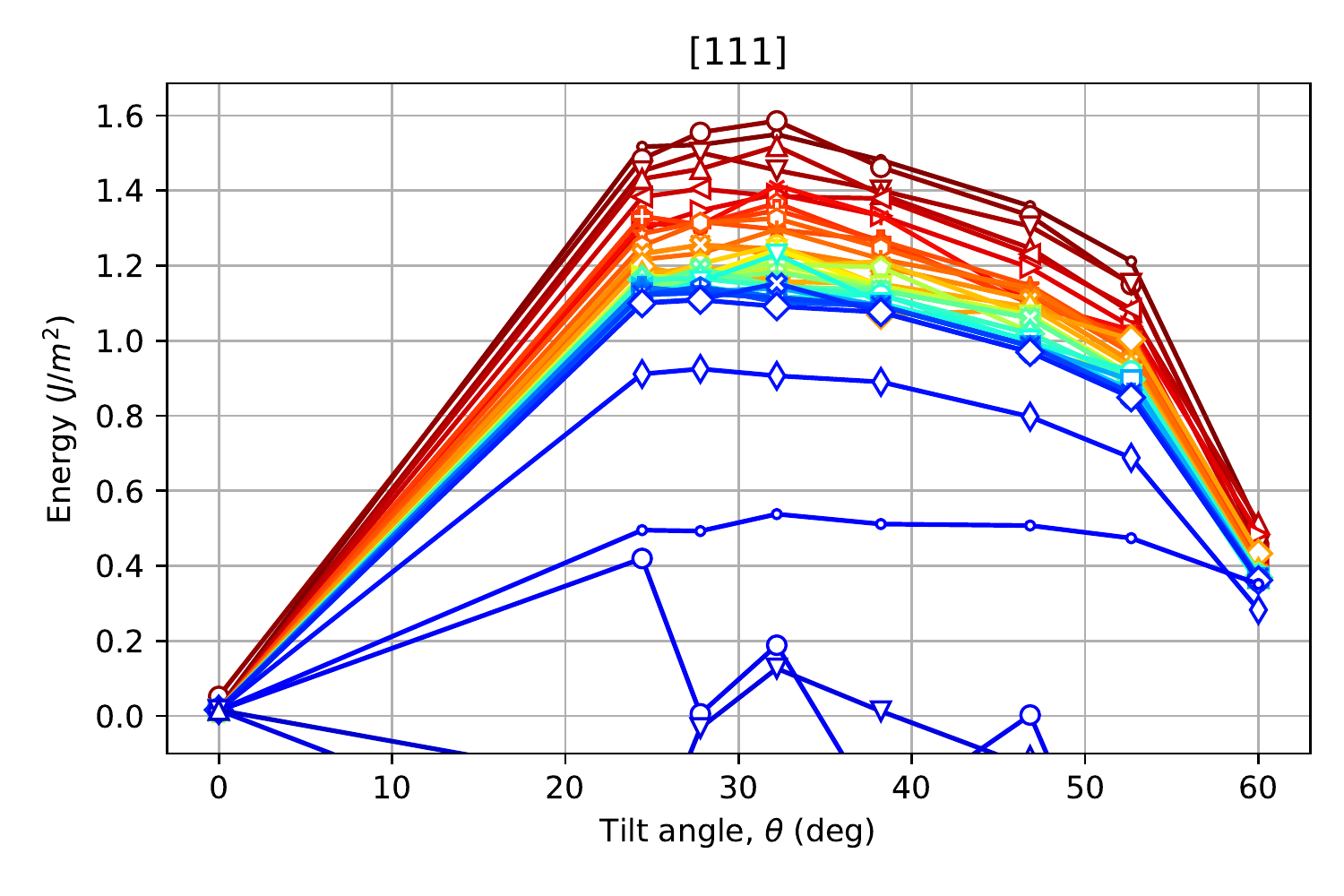}%
  \includegraphics[width=0.5\linewidth]{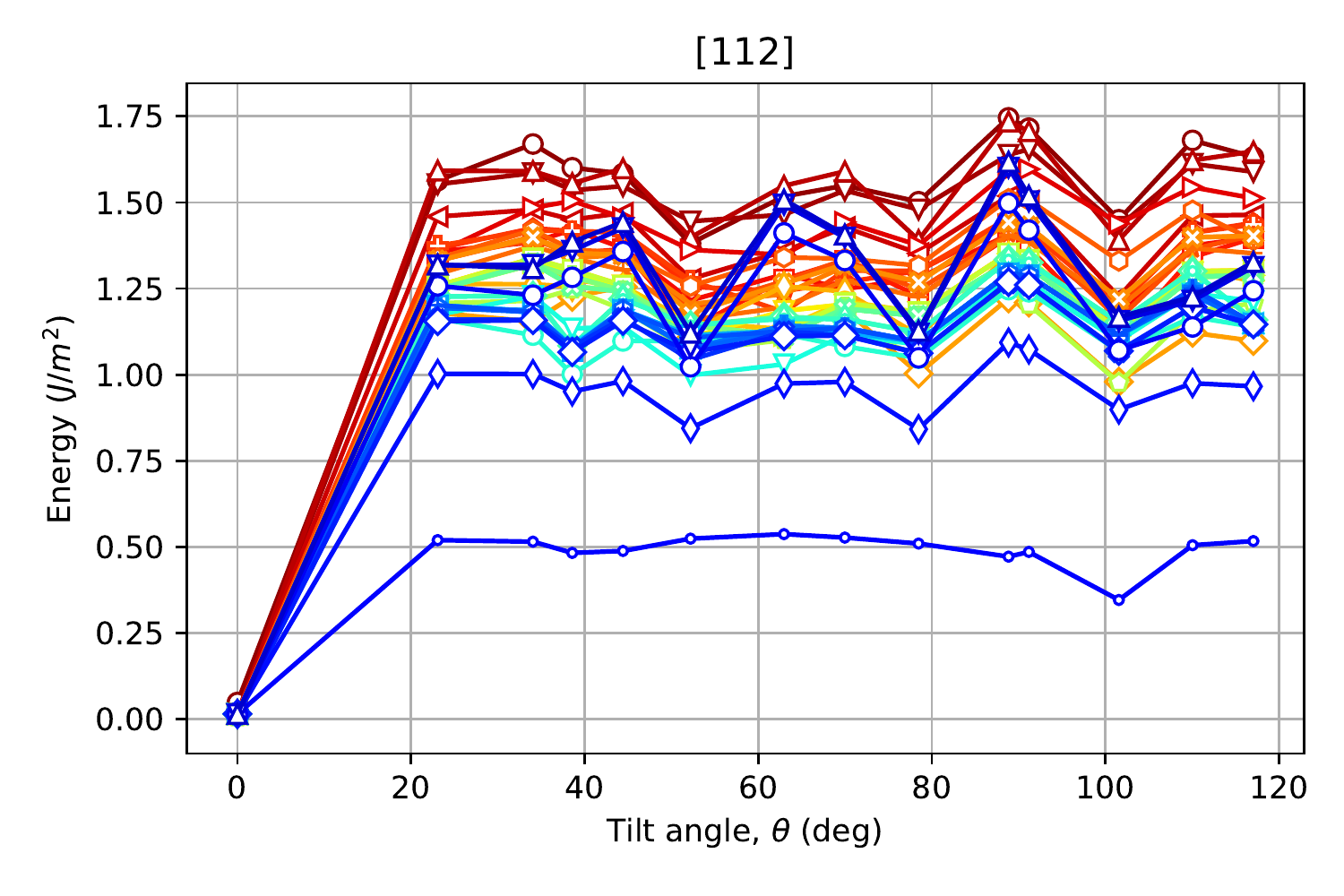}
  \centering\resizebox{\textwidth}{!}{\input{resultsv4/data/GrainBoundaryCubicCrystalSymmetricTiltRelaxedEnergyVsAngle_bcc100_Fe__TE_175540441720_001/legend.pgf}}
  \caption{GB energies for bcc Fe.}
  \label{fig:md_bcc_fe}
\end{figure}

\begin{figure}
  \centering
  \includegraphics[width=0.5\linewidth]{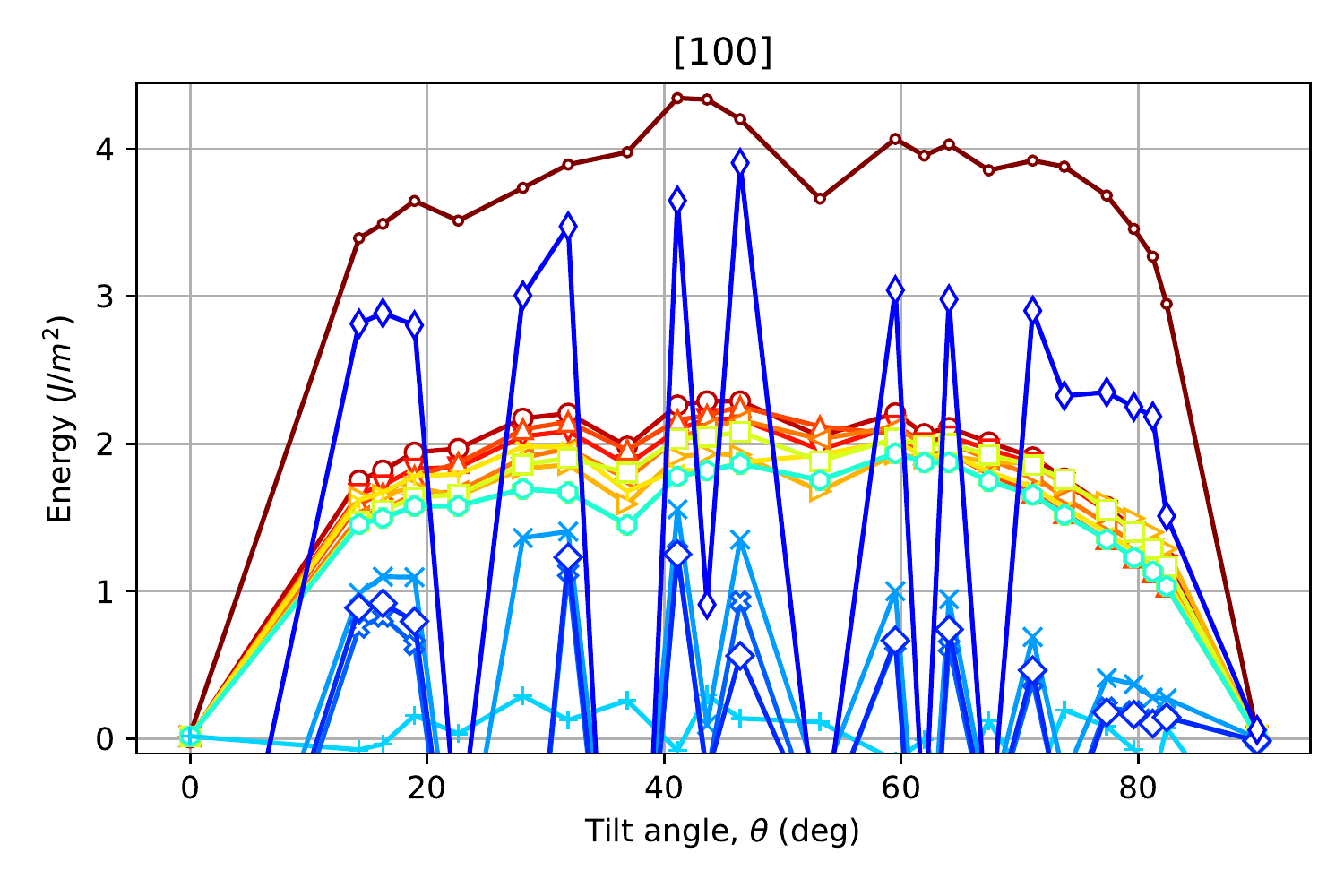}%
  \includegraphics[width=0.5\linewidth]{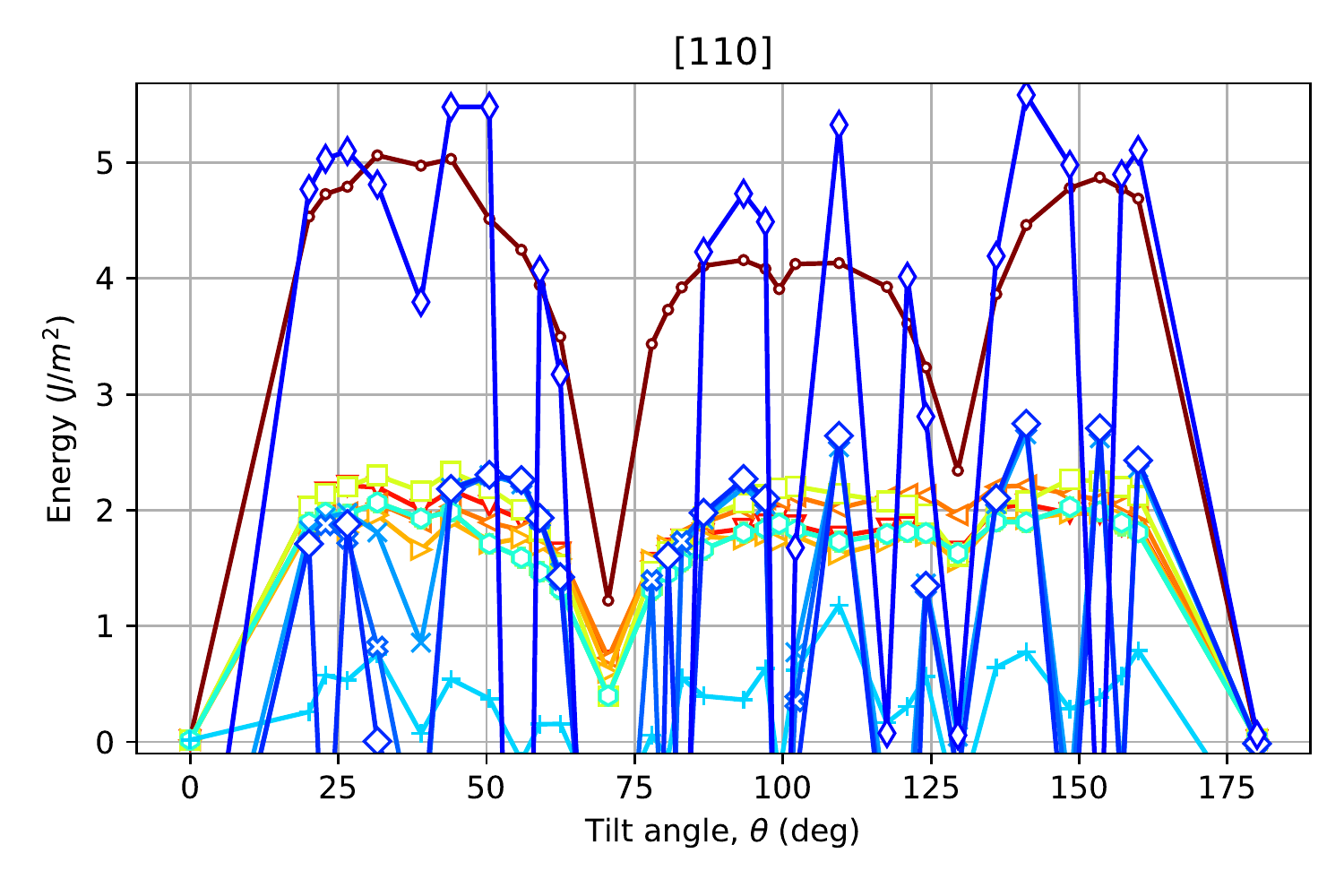}
  \includegraphics[width=0.5\linewidth]{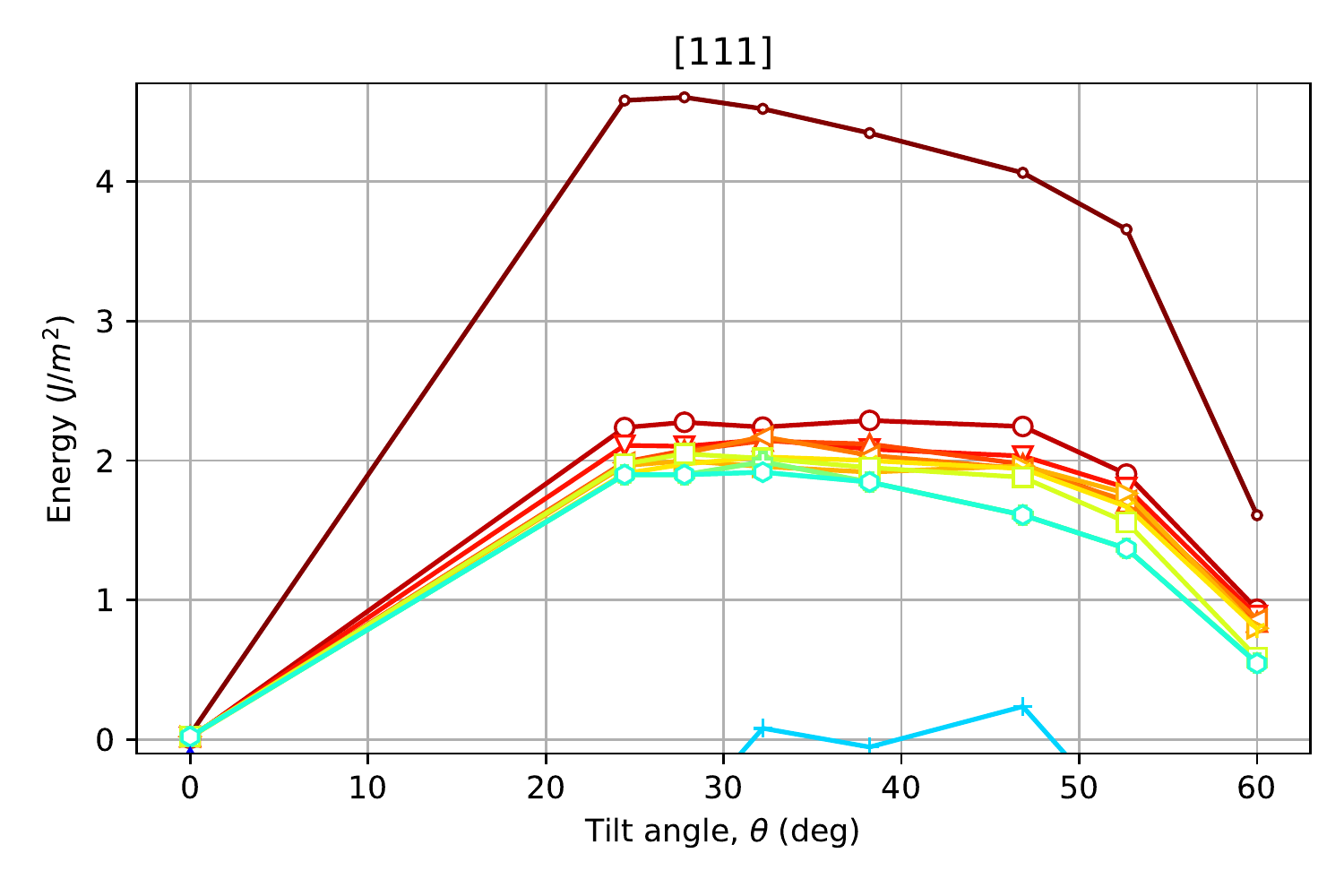}%
  \includegraphics[width=0.5\linewidth]{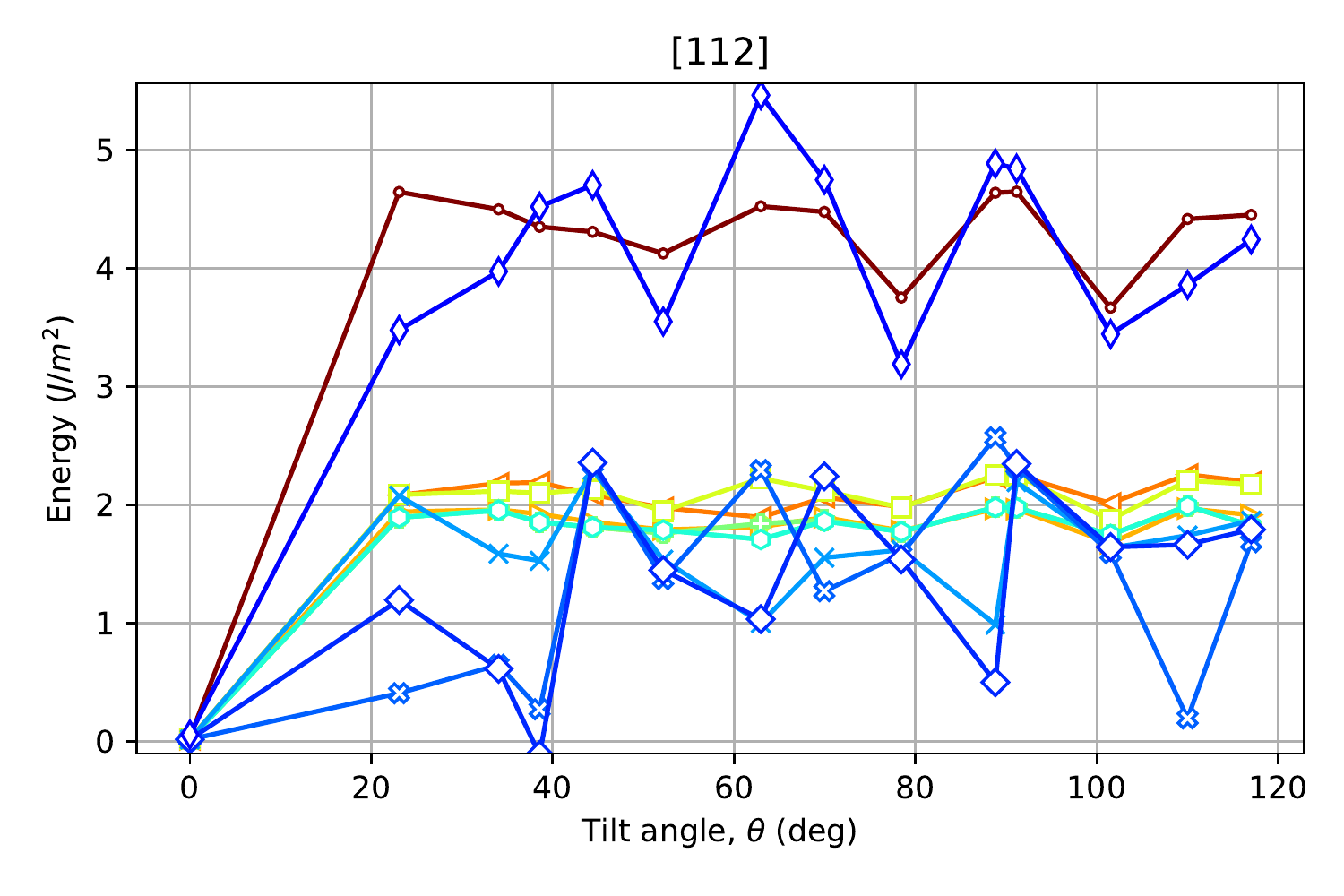}
  \centering\resizebox{\textwidth}{!}{\input{resultsv4/data/GrainBoundaryCubicCrystalSymmetricTiltRelaxedEnergyVsAngle_bcc100_Mo__TE_529178526487_001/legend.pgf}}
  \caption{GB energies for bcc Mo.}
  \label{fig:md_bcc_mo}
\end{figure}

In this section we present calculations for GB energies explored along the four inequivalent tilt axes that give rise to symmetric tilt boundaries in cubic materials~\cite{sutton1995interfaces}:  $[100]$, $[110]$, $[111]$, and $[112]$.
This is far from an exhaustive study of all grain boundaries, but rather provides a representative sampling across a range of boundaries that demonstrates high angle and low angle cusp behavior.
GB energy was calculated using the following \kimtests:
fcc Al \cite{TE918853243284003,TE202986963854001,TE117904176283001,TE641102822364001},
fcc Cu \cite{TE529988253259001,TE708214008908001,TE603516505525001,TE288691353820001},
fcc Ni \cite{TE457754988992001,TE980409230161001,TE035582886963001,TE893686795562001},
fcc Fe, \cite{TE814353485766001,TE729107030375001,TE989116099275001,TE317621478872001}
bcc Fe, \cite{TE175540441720001,TE558145380113001,TE752424681735001,TE187984704771001}
and bcc Mo \cite{TE529178526487001,TE671775543629001,TE107990591816001,TE568499993431001}.
Iron is uniquely present for both fcc and bcc tests, where the only distinction between them is the initial condition of the lattice.
For each chemical element, GB energies were computed for every matching potential in OpenKIM.
The results presented herein are up-to-date as of December 2022, but the \kimtests are being continuously run and the latest results are available on openkim.org.

With such a large number of \IPs fitted for different purposes, it is unavoidable that some of the computations result in errors or unphysical behavior.
For example, a cause of several errors is that the OpenKIM Embedded Atom Model (EAM) \kimmodeldriver~\cite{MD120291908751005,MD120291908751005a,MD120291908751005b,tadmor:elliott:2011,elliott:tadmor:2011} raises an error when the electron density is outside of the fitted range of the embedding function, unlike other implementations of the EAM method that extrapolate.
A full examination of every \IP that caused errors and aberrant behavior is outside of the scope of the present work.
However, some discussion of \IP behavior is provided in \cref{sec:potentials}.
If an \IP encountered errors for some tilt axes but not others, its energy curves are presented only on tilt axis plots corresponding to \kimtests for which no errors occurred.

\subsection{General trends in grain boundary energy}

Despite the diversity and complexity of GB energy calculations, the results are found to be remarkably well-behaved for the vast majority of the tested \IPs.
Included in this work is a series of plots illustrating the range of behaviors for the full spectrum of potentials currently available (as of December 2022) in OpenKIM for the aforementioned atomic species.
To faithfully represent such a large number of energies in a compact way, we have ordered the \IPs by their average energy across all GB tests.
Therefore, the position of an \IP in the legend is a rough indicator of its relative position in the spectrum, which may or may not be clearly visible due to the large number of potentials.
Each \IP in the legend contains its KIM ID (a shorthand designation indicating the potential type, authors, publication year, and supported chemical elements) and two or three citations: one is to the KIM \kimmodel (defining the \IP parameters), one refers to the corresponding KIM \kimmodeldriver (containing the \IP functional form, if applicable), and one is to the primary scientific source in which the \IP was originally published.
We encourage the reader to follow these citations to further investigate the behavior of \IPs of interest and to directly access the data, which is published in the OpenKIM repository.

For the fcc metals, Al (\cref{fig:md_fcc_al}), Cu (\cref{fig:md_fcc_cu}), Ni (\cref{fig:md_fcc_ni}), and Fe (\cref{fig:md_fcc_fe}), similar trends are noted and generally good behavior (i.e., stability of the boundary sufficient to calculate the relaxed GB formation energy) was observed.
Remarkably, there is a variation of up to two or three fold in the average high angle GB energy calculated by different potentials; yet, the angles at which cusps in the energy occur are consistent.
For all fcc materials, especially, the $\Sigma3$ boundary along the $[110]$ axis exhibits close agreement across potentials.
fcc Fe (\cref{fig:md_fcc_fe}) is the most disordered plot by far, with many potentials exhibiting unstable behavior.
As the low-temperature ground state of iron is bcc, it is not surprising that many \IPs rearrange the atoms so as to occupy this state.\footnote{Physically, the ground state of Fe is bcc up to $\sim$1180K. The fact that some potentials produce stable Fe fcc GBs at 0K does not necessarily reflect physical stability.}
Somewhat unexpectedly, however, many of the models are shown to be stable in the fcc phase for a large proportion of the GB configurations relevant to this work.

The GB energy relations computed for bcc Fe (\cref{fig:md_bcc_fe}) exhibit agreement amongst well-behaved potentials that rivals (and somewhat exceeds) that observed in the fcc metals.
bcc Mo (\cref{fig:md_bcc_mo}) has the smallest spread among its well-behaved potentials, although this may simply be due to a lower number of tested potentials.
The Morse pair potentials~\cite{MO331285495617004,MO984358344196004,MD552566534109004,elliott:tadmor:2011} exhibit unstable behavior for both bcc materials because pair potentials tend to favor close-packed structures such as fcc or hexagonal close-packed (hcp).\footnote{We are aware of at least two methods used to stabilize pair potentials in the bcc structure. One is to use a double minimum in the energy curve. Another is to maintain a single minimum, but to use a cutoff that excludes the second nearest neighbor in fcc and design the curve such that the combined sum of first and second bcc neighbors is more stable than the sum of first fcc neighbors. The latter is used in the well-behaved modified Johnson (MJ) pair potential.} 
In the case of Mo, the pathological behavior of the Zhang and Nguyen Tersoff-style potential~\cite{MO152208847456001, MD077075034781005, tadmor:elliott:2011, elliott:tadmor:2011} and the Stillinger--Weber (SW)-style MX2 potentials~\cite{MO201919462778001,MO677328661525000,MD242389978788001} may be attributed to the fact that they are designed for two-dimensional multicomponent systems.

\begin{figure}
  \begin{subfigure}{0.5\linewidth}
    \includegraphics[width=\linewidth]{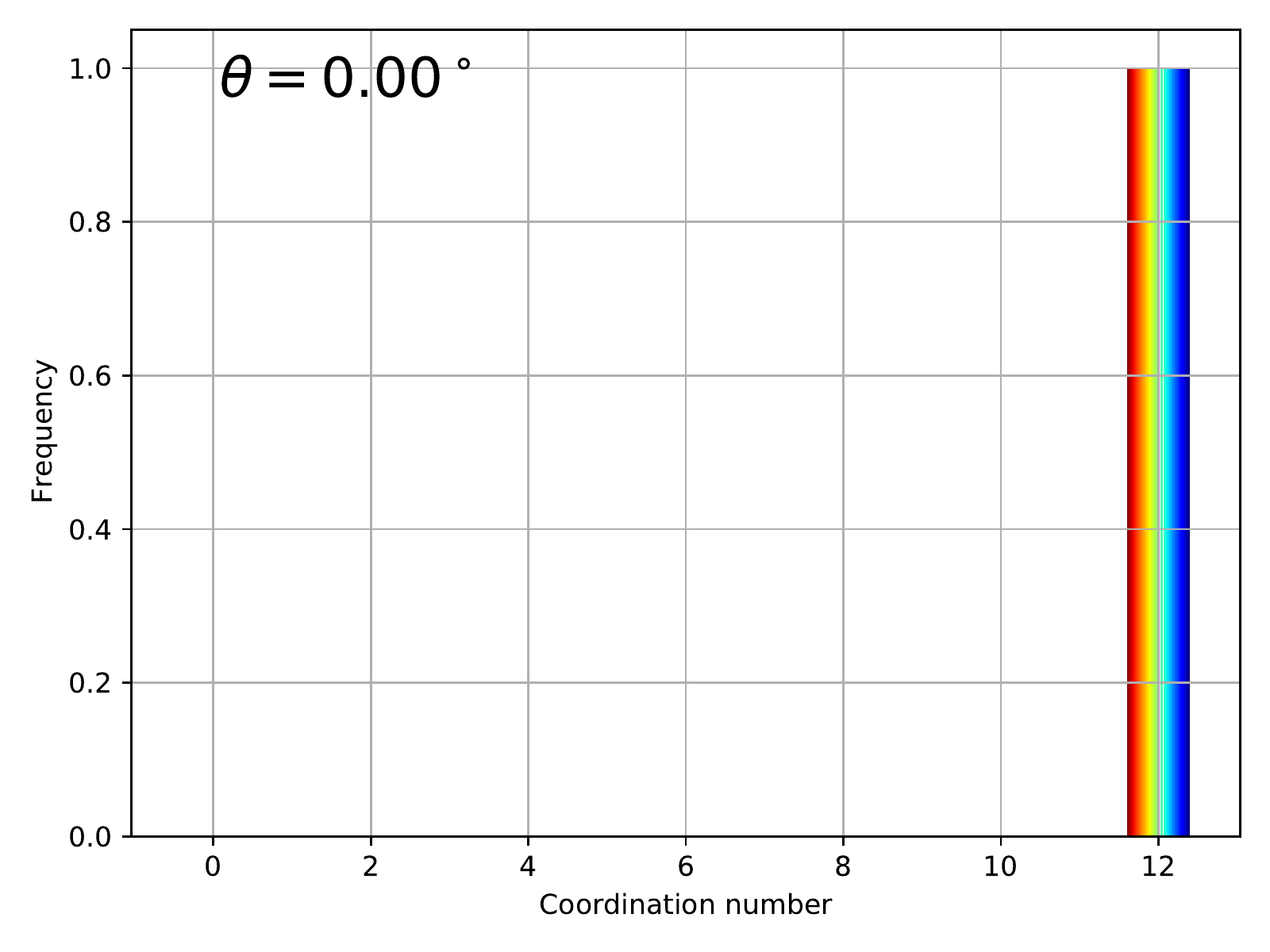}
    \includegraphics[width=\linewidth]{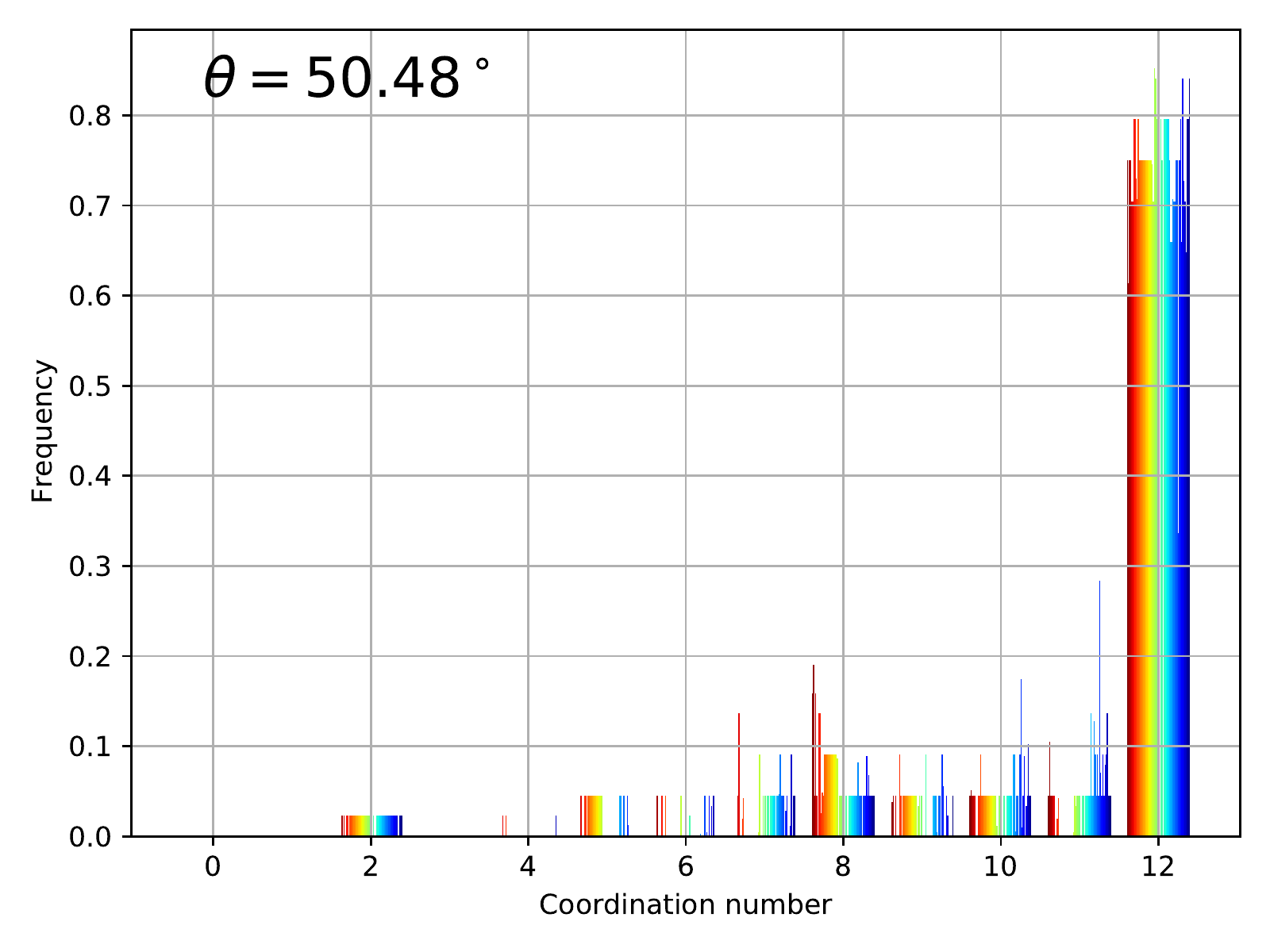}
    \includegraphics[width=\linewidth]{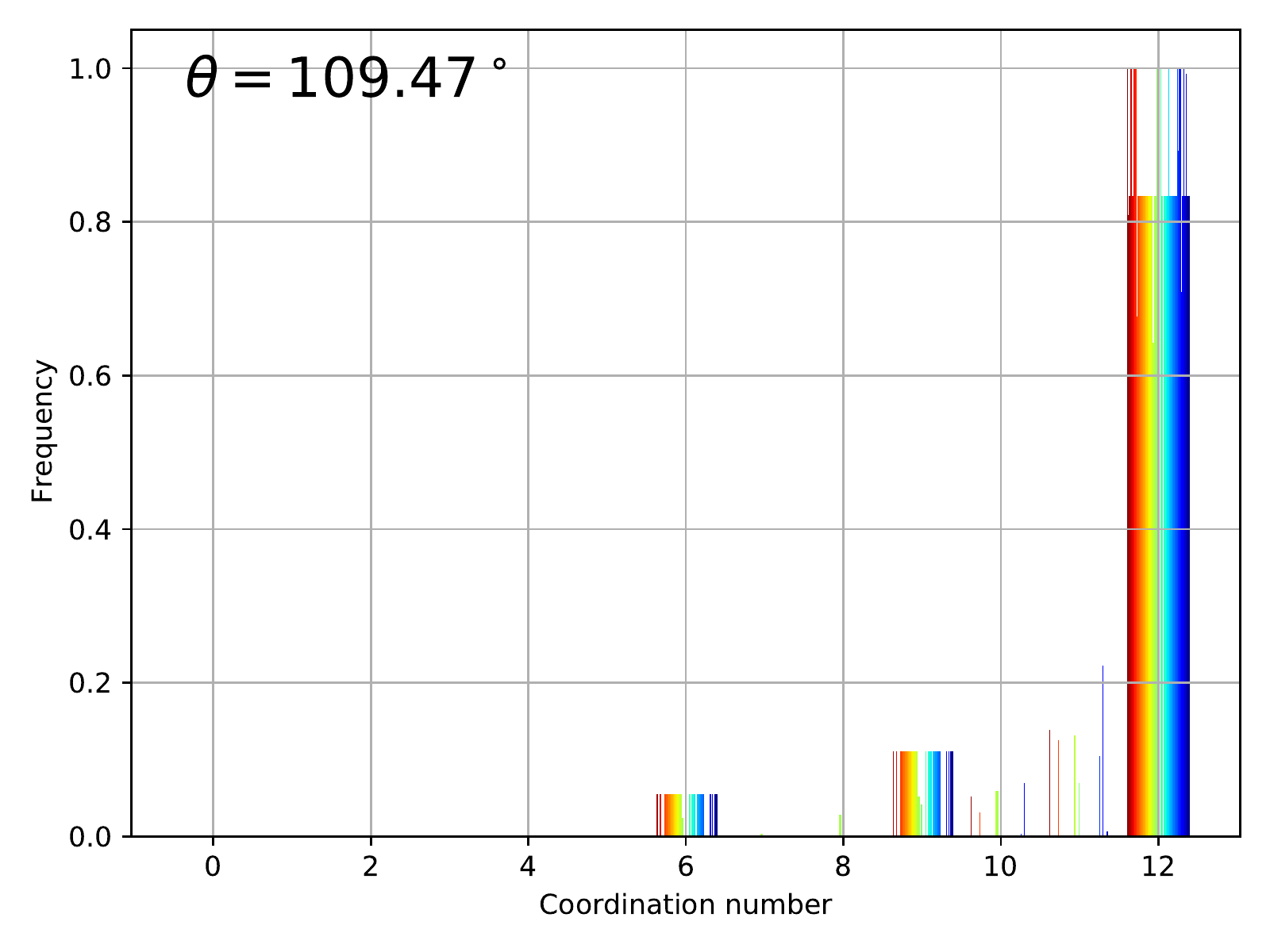}
    \caption{Low-energy boundaries (cusps)}
    \label{fig:hist_fcc_al_cusps}
  \end{subfigure}
  \begin{subfigure}{0.5\linewidth}
    \includegraphics[width=\linewidth]{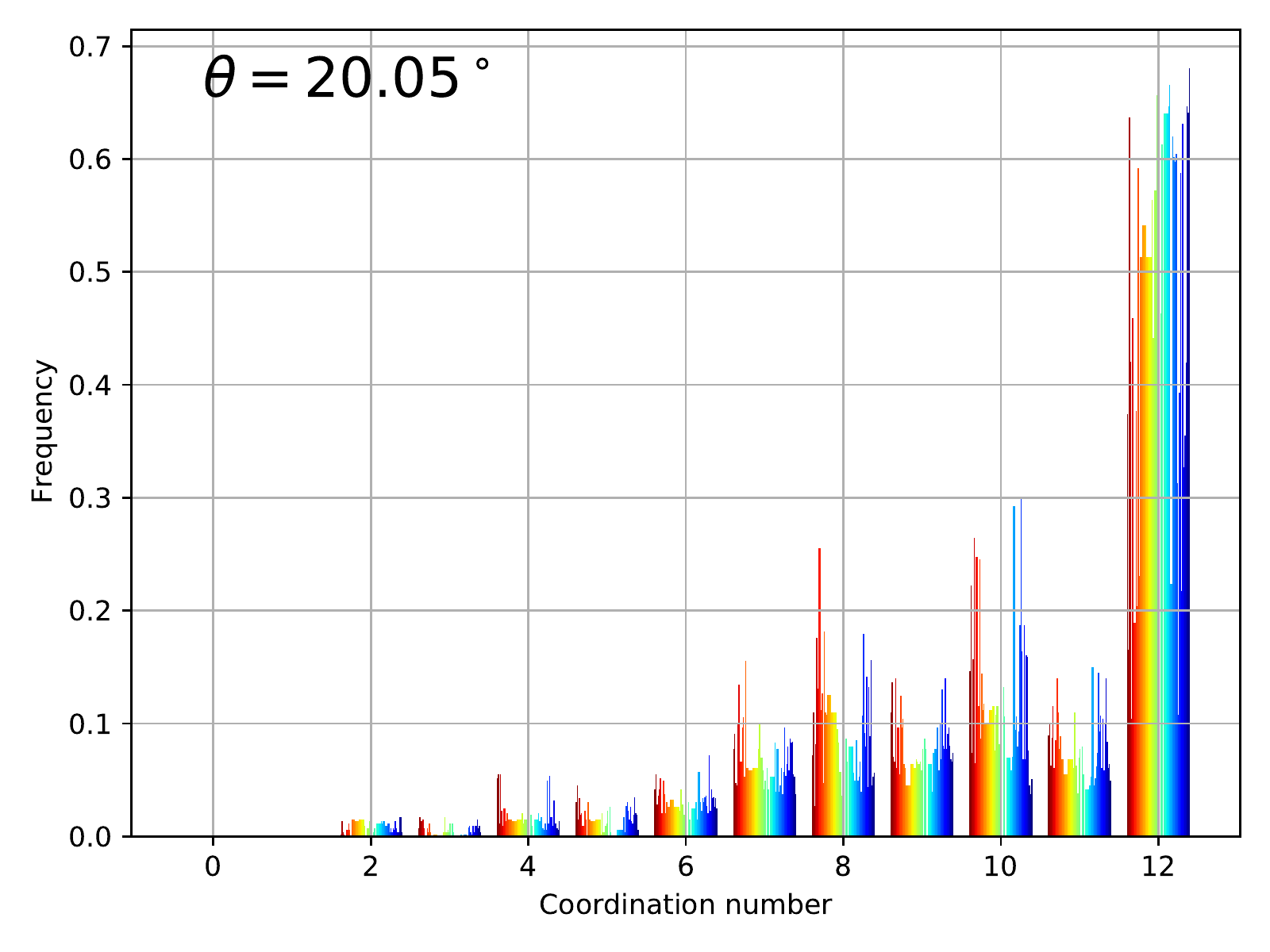}
    \includegraphics[width=\linewidth]{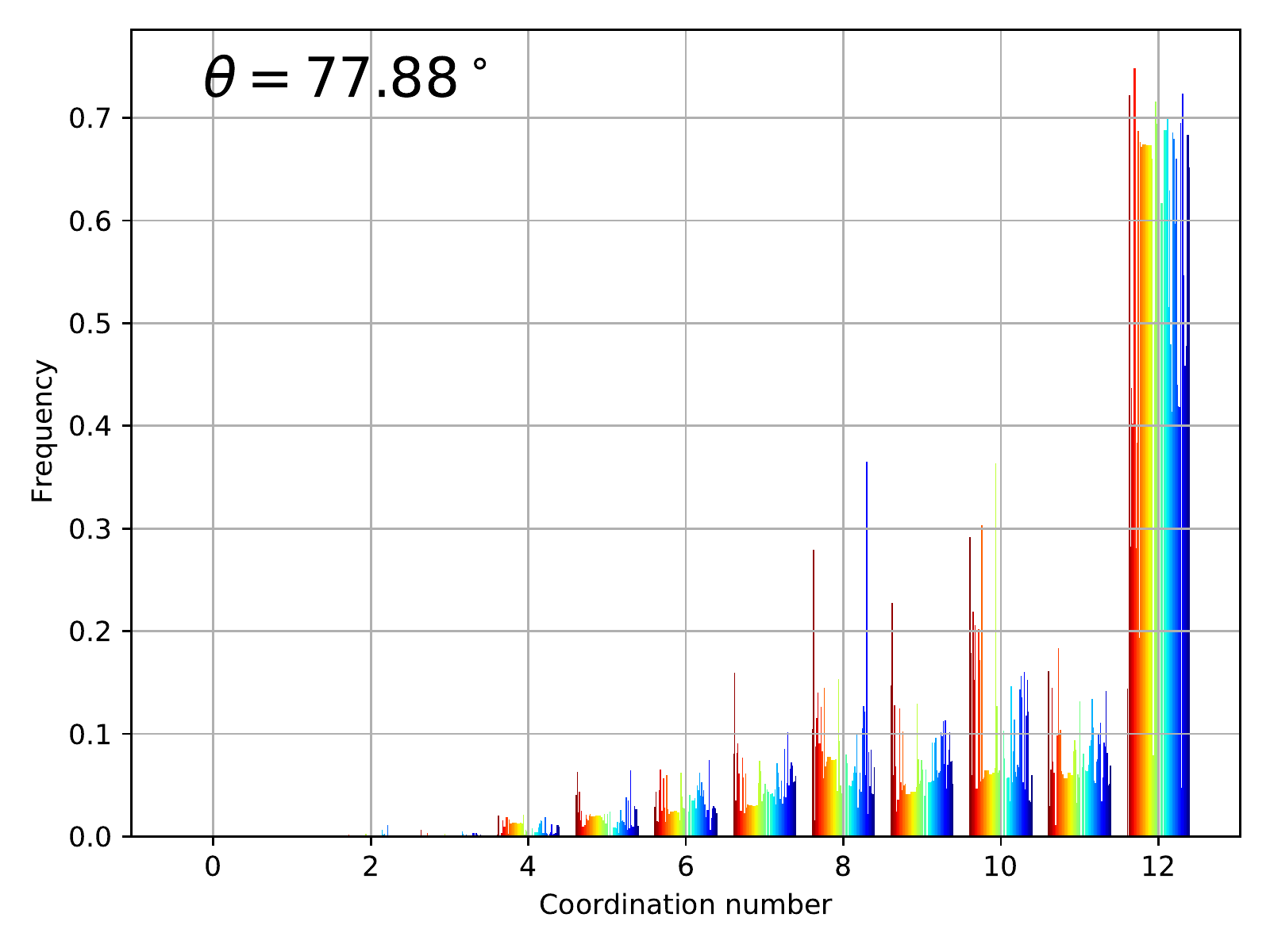}
    \includegraphics[width=\linewidth]{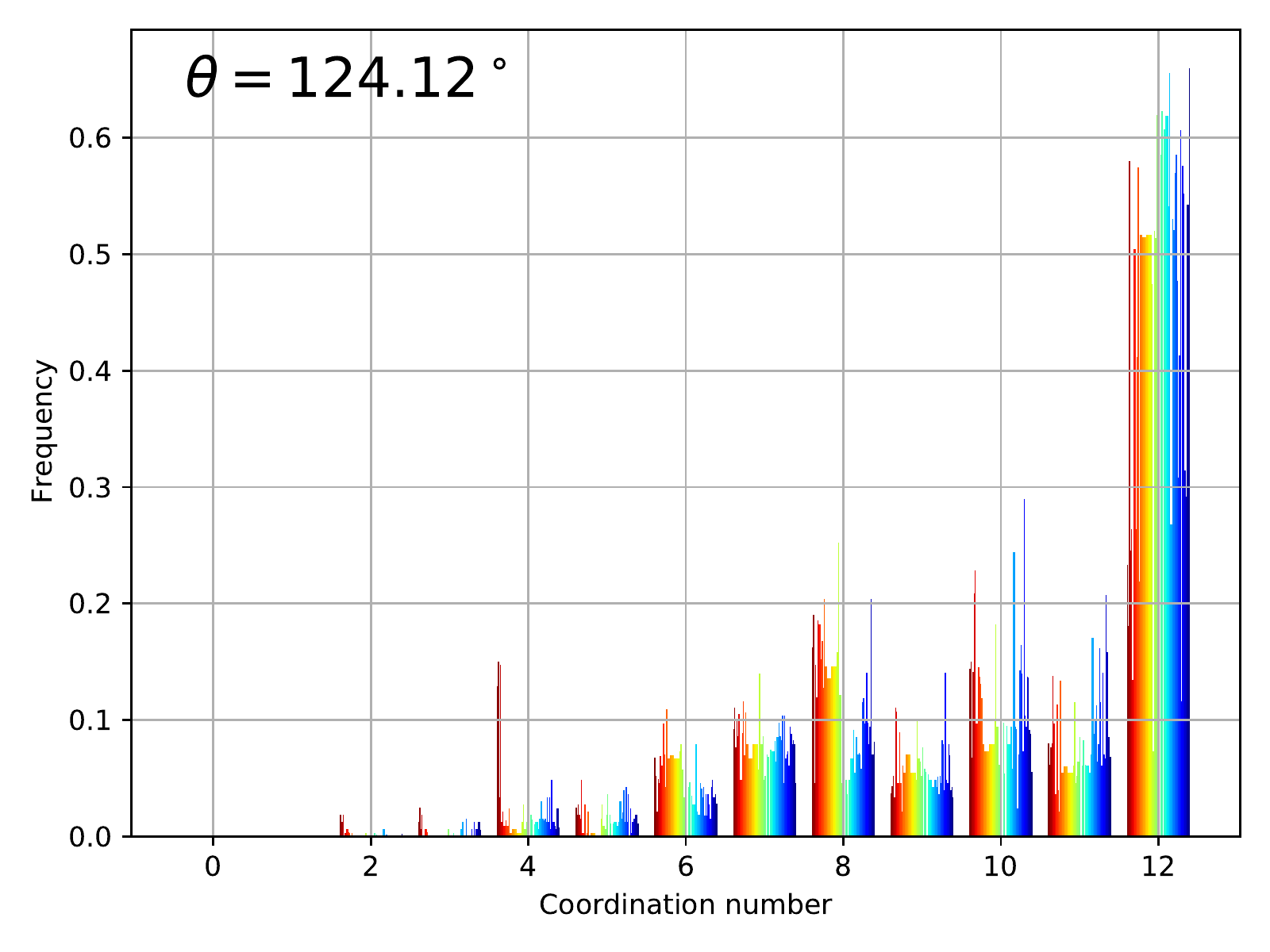}
    \caption{High-energy boundaries}
    \label{fig:hist_fcc_al_hagb}
  \end{subfigure}
  \caption{Combined counts of coordination numbers for all Al \IPs at different angles of the $[110]$ fcc Al boundary.}
  \label{fig:hist_fcc_al}
\end{figure}

To further explore the question of energy variance between \IPs, we consider the spectrum of microscopic degrees of freedom as a function of potential.
Specifically, we consider six $[110]$ tilt boundaries in fcc Al.
Three of the configurations correspond to cusps in the energy ($\theta=0^\circ,50.5^\circ,109.5^\circ$, \cref{fig:hist_fcc_al_cusps}) and three to high-energy boundaries ($\theta=20.1^\circ,77.8^\circ,124.1^\circ$, \cref{fig:hist_fcc_al_hagb}).
Even for this small selection, reporting the full selection of microstates for all $\sim$60 \IPs is not feasible.
Instead, for each boundary, a histogram is presented showing the distribution of coordination numbers for each \IP's lowest energy relaxed structure.
The histograms of all Al \IPs are superimposed, with the bars associated with a given IP colored to correspond to the legend color in \cref{fig:md_fcc_al}.
No differentiation is made between boundary atoms and non-boundary atoms, and so each histogram reflects the coordination numbers for all atoms in the simulation, including atoms in the bulk.
This allows for a more straightforward comparison between \IPs.
However, it also means that the magnitude of the spike at \#=12 depends on the normal simulation domain size, which varies from boundary to boundary.
Therefore, while the histogram magnitudes may be compared potential-to-potential, they should not be compared boundary-to-boundary.

The $\theta=0$ cusp corresponds to the absence of a GB.
As expected, in this case all atoms have a coordination of 12, corresponding to the ideal fcc crystal structure.
For the other two energy cusps, a scattering of additional coordinations appear in addition to 12. However, we note that there is a clear qualitative difference between these cusps. For the $\theta=109.47^\circ$ cusp, for most \IPs, the ground state structure involves primarily three coordinations: 12, 9 and 6. This suggests that most of the \IPs relax to a similar regular geometry, and indeed the scatter in energy at this cusp as seen in \cref{fig:md_fcc_al} for the [110] tilt axis is relatively small. In contrast, for the $\theta=50.48^\circ$ cusp, the spread in coordinations is broader and less uniform across \IPs, suggesting a wider spread in structures, and indeed the spread in energy for this cusp is about 2.5 times larger than for $\theta=109.47^\circ$. This spread includes both IP effects (given the energy spread at $\theta=0^\circ$ for an identical structure), and configuration effects. Turning to the high-energy boundaries, we see that these exhibit a significantly larger spread in both coordination and energies (cf.\ \cref{fig:md_fcc_al}), which indicates a broad spread in ground state structures. The spread in energy in this case is due both to the \IP itself and the ground state structure (which depends indirectly on the IP), as also observed for the low-energy cusps.\footnote{Careful examination of \cref{fig:hist_fcc_al} shows that there is not a direct relationship between configuration and energy across \IPs. Were this true, it would be manifested in \cref{fig:hist_fcc_al} as near-identical frequencies at each coordination number for any given energy band of \IPs, e.g., all of the dark blue bands would have a similar frequency at any given coordination number for each subfigure. That is, the frequencies shown at a given coordination number in any subfigure would appear to vary (piecewise) continuously as the energy decreases from left to right. To the contrary, the rightmost column shows significant variance in coordination distribution for \IPs that predict similar energies for the same GB. Hence, we conclude that different \IPs may drive the minimizer to different configurations, but this does not fully explain the observed energy differences, one must also account for the differences between the \IPs themselves.}


\subsection{Examples of individual potential behavior}
\label{sec:potentials}
The data that has been presented may be used to compare \IPs with respect to GB predictions, and to eventually design \IPs to predict accurate boundary energies.
Although an exhaustive analysis along these lines is outside the scope of this work, it is illuminating to examine the behavior of some specific \IPs.

\begin{figure}
    \centering
    \includegraphics[width=\linewidth]{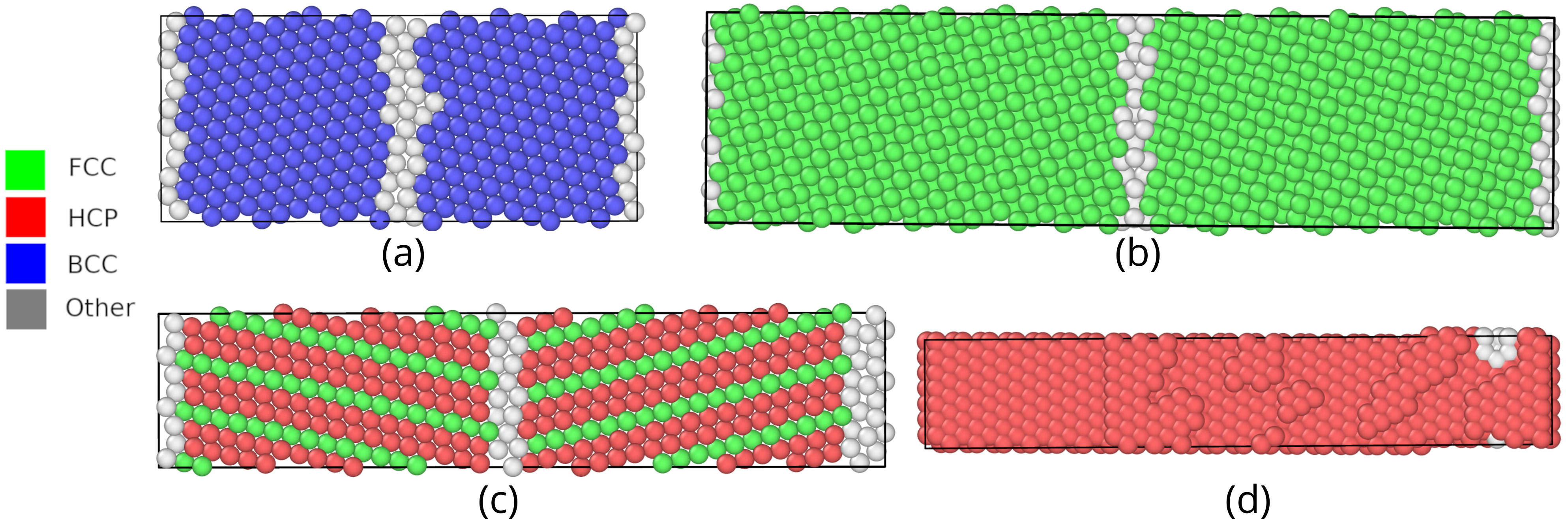}
    \caption{
      Some notable post-relaxation GB configurations.
      Colors shown correspond to the local crystal structure as determined by the common neighbor analysis (CNA)~\cite{honeycutt1987cna} modifier in the OVITO visualization tool~\cite{ovito}.
      (a) shows a typical bcc Fe boundary using the EAM potential developed by Zhou, Johnson, and Wadley~\cite{MO650279905230005} tilted approximately 32 degrees about the $[111]$ axis.
      (b) shows a typical fcc Al GB tilted approximately 31 degrees about the $[100]$ axis using an EAM Dynamo potential by Mishin, Farkas, and Mehl~\cite{MO651801486679005}.
      (c) shows an unstable GB using an EAM IMD potential by Schopf, Brommer, and Frigan~\cite{MO878712978062003} that was initialized as fcc Al tilted 70 degrees about the $[110]$ axis, but underwent a transition to an arrangement of alternating fcc and hcp layers.
      (d) shows what was initially a 55-degree GB about the $[110]$ axis of bcc Mo using a Stillinger--Weber MX2 potential by Kurniawan, Petrie, and Williams~\cite{MO677328661525000}, where the two grain boundaries required for periodic boundary conditions have annihilated and caused almost the entire configuration to transform into hcp.
    }
    \label{fig:structure_visualization}
\end{figure}

The GB energy tests produce for most \IPs a ground state atomic structure that is well-behaved and has an expected form: the domain is divided into two symmetric grains separated in the center and across the periodic boundary by two well-defined GBs.
For example, the result for bcc Fe for a $32^\circ$ tilt about the $[111]$ axis, run using the Zhou, Johnson, and Wadley EAM potential~\cite{MO650279905230005}, is shown in \cref{fig:structure_visualization}a, and clearly has a well-behaved structure.
Moreover, all of the potentials used in benchmark GB literature are well-behaved and produce roughly average results (with respect to the ensemble of \IPs considered here).
Of particular importance are the results for the two potentials used in the well-known Olmsted dataset~\cite{olmsted2009survey} --- the Foiles and Hoyt EAM potential~\cite{MO776437554506000,MD120291908751005} and the Ercolessi--Adams EAM potential~\cite{MO324507536345003,MO324507536345003a}.
Reasonable results and average energies are also found for the Mishin, Farkas, and Mehl EAM potential~\cite{MO400591584784005,MD120291908751005,mishin1999interatomic} used in Homer's recent work~\cite{homer2022examination}, as presented in \cref{fig:structure_visualization}b.

We now consider examples of \IPs that produce unexpected results.
The OpenKIM website provides easy access to descriptions of \IPs, the original citation where the potential was developed, and selected comparisons of canonical properties (e.g., lattice and elastic constants). In the following examples, we demonstrate how such information can be used to explain the behavior of several \IPs that were outliers, unstable, and/or failed to relax. 

The EAM IMD potential of Schopf, Brommer, and Frigan for AlMnPd~\cite{MO878712978062003,MD113599595631003,MO878712978062003a} was found to be unstable for fcc Al at high angle [110] boundaries.
This is a potential developed for a complex metallic alloy (CMA) -- the $\Xi$ phase of AlMnPd, which is a quasicrystal approximant.
It is, therefore, not surprising that it does not provide an accurate description of fcc Al.
Indeed, a further examination of the results of other \kimtests indicates that the corresponding lattice constant and elastic constants predicted by this \IP are extreme in magnitude compared to the other potentials considered here and, furthermore, that its predicted ground state for pure Al is the hcp structure.
\cref{fig:structure_visualization}c shows the relaxed configuration reached for a representative unstable fcc Al [110] boundary, where it can be seen that the initially fcc bulk has developed interspersed bands of atomic environments similar to those found in the hcp structure.

As mentioned in the previous section, the SW MX2 potential of Kurniawan, Petrie, and Williams~\cite{MO677328661525000} was developed for MoS$_2$ layers, and therefore is unsuitable for modeling bulk Mo.
Indeed, there is a disclaimer on this potential's OpenKIM page stating as much, and its ground state can be seen to be hcp.
\Cref{fig:structure_visualization}d shows that at the angles where this potential exhibits unstable behavior, the pair of grain boundaries annihilates and the material relaxes to a nearly perfect hcp structure.
The two EAM IMD potentials by Bromer and Gaehler (``A''~\cite{brommer2006effective,MO122703700223003,MD113599595631003} and ``B''~\cite{brommer2006effective,MO128037485276003,MD113599595631003}) were developed for AlNiCo quasicrystals and both have extreme lattice and elastic constants for Ni.
The ``B'' potential is an outlier for Ni, while the ``A'' potential had massively positive energies and numerous failures to relax, and is not included in the plots.
On the other hand, the lattice and elastic constants for Al are less extreme for both potentials, and their GB energy relations for fcc Al exhibit unremarkable behavior.
Finally, the EAM Dynamo potential of Sun, Zhang, and Mendelev for Fe~\cite{MO044341472608000,MD120291908751005,MO044341472608000a} is an outlier for both fcc and bcc, although less so for fcc (\cref{fig:md_fcc_fe,fig:md_bcc_fe}).
This potential is designed for high-pressure simulations of the Earth's core, and has extreme material constants. 

In general, we found that the relaxed structures of energetically stable outliers were the same as those of well-behaved potentials. 
This should be expected based on their tendency to follow the trends of the energy-angle dependence, and only be incorrect in the magnitude of the GB energy. 

\subsection{Comparison to theoretical benchmark}

\begin{figure}
  \centering
  \begin{subfigure}{0.5\linewidth}
    \includegraphics[width=\linewidth]{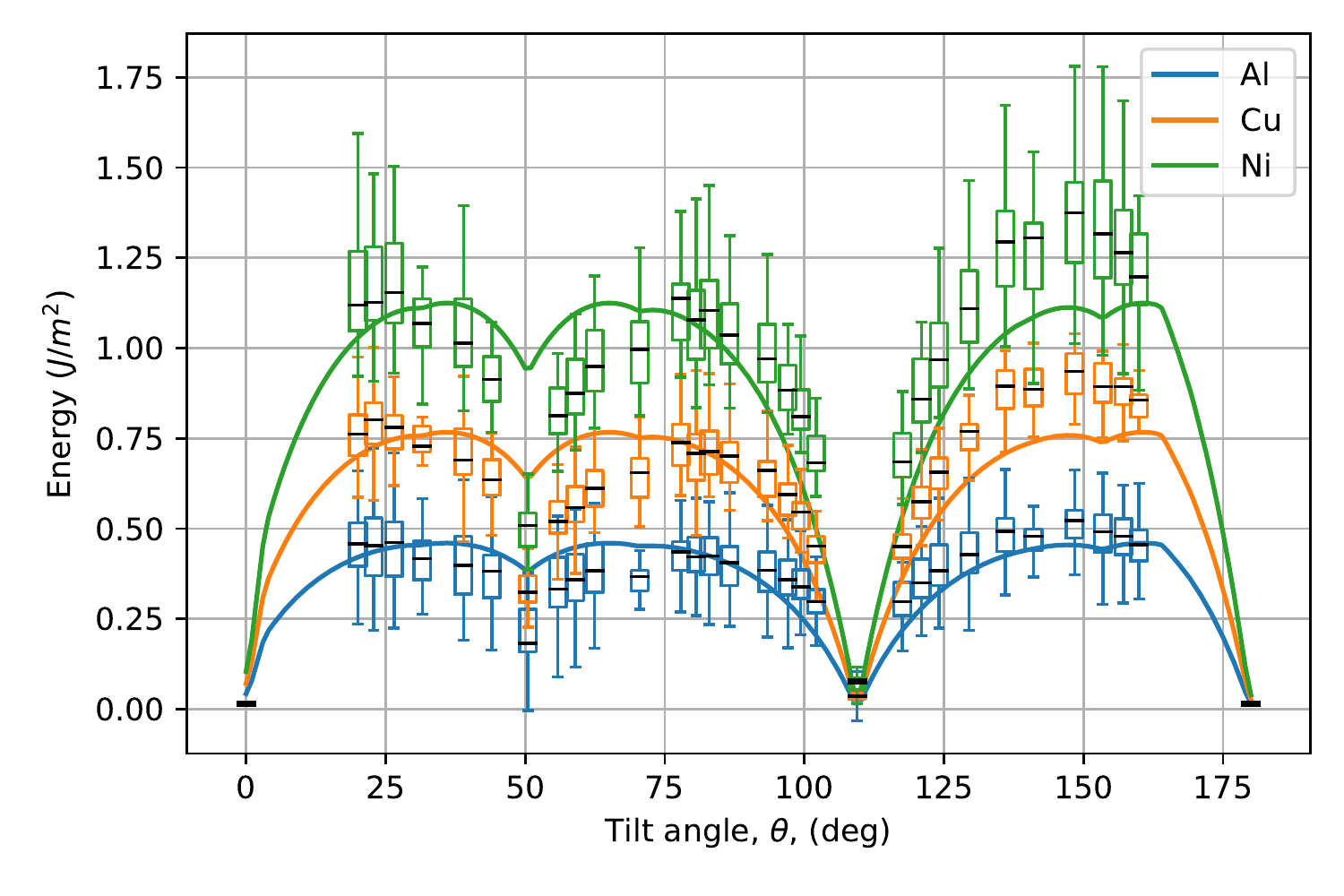}
    \caption{fcc}
    \label{fig:ave_fcc}
  \end{subfigure}%
  \begin{subfigure}{0.5\linewidth}
    \includegraphics[width=\linewidth]{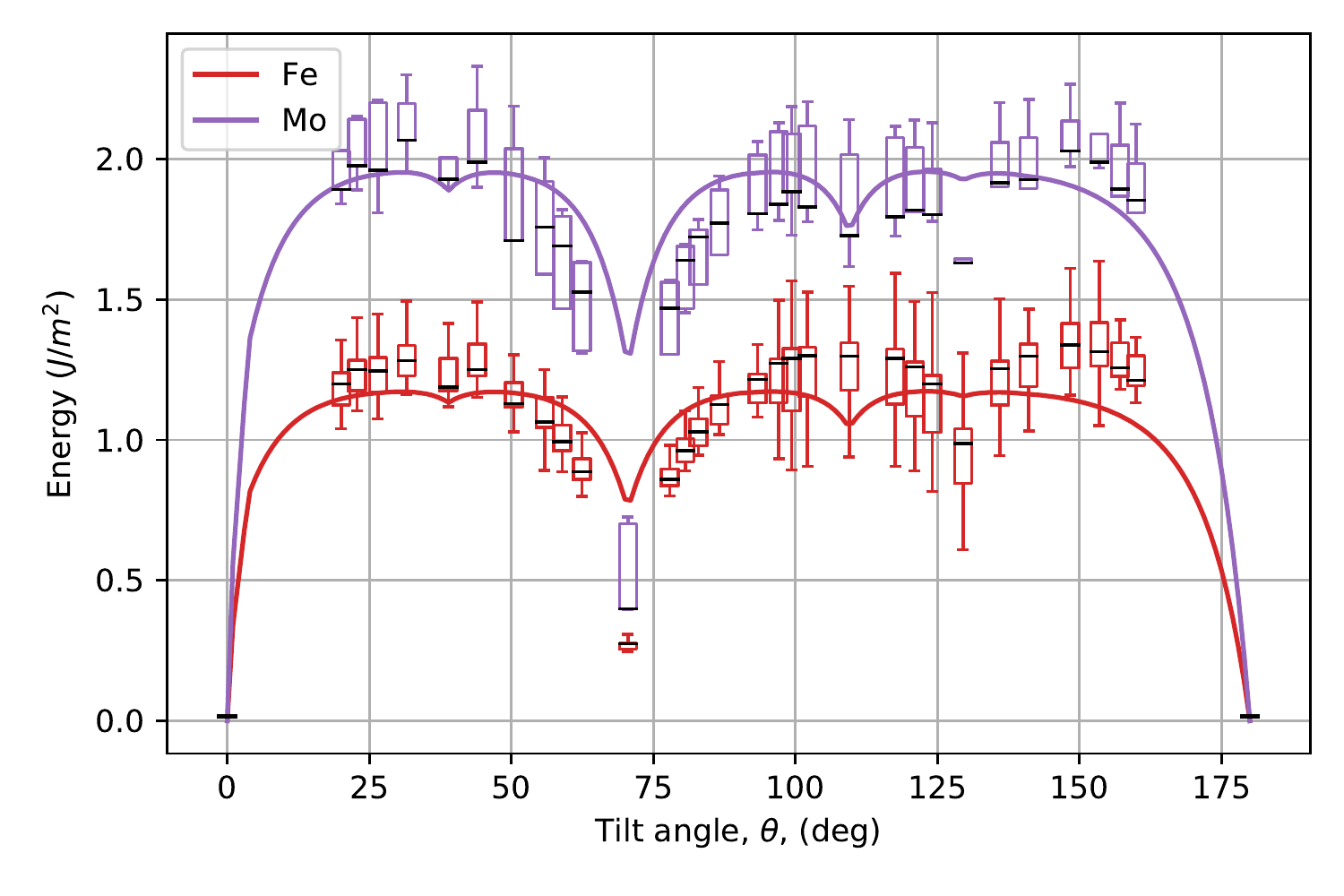}
    \caption{bcc}
    \label{fig:ave_bcc}
  \end{subfigure}
  \caption{Statistical distribution of reported GB energy for all species and potentials over $\hkl[110]$ tilt axes (box plots) compared to lattice matching (solid lines).}
\end{figure}

Given the lack of experimental GB data for the full range of tilt boundaries, we use the results from a theoretical model for comparison.
Combining the lattice matching model of~\cite{runnels2016analytical} with the application of a facet-relaxation scheme~\cite{runnels2015relaxation} provides a benchmark for comparison that does not depend on an extensive set of fitting parameters.
As a geometric/crystallographic method, the lattice matching model depends on only three parameters: (i) a window function parameter $\varepsilon$ that correlates to cusp width, (ii) a thermalization parameter $\sigma$ that corresponds to temperature, and (iii) a scaling factor $E_0$.
Determination of $\varepsilon$ and $\sigma$ is relatively straightforward, and is discussed in great detail (along with a full presentation of the theory) in \cite{runnels2016model}.
The parameter $E_0$, however, is a fitting parameter that linearly scales the reported energy.
In prior work, it was found by calibration to existing atomistic data sets.
These results indicate that there is considerable disagreement between potentials in what the correct value of $E_0$ may be.

An average over the potentials was taken and plotted alongside the lattice matching model fitted to that average for all available species (\cref{fig:ave_fcc,fig:ave_bcc}). 
Although the \IPs display a wide range of energy values, they also demonstrate a remarkable consistency in the general features of the GB energy landscape, correlating closely to the prediction of the geometric lattice matching energy model.
This reinforces the hypothesis that GB energy is primarily driven by the geometry and crystallography, and is consistent across potentials and species up to a scaling factor.
Yet, the value of the scaling factor itself remains undetermined, as the variance between models is large.

\begin{figure}
  \begin{subfigure}{0.5\linewidth}
    \includegraphics[width=\linewidth]{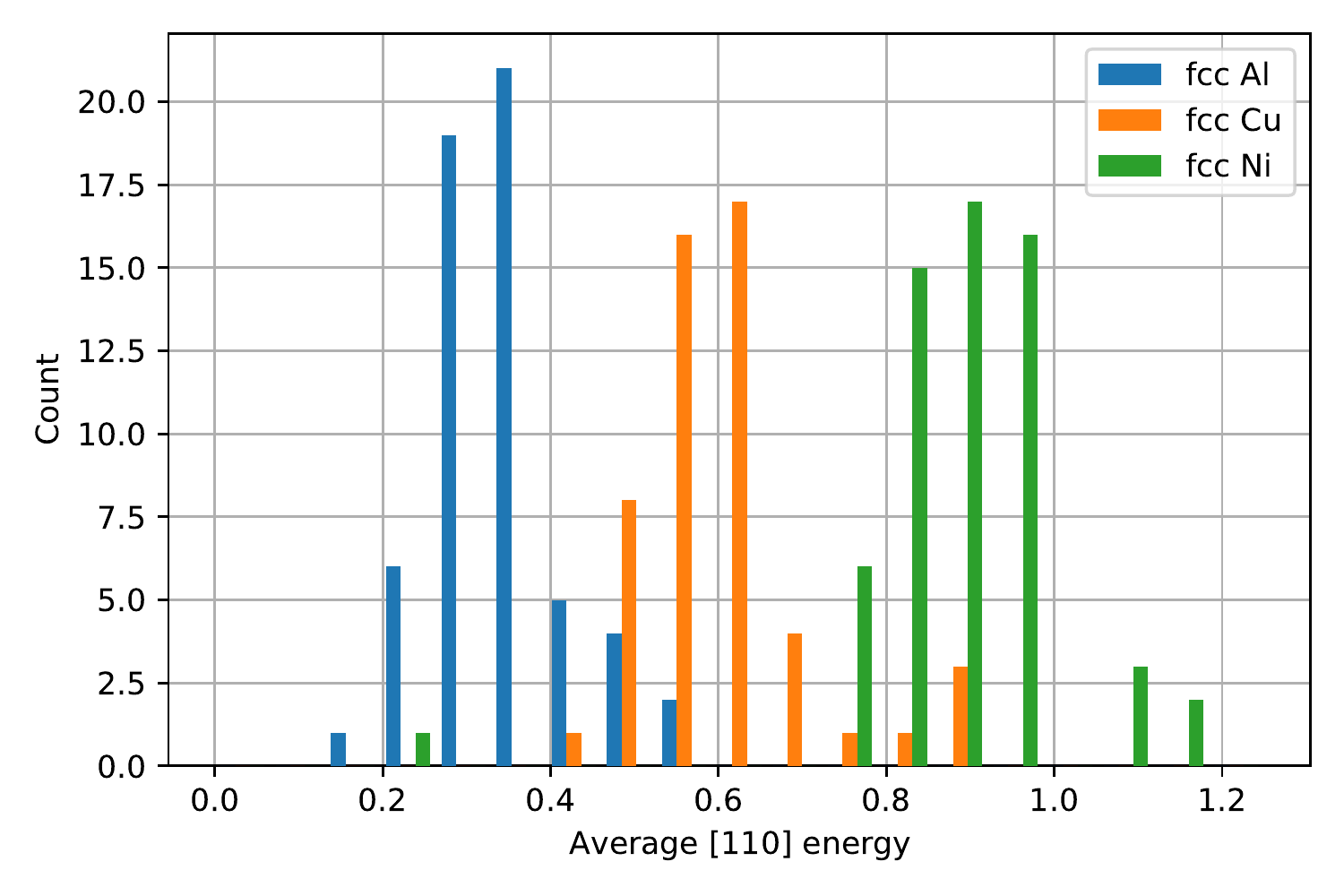}
    \caption{}
  \end{subfigure}
  \begin{subfigure}{0.5\linewidth}
    \includegraphics[width=\linewidth]{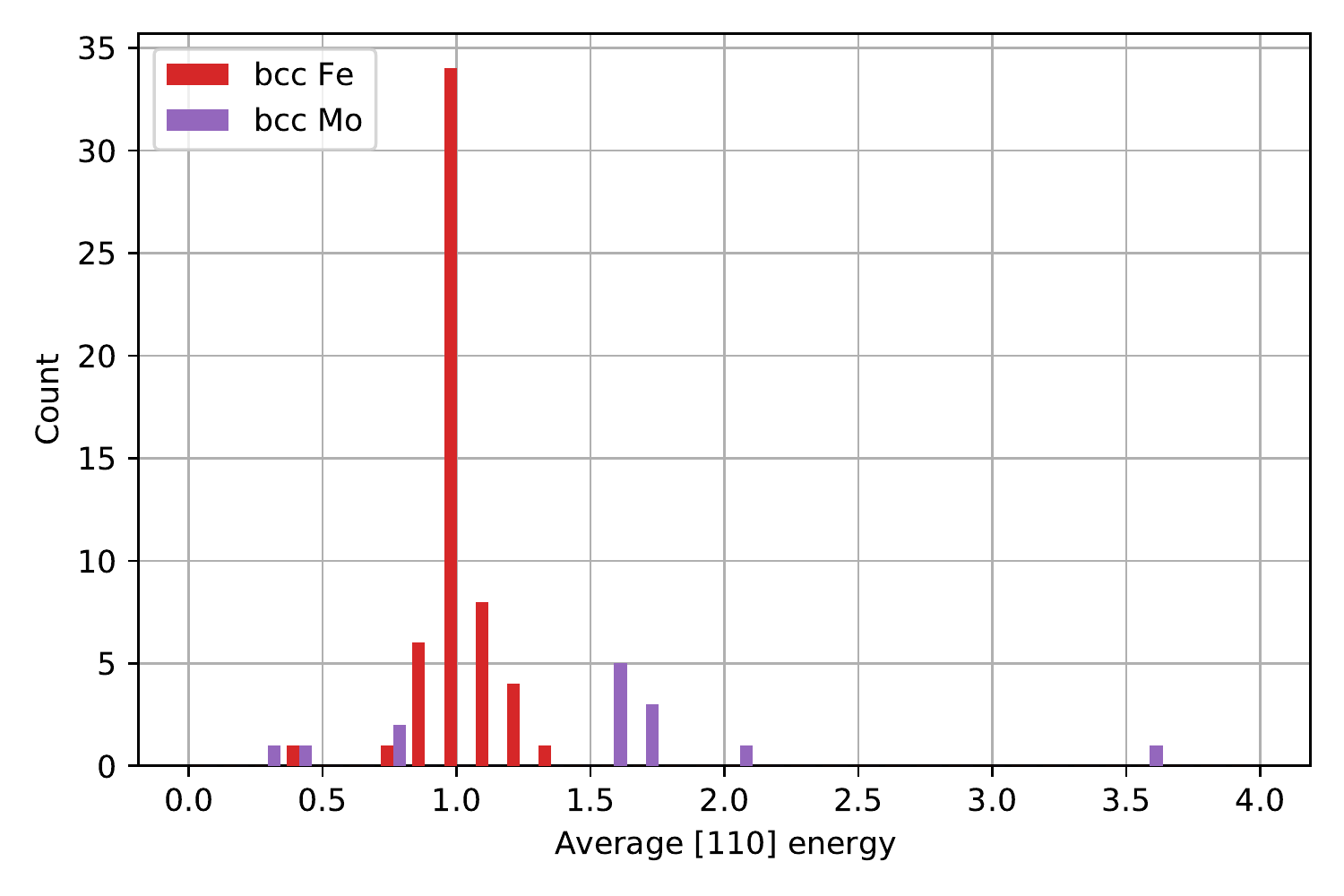}
    \caption{}
  \end{subfigure}
  \caption{Distribution of energies averaged over [110] tilt axis for fcc tests (a) and bcc tests (b)}
  \label{fig:energy_histogram}
\end{figure}

In light of the shortage of available experimental or ab initio data for validation, the large discrepancy in reported energy for different potentials causes some concern regarding the legitimacy of molecular dynamics GB energy calculations for any particular \IP.
\Cref{fig:energy_histogram} shows the distribution of average energies over the [110] tilt axes.
In the case of Al potentials there appears to be some degree of agreement among many \IPs with the distribution mostly concentrated near the mean, although this is not necessarily an indicator of accuracy.
The distributions of fcc Cu and Ni are more lopsided,
with the Fe and Mo series more erratic still.
Overall, these results provide a sense of the uncertainty in GB energy predictions by classical \IPs. The set of \IPs that provide a prediction for a given GB comprise a \emph{multi-model ensemble}. Studies in other physical domains, such as climate change \cite{tebaldi2007UQmultimodel}, have shown that multi-model ensemble predictions generally outperform those of the ``best'' model, and are necessary to account for all aspects of model uncertainty (initial and boundary conditions, parametric and structural). However, care must be used when considering the prediction of a multi-model ensemble due to the possibilities for inter-model dependencies and biases, which can lead to double counting. For example, the high agreement between Al potentials may be due to the fact that many are EAM-type potentials that share a common structural bias and, in some cases, are not independent. Nevertheless the spread in \IP predictions does provide an estimate of uncertainty. Future work should focus on methods for \emph{weighting} ensemble members based on metrics for assessing their likely accuracy (e.g., by considering correlations between GB energy and other property predictions for which reference data is available), as done for example in the reliability ensemble average (REA) approach \cite{giorgi2002UQmultimodel}. The OpenKIM framework is very amenable to such analysis, but a full exploration along these lines is left to future work.

\subsection{Ongoing testing on OpenKIM.org}

The framework presented here is accessible online, and the GB energy test result database is constantly updated as new potentials are added.
The general and systematic framework used to generate these \kimtests makes it possible to add new \kimtests for more materials and \IPs automatically.
The current, raw data for the \kimtests presented here is available at \url{https://openkim.org}.

\section{Conclusion}\label{sec:conclusion}

We have leveraged the OpenKIM framework to systematically compute atomistic symmetric tilt GB energies for multiple \IPs and materials using a newly developed OpenKIM \kimtestdriver.
This system will continue to be applied to all future \IPs that are uploaded to OpenKIM.
The results from all simulations conducted as part of this study (as well as all future calculations) are available online at \url{https://openkim.org}, where they can be explored using text- and graphics-based visualization tools.
The results of these calculations were compared and common trends identified, leading to the conclusion that although the overall shape of the energy landscape appears to be \IP-agnostic, the magnitude of the energy differs significantly between \IPs and necessitates additional validation.

The foremost limitation of the results presented here is the relatively limited global optimization.
Generally, determining the absolute ground state requires a high-resolution grid search over the microscopic GB degrees of freedom; here, only a relatively coarse search is performed
\replaced[id=R2,comment={2.3b}]{
  due to the prohibitively excessive computational time required for a high-throughput framework.
}{
  There are two reasons for this.
  First, the computational time required for a full resolution grid search is prohibitive for a high-throughput framework.
  Second, it is unlikely that ground states will be present in realistic calculations, especially when they are not reachable except by high-resolution grid search.
  Therefore, the energies found by the OpenKIM \kimtest are more likely to be representative of actual energies in large simulations, and therefore of greater importance to most practitioners when simulating systems in which grain boundaries play a role.
}

Future work will focus on extending this work to more general GBs and crystal structures, and to seeking connections between the predicted GB energy and other properties predicted by an \IP in order to better assess the uncertainty in its GB energy predictions within a multi-model ensemble paradigm.
\added[id=R2,comment={2.3a}]{
  To better approximate the ground state in futer iterations of this KIM test, genetic methods such as the USPEX method \cite{cantwell2020grain,mazitov2021grain} or the Monte Carlo method \cite{guziewski2020application} may be used.
}

\section{Acknowledgments}

The authors acknowledge partial support by the National Science Foundation (NSF) under grants No.\ DMR-1834251, DMR-1834332, and OAC-1931304.
BR acknowledges support from the National Science Foundation, grant No.~MOMS-2142164.

Computational support is acknowledged for the following resources: (1) Extreme Science and Engineering Discovery Environment (XSEDE) program allocation TG-PHY130007, which is supported by National Science Foundation grant No.~ACI-1053575~\cite{XSEDE}; (2) Stampede2 supercomputer at the Texas Advanced Computing Center (TACC) as well as the Jetstream2 cloud computing environment at Indiana University through allocation MAT200008 from the Advanced Cyberinfrastructure Coordination Ecosystem: Services and Support (ACCESS) program, which is supported by National Science Foundation grants No.~0941493; (3) Minnesota Supercomputing Institute (MSI) at the University of Minnesota;
and (4) INCLINE cluster at the University of Colorado Colorado Springs, which is supported by the National Science Foundation, grant No.~OAC-2017917.

The authors thank Mark Transtrum for helpful discussions.

\section{Data Availability}

All data required to reproduce these findings are available to download from \url{https://openkim.org/}.

\FloatBarrier
\bibliographystyle{ieeetr}
\bibliography{library,kim}
\end{document}

%% file: resultsv4/data/GrainBoundaryCubicCrystalSymmetricTiltRelaxedEnergyVsAngle_bcc100_Mo__TE_529178526487_001/legend.pgf
\begingroup%
\makeatletter%
\begin{pgfpicture}%
\pgfpathrectangle{\pgfpointorigin}{\pgfqpoint{16.000000in}{4.500000in}}%
\pgfusepath{use as bounding box, clip}%
\begin{pgfscope}%
\pgfsetbuttcap%
\pgfsetmiterjoin%
\pgfsetlinewidth{0.000000pt}%
\definecolor{currentstroke}{rgb}{1.000000,1.000000,1.000000}%
\pgfsetstrokecolor{currentstroke}%
\pgfsetstrokeopacity{0.000000}%
\pgfsetdash{}{0pt}%
\pgfpathmoveto{\pgfqpoint{0.000000in}{0.000000in}}%
\pgfpathlineto{\pgfqpoint{16.000000in}{0.000000in}}%
\pgfpathlineto{\pgfqpoint{16.000000in}{4.500000in}}%
\pgfpathlineto{\pgfqpoint{0.000000in}{4.500000in}}%
\pgfpathclose%
\pgfusepath{}%
\end{pgfscope}%
\begin{pgfscope}%
\pgfsetbuttcap%
\pgfsetmiterjoin%
\definecolor{currentfill}{rgb}{1.000000,1.000000,1.000000}%
\pgfsetfillcolor{currentfill}%
\pgfsetfillopacity{0.800000}%
\pgfsetlinewidth{1.003750pt}%
\definecolor{currentstroke}{rgb}{0.800000,0.800000,0.800000}%
\pgfsetstrokecolor{currentstroke}%
\pgfsetstrokeopacity{0.800000}%
\pgfsetdash{}{0pt}%
\pgfpathmoveto{\pgfqpoint{1.159385in}{1.279584in}}%
\pgfpathlineto{\pgfqpoint{14.840615in}{1.279584in}}%
\pgfpathquadraticcurveto{\pgfqpoint{14.868392in}{1.279584in}}{\pgfqpoint{14.868392in}{1.307361in}}%
\pgfpathlineto{\pgfqpoint{14.868392in}{3.192639in}}%
\pgfpathquadraticcurveto{\pgfqpoint{14.868392in}{3.220416in}}{\pgfqpoint{14.840615in}{3.220416in}}%
\pgfpathlineto{\pgfqpoint{1.159385in}{3.220416in}}%
\pgfpathquadraticcurveto{\pgfqpoint{1.131608in}{3.220416in}}{\pgfqpoint{1.131608in}{3.192639in}}%
\pgfpathlineto{\pgfqpoint{1.131608in}{1.307361in}}%
\pgfpathquadraticcurveto{\pgfqpoint{1.131608in}{1.279584in}}{\pgfqpoint{1.159385in}{1.279584in}}%
\pgfpathclose%
\pgfusepath{stroke,fill}%
\end{pgfscope}%
\begin{pgfscope}%
\pgfsetrectcap%
\pgfsetroundjoin%
\pgfsetlinewidth{1.505625pt}%
\definecolor{currentstroke}{rgb}{0.500000,0.000000,0.000000}%
\pgfsetstrokecolor{currentstroke}%
\pgfsetdash{}{0pt}%
\pgfpathmoveto{\pgfqpoint{1.409385in}{2.887590in}}%
\pgfpathlineto{\pgfqpoint{1.687163in}{2.887590in}}%
\pgfusepath{stroke}%
\end{pgfscope}%
\begin{pgfscope}%
\pgfsetbuttcap%
\pgfsetroundjoin%
\definecolor{currentfill}{rgb}{1.000000,1.000000,1.000000}%
\pgfsetfillcolor{currentfill}%
\pgfsetlinewidth{1.003750pt}%
\definecolor{currentstroke}{rgb}{0.500000,0.000000,0.000000}%
\pgfsetstrokecolor{currentstroke}%
\pgfsetdash{}{0pt}%
\pgfsys@defobject{currentmarker}{\pgfqpoint{-0.020833in}{-0.020833in}}{\pgfqpoint{0.020833in}{0.020833in}}{%
\pgfpathmoveto{\pgfqpoint{0.000000in}{-0.020833in}}%
\pgfpathcurveto{\pgfqpoint{0.005525in}{-0.020833in}}{\pgfqpoint{0.010825in}{-0.018638in}}{\pgfqpoint{0.014731in}{-0.014731in}}%
\pgfpathcurveto{\pgfqpoint{0.018638in}{-0.010825in}}{\pgfqpoint{0.020833in}{-0.005525in}}{\pgfqpoint{0.020833in}{0.000000in}}%
\pgfpathcurveto{\pgfqpoint{0.020833in}{0.005525in}}{\pgfqpoint{0.018638in}{0.010825in}}{\pgfqpoint{0.014731in}{0.014731in}}%
\pgfpathcurveto{\pgfqpoint{0.010825in}{0.018638in}}{\pgfqpoint{0.005525in}{0.020833in}}{\pgfqpoint{0.000000in}{0.020833in}}%
\pgfpathcurveto{\pgfqpoint{-0.005525in}{0.020833in}}{\pgfqpoint{-0.010825in}{0.018638in}}{\pgfqpoint{-0.014731in}{0.014731in}}%
\pgfpathcurveto{\pgfqpoint{-0.018638in}{0.010825in}}{\pgfqpoint{-0.020833in}{0.005525in}}{\pgfqpoint{-0.020833in}{0.000000in}}%
\pgfpathcurveto{\pgfqpoint{-0.020833in}{-0.005525in}}{\pgfqpoint{-0.018638in}{-0.010825in}}{\pgfqpoint{-0.014731in}{-0.014731in}}%
\pgfpathcurveto{\pgfqpoint{-0.010825in}{-0.018638in}}{\pgfqpoint{-0.005525in}{-0.020833in}}{\pgfqpoint{0.000000in}{-0.020833in}}%
\pgfpathclose%
\pgfusepath{stroke,fill}%
}%
\begin{pgfscope}%
\pgfsys@transformshift{1.548274in}{2.887590in}%
\pgfsys@useobject{currentmarker}{}%
\end{pgfscope}%
\end{pgfscope}%
\begin{pgfscope}%
\definecolor{textcolor}{rgb}{0.000000,0.000000,0.000000}%
\pgfsetstrokecolor{textcolor}%
\pgfsetfillcolor{textcolor}%
\pgftext[x=1.798274in,y=2.838979in,left,base]{\color{textcolor}\sffamily\fontsize{10.000000}{12.000000}\selectfont SW\_MX2\_WenShirodkarPlechac\_2017\_MoS \protect{\cite{wen2017force,MO201919462778001,MD242389978788001}}}%
\end{pgfscope}%
\begin{pgfscope}%
\pgfsetrectcap%
\pgfsetroundjoin%
\pgfsetlinewidth{1.505625pt}%
\definecolor{currentstroke}{rgb}{0.749554,0.000000,0.000000}%
\pgfsetstrokecolor{currentstroke}%
\pgfsetdash{}{0pt}%
\pgfpathmoveto{\pgfqpoint{1.409385in}{2.679773in}}%
\pgfpathlineto{\pgfqpoint{1.687163in}{2.679773in}}%
\pgfusepath{stroke}%
\end{pgfscope}%
\begin{pgfscope}%
\pgfsetbuttcap%
\pgfsetroundjoin%
\definecolor{currentfill}{rgb}{1.000000,1.000000,1.000000}%
\pgfsetfillcolor{currentfill}%
\pgfsetlinewidth{1.003750pt}%
\definecolor{currentstroke}{rgb}{0.749554,0.000000,0.000000}%
\pgfsetstrokecolor{currentstroke}%
\pgfsetdash{}{0pt}%
\pgfsys@defobject{currentmarker}{\pgfqpoint{-0.041667in}{-0.041667in}}{\pgfqpoint{0.041667in}{0.041667in}}{%
\pgfpathmoveto{\pgfqpoint{0.000000in}{-0.041667in}}%
\pgfpathcurveto{\pgfqpoint{0.011050in}{-0.041667in}}{\pgfqpoint{0.021649in}{-0.037276in}}{\pgfqpoint{0.029463in}{-0.029463in}}%
\pgfpathcurveto{\pgfqpoint{0.037276in}{-0.021649in}}{\pgfqpoint{0.041667in}{-0.011050in}}{\pgfqpoint{0.041667in}{0.000000in}}%
\pgfpathcurveto{\pgfqpoint{0.041667in}{0.011050in}}{\pgfqpoint{0.037276in}{0.021649in}}{\pgfqpoint{0.029463in}{0.029463in}}%
\pgfpathcurveto{\pgfqpoint{0.021649in}{0.037276in}}{\pgfqpoint{0.011050in}{0.041667in}}{\pgfqpoint{0.000000in}{0.041667in}}%
\pgfpathcurveto{\pgfqpoint{-0.011050in}{0.041667in}}{\pgfqpoint{-0.021649in}{0.037276in}}{\pgfqpoint{-0.029463in}{0.029463in}}%
\pgfpathcurveto{\pgfqpoint{-0.037276in}{0.021649in}}{\pgfqpoint{-0.041667in}{0.011050in}}{\pgfqpoint{-0.041667in}{0.000000in}}%
\pgfpathcurveto{\pgfqpoint{-0.041667in}{-0.011050in}}{\pgfqpoint{-0.037276in}{-0.021649in}}{\pgfqpoint{-0.029463in}{-0.029463in}}%
\pgfpathcurveto{\pgfqpoint{-0.021649in}{-0.037276in}}{\pgfqpoint{-0.011050in}{-0.041667in}}{\pgfqpoint{0.000000in}{-0.041667in}}%
\pgfpathclose%
\pgfusepath{stroke,fill}%
}%
\begin{pgfscope}%
\pgfsys@transformshift{1.548274in}{2.679773in}%
\pgfsys@useobject{currentmarker}{}%
\end{pgfscope}%
\end{pgfscope}%
\begin{pgfscope}%
\definecolor{textcolor}{rgb}{0.000000,0.000000,0.000000}%
\pgfsetstrokecolor{textcolor}%
\pgfsetfillcolor{textcolor}%
\pgftext[x=1.798274in,y=2.631162in,left,base]{\color{textcolor}\sffamily\fontsize{10.000000}{12.000000}\selectfont SNAP\_LiHuChen\_2018\_NiMo \protect{\cite{li2018quantum,MO468686727341000,MD536750310735000}}}%
\end{pgfscope}%
\begin{pgfscope}%
\pgfsetrectcap%
\pgfsetroundjoin%
\pgfsetlinewidth{1.505625pt}%
\definecolor{currentstroke}{rgb}{0.999109,0.073348,0.000000}%
\pgfsetstrokecolor{currentstroke}%
\pgfsetdash{}{0pt}%
\pgfpathmoveto{\pgfqpoint{1.409385in}{2.471956in}}%
\pgfpathlineto{\pgfqpoint{1.687163in}{2.471956in}}%
\pgfusepath{stroke}%
\end{pgfscope}%
\begin{pgfscope}%
\pgfsetbuttcap%
\pgfsetmiterjoin%
\definecolor{currentfill}{rgb}{1.000000,1.000000,1.000000}%
\pgfsetfillcolor{currentfill}%
\pgfsetlinewidth{1.003750pt}%
\definecolor{currentstroke}{rgb}{0.999109,0.073348,0.000000}%
\pgfsetstrokecolor{currentstroke}%
\pgfsetdash{}{0pt}%
\pgfsys@defobject{currentmarker}{\pgfqpoint{-0.041667in}{-0.041667in}}{\pgfqpoint{0.041667in}{0.041667in}}{%
\pgfpathmoveto{\pgfqpoint{-0.000000in}{-0.041667in}}%
\pgfpathlineto{\pgfqpoint{0.041667in}{0.041667in}}%
\pgfpathlineto{\pgfqpoint{-0.041667in}{0.041667in}}%
\pgfpathclose%
\pgfusepath{stroke,fill}%
}%
\begin{pgfscope}%
\pgfsys@transformshift{1.548274in}{2.471956in}%
\pgfsys@useobject{currentmarker}{}%
\end{pgfscope}%
\end{pgfscope}%
\begin{pgfscope}%
\definecolor{textcolor}{rgb}{0.000000,0.000000,0.000000}%
\pgfsetstrokecolor{textcolor}%
\pgfsetfillcolor{textcolor}%
\pgftext[x=1.798274in,y=2.423345in,left,base]{\color{textcolor}\sffamily\fontsize{10.000000}{12.000000}\selectfont SNAP\_ChenDengTran\_2017\_Mo \protect{\cite{MO698578166685000a,MO698578166685000,MD536750310735000}}}%
\end{pgfscope}%
\begin{pgfscope}%
\pgfsetrectcap%
\pgfsetroundjoin%
\pgfsetlinewidth{1.505625pt}%
\definecolor{currentstroke}{rgb}{1.000000,0.276688,0.000000}%
\pgfsetstrokecolor{currentstroke}%
\pgfsetdash{}{0pt}%
\pgfpathmoveto{\pgfqpoint{1.409385in}{2.264138in}}%
\pgfpathlineto{\pgfqpoint{1.687163in}{2.264138in}}%
\pgfusepath{stroke}%
\end{pgfscope}%
\begin{pgfscope}%
\pgfsetbuttcap%
\pgfsetmiterjoin%
\definecolor{currentfill}{rgb}{1.000000,1.000000,1.000000}%
\pgfsetfillcolor{currentfill}%
\pgfsetlinewidth{1.003750pt}%
\definecolor{currentstroke}{rgb}{1.000000,0.276688,0.000000}%
\pgfsetstrokecolor{currentstroke}%
\pgfsetdash{}{0pt}%
\pgfsys@defobject{currentmarker}{\pgfqpoint{-0.041667in}{-0.041667in}}{\pgfqpoint{0.041667in}{0.041667in}}{%
\pgfpathmoveto{\pgfqpoint{0.000000in}{0.041667in}}%
\pgfpathlineto{\pgfqpoint{-0.041667in}{-0.041667in}}%
\pgfpathlineto{\pgfqpoint{0.041667in}{-0.041667in}}%
\pgfpathclose%
\pgfusepath{stroke,fill}%
}%
\begin{pgfscope}%
\pgfsys@transformshift{1.548274in}{2.264138in}%
\pgfsys@useobject{currentmarker}{}%
\end{pgfscope}%
\end{pgfscope}%
\begin{pgfscope}%
\definecolor{textcolor}{rgb}{0.000000,0.000000,0.000000}%
\pgfsetstrokecolor{textcolor}%
\pgfsetfillcolor{textcolor}%
\pgftext[x=1.798274in,y=2.215527in,left,base]{\color{textcolor}\sffamily\fontsize{10.000000}{12.000000}\selectfont SNAP\_LiChenZheng\_2019\_NbTaWMo \protect{\cite{MO560387080449000a,MO560387080449000,MD536750310735000}}}%
\end{pgfscope}%
\begin{pgfscope}%
\pgfsetrectcap%
\pgfsetroundjoin%
\pgfsetlinewidth{1.505625pt}%
\definecolor{currentstroke}{rgb}{1.000000,0.480029,0.000000}%
\pgfsetstrokecolor{currentstroke}%
\pgfsetdash{}{0pt}%
\pgfpathmoveto{\pgfqpoint{1.409385in}{2.056321in}}%
\pgfpathlineto{\pgfqpoint{1.687163in}{2.056321in}}%
\pgfusepath{stroke}%
\end{pgfscope}%
\begin{pgfscope}%
\pgfsetbuttcap%
\pgfsetmiterjoin%
\definecolor{currentfill}{rgb}{1.000000,1.000000,1.000000}%
\pgfsetfillcolor{currentfill}%
\pgfsetlinewidth{1.003750pt}%
\definecolor{currentstroke}{rgb}{1.000000,0.480029,0.000000}%
\pgfsetstrokecolor{currentstroke}%
\pgfsetdash{}{0pt}%
\pgfsys@defobject{currentmarker}{\pgfqpoint{-0.041667in}{-0.041667in}}{\pgfqpoint{0.041667in}{0.041667in}}{%
\pgfpathmoveto{\pgfqpoint{-0.041667in}{0.000000in}}%
\pgfpathlineto{\pgfqpoint{0.041667in}{-0.041667in}}%
\pgfpathlineto{\pgfqpoint{0.041667in}{0.041667in}}%
\pgfpathclose%
\pgfusepath{stroke,fill}%
}%
\begin{pgfscope}%
\pgfsys@transformshift{1.548274in}{2.056321in}%
\pgfsys@useobject{currentmarker}{}%
\end{pgfscope}%
\end{pgfscope}%
\begin{pgfscope}%
\definecolor{textcolor}{rgb}{0.000000,0.000000,0.000000}%
\pgfsetstrokecolor{textcolor}%
\pgfsetfillcolor{textcolor}%
\pgftext[x=1.798274in,y=2.007710in,left,base]{\color{textcolor}\sffamily\fontsize{10.000000}{12.000000}\selectfont MEAM\_LAMMPS\_ParkFellingerLenosky\_2012\_Mo \protect{\cite{MO269937397263001a,MO269937397263001,MD249792265679001}}}%
\end{pgfscope}%
\begin{pgfscope}%
\pgfsetrectcap%
\pgfsetroundjoin%
\pgfsetlinewidth{1.505625pt}%
\definecolor{currentstroke}{rgb}{1.000000,0.697894,0.000000}%
\pgfsetstrokecolor{currentstroke}%
\pgfsetdash{}{0pt}%
\pgfpathmoveto{\pgfqpoint{1.409385in}{1.848504in}}%
\pgfpathlineto{\pgfqpoint{1.687163in}{1.848504in}}%
\pgfusepath{stroke}%
\end{pgfscope}%
\begin{pgfscope}%
\pgfsetbuttcap%
\pgfsetmiterjoin%
\definecolor{currentfill}{rgb}{1.000000,1.000000,1.000000}%
\pgfsetfillcolor{currentfill}%
\pgfsetlinewidth{1.003750pt}%
\definecolor{currentstroke}{rgb}{1.000000,0.697894,0.000000}%
\pgfsetstrokecolor{currentstroke}%
\pgfsetdash{}{0pt}%
\pgfsys@defobject{currentmarker}{\pgfqpoint{-0.041667in}{-0.041667in}}{\pgfqpoint{0.041667in}{0.041667in}}{%
\pgfpathmoveto{\pgfqpoint{0.041667in}{-0.000000in}}%
\pgfpathlineto{\pgfqpoint{-0.041667in}{0.041667in}}%
\pgfpathlineto{\pgfqpoint{-0.041667in}{-0.041667in}}%
\pgfpathclose%
\pgfusepath{stroke,fill}%
}%
\begin{pgfscope}%
\pgfsys@transformshift{1.548274in}{1.848504in}%
\pgfsys@useobject{currentmarker}{}%
\end{pgfscope}%
\end{pgfscope}%
\begin{pgfscope}%
\definecolor{textcolor}{rgb}{0.000000,0.000000,0.000000}%
\pgfsetstrokecolor{textcolor}%
\pgfsetfillcolor{textcolor}%
\pgftext[x=1.798274in,y=1.799893in,left,base]{\color{textcolor}\sffamily\fontsize{10.000000}{12.000000}\selectfont Sim\_LAMMPS\_ADP\_StarikovKolotovaKuksin\_2017\_UMo \protect{\cite{SM682749584055000a,SM682749584055000}}}%
\end{pgfscope}%
\begin{pgfscope}%
\pgfsetrectcap%
\pgfsetroundjoin%
\pgfsetlinewidth{1.505625pt}%
\definecolor{currentstroke}{rgb}{1.000000,0.901235,0.000000}%
\pgfsetstrokecolor{currentstroke}%
\pgfsetdash{}{0pt}%
\pgfpathmoveto{\pgfqpoint{1.409385in}{1.640686in}}%
\pgfpathlineto{\pgfqpoint{1.687163in}{1.640686in}}%
\pgfusepath{stroke}%
\end{pgfscope}%
\begin{pgfscope}%
\pgfsetbuttcap%
\pgfsetroundjoin%
\definecolor{currentfill}{rgb}{1.000000,1.000000,1.000000}%
\pgfsetfillcolor{currentfill}%
\pgfsetlinewidth{1.003750pt}%
\definecolor{currentstroke}{rgb}{1.000000,0.901235,0.000000}%
\pgfsetstrokecolor{currentstroke}%
\pgfsetdash{}{0pt}%
\pgfsys@defobject{currentmarker}{\pgfqpoint{-0.033333in}{-0.041667in}}{\pgfqpoint{0.033333in}{0.020833in}}{%
\pgfpathmoveto{\pgfqpoint{0.000000in}{0.000000in}}%
\pgfpathlineto{\pgfqpoint{0.000000in}{-0.041667in}}%
\pgfpathmoveto{\pgfqpoint{0.000000in}{0.000000in}}%
\pgfpathlineto{\pgfqpoint{0.033333in}{0.020833in}}%
\pgfpathmoveto{\pgfqpoint{0.000000in}{0.000000in}}%
\pgfpathlineto{\pgfqpoint{-0.033333in}{0.020833in}}%
\pgfusepath{stroke,fill}%
}%
\begin{pgfscope}%
\pgfsys@transformshift{1.548274in}{1.640686in}%
\pgfsys@useobject{currentmarker}{}%
\end{pgfscope}%
\end{pgfscope}%
\begin{pgfscope}%
\definecolor{textcolor}{rgb}{0.000000,0.000000,0.000000}%
\pgfsetstrokecolor{textcolor}%
\pgfsetfillcolor{textcolor}%
\pgftext[x=1.798274in,y=1.592075in,left,base]{\color{textcolor}\sffamily\fontsize{10.000000}{12.000000}\selectfont SNAP\_ZuoChenLi\_2019\_Mo \protect{\cite{zuo2020performance,MO014123846623000,MD536750310735000}}}%
\end{pgfscope}%
\begin{pgfscope}%
\pgfsetrectcap%
\pgfsetroundjoin%
\pgfsetlinewidth{1.505625pt}%
\definecolor{currentstroke}{rgb}{0.844402,1.000000,0.123340}%
\pgfsetstrokecolor{currentstroke}%
\pgfsetdash{}{0pt}%
\pgfpathmoveto{\pgfqpoint{5.998512in}{2.887590in}}%
\pgfpathlineto{\pgfqpoint{6.276290in}{2.887590in}}%
\pgfusepath{stroke}%
\end{pgfscope}%
\begin{pgfscope}%
\pgfsetbuttcap%
\pgfsetmiterjoin%
\definecolor{currentfill}{rgb}{1.000000,1.000000,1.000000}%
\pgfsetfillcolor{currentfill}%
\pgfsetlinewidth{1.003750pt}%
\definecolor{currentstroke}{rgb}{0.844402,1.000000,0.123340}%
\pgfsetstrokecolor{currentstroke}%
\pgfsetdash{}{0pt}%
\pgfsys@defobject{currentmarker}{\pgfqpoint{-0.041667in}{-0.041667in}}{\pgfqpoint{0.041667in}{0.041667in}}{%
\pgfpathmoveto{\pgfqpoint{-0.041667in}{-0.041667in}}%
\pgfpathlineto{\pgfqpoint{0.041667in}{-0.041667in}}%
\pgfpathlineto{\pgfqpoint{0.041667in}{0.041667in}}%
\pgfpathlineto{\pgfqpoint{-0.041667in}{0.041667in}}%
\pgfpathclose%
\pgfusepath{stroke,fill}%
}%
\begin{pgfscope}%
\pgfsys@transformshift{6.137401in}{2.887590in}%
\pgfsys@useobject{currentmarker}{}%
\end{pgfscope}%
\end{pgfscope}%
\begin{pgfscope}%
\definecolor{textcolor}{rgb}{0.000000,0.000000,0.000000}%
\pgfsetstrokecolor{textcolor}%
\pgfsetfillcolor{textcolor}%
\pgftext[x=6.387401in,y=2.838979in,left,base]{\color{textcolor}\sffamily\fontsize{10.000000}{12.000000}\selectfont EAM\_MagneticCubic\_DerletNguyenDudarev\_2007\_Mo \protect{\cite{MO424746498193002a,MO424746498193002,MD620624592962002}}}%
\end{pgfscope}%
\begin{pgfscope}%
\pgfsetrectcap%
\pgfsetroundjoin%
\pgfsetlinewidth{1.505625pt}%
\definecolor{currentstroke}{rgb}{0.667299,1.000000,0.300443}%
\pgfsetstrokecolor{currentstroke}%
\pgfsetdash{}{0pt}%
\pgfpathmoveto{\pgfqpoint{5.998512in}{2.679773in}}%
\pgfpathlineto{\pgfqpoint{6.276290in}{2.679773in}}%
\pgfusepath{stroke}%
\end{pgfscope}%
\begin{pgfscope}%
\pgfsetbuttcap%
\pgfsetmiterjoin%
\definecolor{currentfill}{rgb}{1.000000,1.000000,1.000000}%
\pgfsetfillcolor{currentfill}%
\pgfsetlinewidth{1.003750pt}%
\definecolor{currentstroke}{rgb}{0.667299,1.000000,0.300443}%
\pgfsetstrokecolor{currentstroke}%
\pgfsetdash{}{0pt}%
\pgfsys@defobject{currentmarker}{\pgfqpoint{-0.039627in}{-0.033709in}}{\pgfqpoint{0.039627in}{0.041667in}}{%
\pgfpathmoveto{\pgfqpoint{0.000000in}{0.041667in}}%
\pgfpathlineto{\pgfqpoint{-0.039627in}{0.012876in}}%
\pgfpathlineto{\pgfqpoint{-0.024491in}{-0.033709in}}%
\pgfpathlineto{\pgfqpoint{0.024491in}{-0.033709in}}%
\pgfpathlineto{\pgfqpoint{0.039627in}{0.012876in}}%
\pgfpathclose%
\pgfusepath{stroke,fill}%
}%
\begin{pgfscope}%
\pgfsys@transformshift{6.137401in}{2.679773in}%
\pgfsys@useobject{currentmarker}{}%
\end{pgfscope}%
\end{pgfscope}%
\begin{pgfscope}%
\definecolor{textcolor}{rgb}{0.000000,0.000000,0.000000}%
\pgfsetstrokecolor{textcolor}%
\pgfsetfillcolor{textcolor}%
\pgftext[x=6.387401in,y=2.631162in,left,base]{\color{textcolor}\sffamily\fontsize{10.000000}{12.000000}\selectfont MEAM\_LAMMPS\_KimSeolJi\_2017\_PtMo \protect{\cite{kim2017second,MO831380044253001,MD249792265679001}}}%
\end{pgfscope}%
\begin{pgfscope}%
\pgfsetrectcap%
\pgfsetroundjoin%
\pgfsetlinewidth{1.505625pt}%
\definecolor{currentstroke}{rgb}{0.490196,1.000000,0.477546}%
\pgfsetstrokecolor{currentstroke}%
\pgfsetdash{}{0pt}%
\pgfpathmoveto{\pgfqpoint{5.998512in}{2.471956in}}%
\pgfpathlineto{\pgfqpoint{6.276290in}{2.471956in}}%
\pgfusepath{stroke}%
\end{pgfscope}%
\begin{pgfscope}%
\pgfsetbuttcap%
\pgfsetmiterjoin%
\definecolor{currentfill}{rgb}{1.000000,1.000000,1.000000}%
\pgfsetfillcolor{currentfill}%
\pgfsetlinewidth{1.003750pt}%
\definecolor{currentstroke}{rgb}{0.490196,1.000000,0.477546}%
\pgfsetstrokecolor{currentstroke}%
\pgfsetdash{}{0pt}%
\pgfsys@defobject{currentmarker}{\pgfqpoint{-0.041667in}{-0.041667in}}{\pgfqpoint{0.041667in}{0.041667in}}{%
\pgfpathmoveto{\pgfqpoint{-0.013889in}{-0.041667in}}%
\pgfpathlineto{\pgfqpoint{0.013889in}{-0.041667in}}%
\pgfpathlineto{\pgfqpoint{0.013889in}{-0.013889in}}%
\pgfpathlineto{\pgfqpoint{0.041667in}{-0.013889in}}%
\pgfpathlineto{\pgfqpoint{0.041667in}{0.013889in}}%
\pgfpathlineto{\pgfqpoint{0.013889in}{0.013889in}}%
\pgfpathlineto{\pgfqpoint{0.013889in}{0.041667in}}%
\pgfpathlineto{\pgfqpoint{-0.013889in}{0.041667in}}%
\pgfpathlineto{\pgfqpoint{-0.013889in}{0.013889in}}%
\pgfpathlineto{\pgfqpoint{-0.041667in}{0.013889in}}%
\pgfpathlineto{\pgfqpoint{-0.041667in}{-0.013889in}}%
\pgfpathlineto{\pgfqpoint{-0.013889in}{-0.013889in}}%
\pgfpathclose%
\pgfusepath{stroke,fill}%
}%
\begin{pgfscope}%
\pgfsys@transformshift{6.137401in}{2.471956in}%
\pgfsys@useobject{currentmarker}{}%
\end{pgfscope}%
\end{pgfscope}%
\begin{pgfscope}%
\definecolor{textcolor}{rgb}{0.000000,0.000000,0.000000}%
\pgfsetstrokecolor{textcolor}%
\pgfsetfillcolor{textcolor}%
\pgftext[x=6.387401in,y=2.423345in,left,base]{\color{textcolor}\sffamily\fontsize{10.000000}{12.000000}\selectfont MEAM\_LAMMPS\_JeongParkDo\_2018\_PdMo \protect{\cite{jeong2018second,MO356501945107001,MD249792265679001}}}%
\end{pgfscope}%
\begin{pgfscope}%
\pgfsetrectcap%
\pgfsetroundjoin%
\pgfsetlinewidth{1.505625pt}%
\definecolor{currentstroke}{rgb}{0.300443,1.000000,0.667299}%
\pgfsetstrokecolor{currentstroke}%
\pgfsetdash{}{0pt}%
\pgfpathmoveto{\pgfqpoint{5.998512in}{2.264138in}}%
\pgfpathlineto{\pgfqpoint{6.276290in}{2.264138in}}%
\pgfusepath{stroke}%
\end{pgfscope}%
\begin{pgfscope}%
\pgfsetbuttcap%
\pgfsetbeveljoin%
\definecolor{currentfill}{rgb}{1.000000,1.000000,1.000000}%
\pgfsetfillcolor{currentfill}%
\pgfsetlinewidth{1.003750pt}%
\definecolor{currentstroke}{rgb}{0.300443,1.000000,0.667299}%
\pgfsetstrokecolor{currentstroke}%
\pgfsetdash{}{0pt}%
\pgfsys@defobject{currentmarker}{\pgfqpoint{-0.039627in}{-0.033709in}}{\pgfqpoint{0.039627in}{0.041667in}}{%
\pgfpathmoveto{\pgfqpoint{0.000000in}{0.041667in}}%
\pgfpathlineto{\pgfqpoint{-0.009355in}{0.012876in}}%
\pgfpathlineto{\pgfqpoint{-0.039627in}{0.012876in}}%
\pgfpathlineto{\pgfqpoint{-0.015136in}{-0.004918in}}%
\pgfpathlineto{\pgfqpoint{-0.024491in}{-0.033709in}}%
\pgfpathlineto{\pgfqpoint{-0.000000in}{-0.015915in}}%
\pgfpathlineto{\pgfqpoint{0.024491in}{-0.033709in}}%
\pgfpathlineto{\pgfqpoint{0.015136in}{-0.004918in}}%
\pgfpathlineto{\pgfqpoint{0.039627in}{0.012876in}}%
\pgfpathlineto{\pgfqpoint{0.009355in}{0.012876in}}%
\pgfpathclose%
\pgfusepath{stroke,fill}%
}%
\begin{pgfscope}%
\pgfsys@transformshift{6.137401in}{2.264138in}%
\pgfsys@useobject{currentmarker}{}%
\end{pgfscope}%
\end{pgfscope}%
\begin{pgfscope}%
\definecolor{textcolor}{rgb}{0.000000,0.000000,0.000000}%
\pgfsetstrokecolor{textcolor}%
\pgfsetfillcolor{textcolor}%
\pgftext[x=6.387401in,y=2.215527in,left,base]{\color{textcolor}\sffamily\fontsize{10.000000}{12.000000}\selectfont MEAM\_LAMMPS\_WangOhLee\_2020\_CuMo \protect{\cite{wang2020second,MO486450342170001,MD249792265679001}}}%
\end{pgfscope}%
\begin{pgfscope}%
\pgfsetrectcap%
\pgfsetroundjoin%
\pgfsetlinewidth{1.505625pt}%
\definecolor{currentstroke}{rgb}{0.123340,1.000000,0.844402}%
\pgfsetstrokecolor{currentstroke}%
\pgfsetdash{}{0pt}%
\pgfpathmoveto{\pgfqpoint{5.998512in}{2.056321in}}%
\pgfpathlineto{\pgfqpoint{6.276290in}{2.056321in}}%
\pgfusepath{stroke}%
\end{pgfscope}%
\begin{pgfscope}%
\pgfsetbuttcap%
\pgfsetmiterjoin%
\definecolor{currentfill}{rgb}{1.000000,1.000000,1.000000}%
\pgfsetfillcolor{currentfill}%
\pgfsetlinewidth{1.003750pt}%
\definecolor{currentstroke}{rgb}{0.123340,1.000000,0.844402}%
\pgfsetstrokecolor{currentstroke}%
\pgfsetdash{}{0pt}%
\pgfsys@defobject{currentmarker}{\pgfqpoint{-0.036084in}{-0.041667in}}{\pgfqpoint{0.036084in}{0.041667in}}{%
\pgfpathmoveto{\pgfqpoint{0.000000in}{0.041667in}}%
\pgfpathlineto{\pgfqpoint{-0.036084in}{0.020833in}}%
\pgfpathlineto{\pgfqpoint{-0.036084in}{-0.020833in}}%
\pgfpathlineto{\pgfqpoint{-0.000000in}{-0.041667in}}%
\pgfpathlineto{\pgfqpoint{0.036084in}{-0.020833in}}%
\pgfpathlineto{\pgfqpoint{0.036084in}{0.020833in}}%
\pgfpathclose%
\pgfusepath{stroke,fill}%
}%
\begin{pgfscope}%
\pgfsys@transformshift{6.137401in}{2.056321in}%
\pgfsys@useobject{currentmarker}{}%
\end{pgfscope}%
\end{pgfscope}%
\begin{pgfscope}%
\definecolor{textcolor}{rgb}{0.000000,0.000000,0.000000}%
\pgfsetstrokecolor{textcolor}%
\pgfsetfillcolor{textcolor}%
\pgftext[x=6.387401in,y=2.007710in,left,base]{\color{textcolor}\sffamily\fontsize{10.000000}{12.000000}\selectfont MEAM\_LAMMPS\_WangOhLee\_2020\_CuMo \protect{\cite{wang2020second,MO380272712420001,MD249792265679001}}}%
\end{pgfscope}%
\begin{pgfscope}%
\pgfsetrectcap%
\pgfsetroundjoin%
\pgfsetlinewidth{1.505625pt}%
\definecolor{currentstroke}{rgb}{0.000000,0.833333,1.000000}%
\pgfsetstrokecolor{currentstroke}%
\pgfsetdash{}{0pt}%
\pgfpathmoveto{\pgfqpoint{5.998512in}{1.848504in}}%
\pgfpathlineto{\pgfqpoint{6.276290in}{1.848504in}}%
\pgfusepath{stroke}%
\end{pgfscope}%
\begin{pgfscope}%
\pgfsetbuttcap%
\pgfsetroundjoin%
\definecolor{currentfill}{rgb}{1.000000,1.000000,1.000000}%
\pgfsetfillcolor{currentfill}%
\pgfsetlinewidth{1.003750pt}%
\definecolor{currentstroke}{rgb}{0.000000,0.833333,1.000000}%
\pgfsetstrokecolor{currentstroke}%
\pgfsetdash{}{0pt}%
\pgfsys@defobject{currentmarker}{\pgfqpoint{-0.041667in}{-0.041667in}}{\pgfqpoint{0.041667in}{0.041667in}}{%
\pgfpathmoveto{\pgfqpoint{-0.041667in}{0.000000in}}%
\pgfpathlineto{\pgfqpoint{0.041667in}{0.000000in}}%
\pgfpathmoveto{\pgfqpoint{0.000000in}{-0.041667in}}%
\pgfpathlineto{\pgfqpoint{0.000000in}{0.041667in}}%
\pgfusepath{stroke,fill}%
}%
\begin{pgfscope}%
\pgfsys@transformshift{6.137401in}{1.848504in}%
\pgfsys@useobject{currentmarker}{}%
\end{pgfscope}%
\end{pgfscope}%
\begin{pgfscope}%
\definecolor{textcolor}{rgb}{0.000000,0.000000,0.000000}%
\pgfsetstrokecolor{textcolor}%
\pgfsetfillcolor{textcolor}%
\pgftext[x=6.387401in,y=1.799893in,left,base]{\color{textcolor}\sffamily\fontsize{10.000000}{12.000000}\selectfont Tersoff\_LAMMPS\_ZhangNguyen\_2021\_MoSe \protect{\cite{MO152208847456001,MD077075034781005}}}%
\end{pgfscope}%
\begin{pgfscope}%
\pgfsetrectcap%
\pgfsetroundjoin%
\pgfsetlinewidth{1.505625pt}%
\definecolor{currentstroke}{rgb}{0.000000,0.613725,1.000000}%
\pgfsetstrokecolor{currentstroke}%
\pgfsetdash{}{0pt}%
\pgfpathmoveto{\pgfqpoint{10.515897in}{2.887590in}}%
\pgfpathlineto{\pgfqpoint{10.793675in}{2.887590in}}%
\pgfusepath{stroke}%
\end{pgfscope}%
\begin{pgfscope}%
\pgfsetbuttcap%
\pgfsetroundjoin%
\definecolor{currentfill}{rgb}{1.000000,1.000000,1.000000}%
\pgfsetfillcolor{currentfill}%
\pgfsetlinewidth{1.003750pt}%
\definecolor{currentstroke}{rgb}{0.000000,0.613725,1.000000}%
\pgfsetstrokecolor{currentstroke}%
\pgfsetdash{}{0pt}%
\pgfsys@defobject{currentmarker}{\pgfqpoint{-0.041667in}{-0.041667in}}{\pgfqpoint{0.041667in}{0.041667in}}{%
\pgfpathmoveto{\pgfqpoint{-0.041667in}{-0.041667in}}%
\pgfpathlineto{\pgfqpoint{0.041667in}{0.041667in}}%
\pgfpathmoveto{\pgfqpoint{-0.041667in}{0.041667in}}%
\pgfpathlineto{\pgfqpoint{0.041667in}{-0.041667in}}%
\pgfusepath{stroke,fill}%
}%
\begin{pgfscope}%
\pgfsys@transformshift{10.654786in}{2.887590in}%
\pgfsys@useobject{currentmarker}{}%
\end{pgfscope}%
\end{pgfscope}%
\begin{pgfscope}%
\definecolor{textcolor}{rgb}{0.000000,0.000000,0.000000}%
\pgfsetstrokecolor{textcolor}%
\pgfsetfillcolor{textcolor}%
\pgftext[x=10.904786in,y=2.838979in,left,base]{\color{textcolor}\sffamily\fontsize{10.000000}{12.000000}\selectfont Morse\_Shifted\_GirifalcoWeizer\_1959LowCutoff\_Mo \protect{\cite{girifalco1959application,MO228581001644004,MD552566534109004}}}%
\end{pgfscope}%
\begin{pgfscope}%
\pgfsetrectcap%
\pgfsetroundjoin%
\pgfsetlinewidth{1.505625pt}%
\definecolor{currentstroke}{rgb}{0.000000,0.378431,1.000000}%
\pgfsetstrokecolor{currentstroke}%
\pgfsetdash{}{0pt}%
\pgfpathmoveto{\pgfqpoint{10.515897in}{2.679773in}}%
\pgfpathlineto{\pgfqpoint{10.793675in}{2.679773in}}%
\pgfusepath{stroke}%
\end{pgfscope}%
\begin{pgfscope}%
\pgfsetbuttcap%
\pgfsetmiterjoin%
\definecolor{currentfill}{rgb}{1.000000,1.000000,1.000000}%
\pgfsetfillcolor{currentfill}%
\pgfsetlinewidth{1.003750pt}%
\definecolor{currentstroke}{rgb}{0.000000,0.378431,1.000000}%
\pgfsetstrokecolor{currentstroke}%
\pgfsetdash{}{0pt}%
\pgfsys@defobject{currentmarker}{\pgfqpoint{-0.041667in}{-0.041667in}}{\pgfqpoint{0.041667in}{0.041667in}}{%
\pgfpathmoveto{\pgfqpoint{-0.020833in}{-0.041667in}}%
\pgfpathlineto{\pgfqpoint{0.000000in}{-0.020833in}}%
\pgfpathlineto{\pgfqpoint{0.020833in}{-0.041667in}}%
\pgfpathlineto{\pgfqpoint{0.041667in}{-0.020833in}}%
\pgfpathlineto{\pgfqpoint{0.020833in}{0.000000in}}%
\pgfpathlineto{\pgfqpoint{0.041667in}{0.020833in}}%
\pgfpathlineto{\pgfqpoint{0.020833in}{0.041667in}}%
\pgfpathlineto{\pgfqpoint{0.000000in}{0.020833in}}%
\pgfpathlineto{\pgfqpoint{-0.020833in}{0.041667in}}%
\pgfpathlineto{\pgfqpoint{-0.041667in}{0.020833in}}%
\pgfpathlineto{\pgfqpoint{-0.020833in}{0.000000in}}%
\pgfpathlineto{\pgfqpoint{-0.041667in}{-0.020833in}}%
\pgfpathclose%
\pgfusepath{stroke,fill}%
}%
\begin{pgfscope}%
\pgfsys@transformshift{10.654786in}{2.679773in}%
\pgfsys@useobject{currentmarker}{}%
\end{pgfscope}%
\end{pgfscope}%
\begin{pgfscope}%
\definecolor{textcolor}{rgb}{0.000000,0.000000,0.000000}%
\pgfsetstrokecolor{textcolor}%
\pgfsetfillcolor{textcolor}%
\pgftext[x=10.904786in,y=2.631162in,left,base]{\color{textcolor}\sffamily\fontsize{10.000000}{12.000000}\selectfont Morse\_Shifted\_GirifalcoWeizer\_1959MedCutoff\_Mo \protect{\cite{girifalco1959application,MO534363225491004,MD552566534109004}}}%
\end{pgfscope}%
\begin{pgfscope}%
\pgfsetrectcap%
\pgfsetroundjoin%
\pgfsetlinewidth{1.505625pt}%
\definecolor{currentstroke}{rgb}{0.000000,0.158824,1.000000}%
\pgfsetstrokecolor{currentstroke}%
\pgfsetdash{}{0pt}%
\pgfpathmoveto{\pgfqpoint{10.515897in}{2.471956in}}%
\pgfpathlineto{\pgfqpoint{10.793675in}{2.471956in}}%
\pgfusepath{stroke}%
\end{pgfscope}%
\begin{pgfscope}%
\pgfsetbuttcap%
\pgfsetmiterjoin%
\definecolor{currentfill}{rgb}{1.000000,1.000000,1.000000}%
\pgfsetfillcolor{currentfill}%
\pgfsetlinewidth{1.003750pt}%
\definecolor{currentstroke}{rgb}{0.000000,0.158824,1.000000}%
\pgfsetstrokecolor{currentstroke}%
\pgfsetdash{}{0pt}%
\pgfsys@defobject{currentmarker}{\pgfqpoint{-0.058926in}{-0.058926in}}{\pgfqpoint{0.058926in}{0.058926in}}{%
\pgfpathmoveto{\pgfqpoint{-0.000000in}{-0.058926in}}%
\pgfpathlineto{\pgfqpoint{0.058926in}{0.000000in}}%
\pgfpathlineto{\pgfqpoint{0.000000in}{0.058926in}}%
\pgfpathlineto{\pgfqpoint{-0.058926in}{0.000000in}}%
\pgfpathclose%
\pgfusepath{stroke,fill}%
}%
\begin{pgfscope}%
\pgfsys@transformshift{10.654786in}{2.471956in}%
\pgfsys@useobject{currentmarker}{}%
\end{pgfscope}%
\end{pgfscope}%
\begin{pgfscope}%
\definecolor{textcolor}{rgb}{0.000000,0.000000,0.000000}%
\pgfsetstrokecolor{textcolor}%
\pgfsetfillcolor{textcolor}%
\pgftext[x=10.904786in,y=2.423345in,left,base]{\color{textcolor}\sffamily\fontsize{10.000000}{12.000000}\selectfont Morse\_Shifted\_GirifalcoWeizer\_1959HighCutoff\_Mo \protect{\cite{girifalco1959application,MO666830945336004,MD552566534109004}}}%
\end{pgfscope}%
\begin{pgfscope}%
\pgfsetrectcap%
\pgfsetroundjoin%
\pgfsetlinewidth{1.505625pt}%
\definecolor{currentstroke}{rgb}{0.000000,0.000000,0.999109}%
\pgfsetstrokecolor{currentstroke}%
\pgfsetdash{}{0pt}%
\pgfpathmoveto{\pgfqpoint{10.515897in}{2.264138in}}%
\pgfpathlineto{\pgfqpoint{10.793675in}{2.264138in}}%
\pgfusepath{stroke}%
\end{pgfscope}%
\begin{pgfscope}%
\pgfsetbuttcap%
\pgfsetmiterjoin%
\definecolor{currentfill}{rgb}{1.000000,1.000000,1.000000}%
\pgfsetfillcolor{currentfill}%
\pgfsetlinewidth{1.003750pt}%
\definecolor{currentstroke}{rgb}{0.000000,0.000000,0.999109}%
\pgfsetstrokecolor{currentstroke}%
\pgfsetdash{}{0pt}%
\pgfsys@defobject{currentmarker}{\pgfqpoint{-0.035355in}{-0.058926in}}{\pgfqpoint{0.035355in}{0.058926in}}{%
\pgfpathmoveto{\pgfqpoint{-0.000000in}{-0.058926in}}%
\pgfpathlineto{\pgfqpoint{0.035355in}{0.000000in}}%
\pgfpathlineto{\pgfqpoint{0.000000in}{0.058926in}}%
\pgfpathlineto{\pgfqpoint{-0.035355in}{0.000000in}}%
\pgfpathclose%
\pgfusepath{stroke,fill}%
}%
\begin{pgfscope}%
\pgfsys@transformshift{10.654786in}{2.264138in}%
\pgfsys@useobject{currentmarker}{}%
\end{pgfscope}%
\end{pgfscope}%
\begin{pgfscope}%
\definecolor{textcolor}{rgb}{0.000000,0.000000,0.000000}%
\pgfsetstrokecolor{textcolor}%
\pgfsetfillcolor{textcolor}%
\pgftext[x=10.904786in,y=2.215527in,left,base]{\color{textcolor}\sffamily\fontsize{10.000000}{12.000000}\selectfont SW\_MX2\_KurniawanPetrieWilliams\_2021\_MoS \protect{\cite{MO677328661525000a,wen2017force,MO677328661525000,MD242389978788001}}}%
\end{pgfscope}%
\begin{pgfscope}%
\pgfsetrectcap%
\pgfsetroundjoin%
\pgfsetlinewidth{1.505625pt}%
\definecolor{currentstroke}{rgb}{0.000000,0.000000,0.749554}%
\pgfsetstrokecolor{currentstroke}%
\pgfsetdash{}{0pt}%
\pgfpathmoveto{\pgfqpoint{10.515897in}{2.056321in}}%
\pgfpathlineto{\pgfqpoint{10.793675in}{2.056321in}}%
\pgfusepath{stroke}%
\end{pgfscope}%
\begin{pgfscope}%
\pgfsetbuttcap%
\pgfsetroundjoin%
\definecolor{currentfill}{rgb}{1.000000,1.000000,1.000000}%
\pgfsetfillcolor{currentfill}%
\pgfsetlinewidth{1.003750pt}%
\definecolor{currentstroke}{rgb}{0.000000,0.000000,0.749554}%
\pgfsetstrokecolor{currentstroke}%
\pgfsetdash{}{0pt}%
\pgfsys@defobject{currentmarker}{\pgfqpoint{-0.020833in}{-0.020833in}}{\pgfqpoint{0.020833in}{0.020833in}}{%
\pgfpathmoveto{\pgfqpoint{0.000000in}{-0.020833in}}%
\pgfpathcurveto{\pgfqpoint{0.005525in}{-0.020833in}}{\pgfqpoint{0.010825in}{-0.018638in}}{\pgfqpoint{0.014731in}{-0.014731in}}%
\pgfpathcurveto{\pgfqpoint{0.018638in}{-0.010825in}}{\pgfqpoint{0.020833in}{-0.005525in}}{\pgfqpoint{0.020833in}{0.000000in}}%
\pgfpathcurveto{\pgfqpoint{0.020833in}{0.005525in}}{\pgfqpoint{0.018638in}{0.010825in}}{\pgfqpoint{0.014731in}{0.014731in}}%
\pgfpathcurveto{\pgfqpoint{0.010825in}{0.018638in}}{\pgfqpoint{0.005525in}{0.020833in}}{\pgfqpoint{0.000000in}{0.020833in}}%
\pgfpathcurveto{\pgfqpoint{-0.005525in}{0.020833in}}{\pgfqpoint{-0.010825in}{0.018638in}}{\pgfqpoint{-0.014731in}{0.014731in}}%
\pgfpathcurveto{\pgfqpoint{-0.018638in}{0.010825in}}{\pgfqpoint{-0.020833in}{0.005525in}}{\pgfqpoint{-0.020833in}{0.000000in}}%
\pgfpathcurveto{\pgfqpoint{-0.020833in}{-0.005525in}}{\pgfqpoint{-0.018638in}{-0.010825in}}{\pgfqpoint{-0.014731in}{-0.014731in}}%
\pgfpathcurveto{\pgfqpoint{-0.010825in}{-0.018638in}}{\pgfqpoint{-0.005525in}{-0.020833in}}{\pgfqpoint{0.000000in}{-0.020833in}}%
\pgfpathclose%
\pgfusepath{stroke,fill}%
}%
\begin{pgfscope}%
\pgfsys@transformshift{10.654786in}{2.056321in}%
\pgfsys@useobject{currentmarker}{}%
\end{pgfscope}%
\end{pgfscope}%
\begin{pgfscope}%
\definecolor{textcolor}{rgb}{0.000000,0.000000,0.000000}%
\pgfsetstrokecolor{textcolor}%
\pgfsetfillcolor{textcolor}%
\pgftext[x=10.904786in,y=2.007710in,left,base]{\color{textcolor}\sffamily\fontsize{10.000000}{12.000000}\selectfont MEAM\_LAMMPS\_LeeBaskesKim\_2001\_Mo \protect{\cite{lee2001second,MO805823015127000,MD249792265679001}}}%
\end{pgfscope}%
\begin{pgfscope}%
\pgfsetrectcap%
\pgfsetroundjoin%
\pgfsetlinewidth{1.505625pt}%
\definecolor{currentstroke}{rgb}{0.000000,0.000000,0.500000}%
\pgfsetstrokecolor{currentstroke}%
\pgfsetdash{}{0pt}%
\pgfpathmoveto{\pgfqpoint{10.515897in}{1.848504in}}%
\pgfpathlineto{\pgfqpoint{10.793675in}{1.848504in}}%
\pgfusepath{stroke}%
\end{pgfscope}%
\begin{pgfscope}%
\pgfsetbuttcap%
\pgfsetroundjoin%
\definecolor{currentfill}{rgb}{1.000000,1.000000,1.000000}%
\pgfsetfillcolor{currentfill}%
\pgfsetlinewidth{1.003750pt}%
\definecolor{currentstroke}{rgb}{0.000000,0.000000,0.500000}%
\pgfsetstrokecolor{currentstroke}%
\pgfsetdash{}{0pt}%
\pgfsys@defobject{currentmarker}{\pgfqpoint{-0.041667in}{-0.041667in}}{\pgfqpoint{0.041667in}{0.041667in}}{%
\pgfpathmoveto{\pgfqpoint{0.000000in}{-0.041667in}}%
\pgfpathcurveto{\pgfqpoint{0.011050in}{-0.041667in}}{\pgfqpoint{0.021649in}{-0.037276in}}{\pgfqpoint{0.029463in}{-0.029463in}}%
\pgfpathcurveto{\pgfqpoint{0.037276in}{-0.021649in}}{\pgfqpoint{0.041667in}{-0.011050in}}{\pgfqpoint{0.041667in}{0.000000in}}%
\pgfpathcurveto{\pgfqpoint{0.041667in}{0.011050in}}{\pgfqpoint{0.037276in}{0.021649in}}{\pgfqpoint{0.029463in}{0.029463in}}%
\pgfpathcurveto{\pgfqpoint{0.021649in}{0.037276in}}{\pgfqpoint{0.011050in}{0.041667in}}{\pgfqpoint{0.000000in}{0.041667in}}%
\pgfpathcurveto{\pgfqpoint{-0.011050in}{0.041667in}}{\pgfqpoint{-0.021649in}{0.037276in}}{\pgfqpoint{-0.029463in}{0.029463in}}%
\pgfpathcurveto{\pgfqpoint{-0.037276in}{0.021649in}}{\pgfqpoint{-0.041667in}{0.011050in}}{\pgfqpoint{-0.041667in}{0.000000in}}%
\pgfpathcurveto{\pgfqpoint{-0.041667in}{-0.011050in}}{\pgfqpoint{-0.037276in}{-0.021649in}}{\pgfqpoint{-0.029463in}{-0.029463in}}%
\pgfpathcurveto{\pgfqpoint{-0.021649in}{-0.037276in}}{\pgfqpoint{-0.011050in}{-0.041667in}}{\pgfqpoint{0.000000in}{-0.041667in}}%
\pgfpathclose%
\pgfusepath{stroke,fill}%
}%
\begin{pgfscope}%
\pgfsys@transformshift{10.654786in}{1.848504in}%
\pgfsys@useobject{currentmarker}{}%
\end{pgfscope}%
\end{pgfscope}%
\begin{pgfscope}%
\definecolor{textcolor}{rgb}{0.000000,0.000000,0.000000}%
\pgfsetstrokecolor{textcolor}%
\pgfsetfillcolor{textcolor}%
\pgftext[x=10.904786in,y=1.799893in,left,base]{\color{textcolor}\sffamily\fontsize{10.000000}{12.000000}\selectfont SNAP\_ZuoChenLi\_2019quadratic\_Mo \protect{\cite{zuo2020performance,MO692442138123000,MD536750310735000}}}%
\end{pgfscope}%
\end{pgfpicture}%
\makeatother%
\endgroup%